\documentclass[a4paper,11pt]{article}
\pdfoutput=1  

\usepackage{jheppub}  

\usepackage[T1]{fontenc}  
\usepackage{graphicx}
\usepackage{epstopdf}
\usepackage{bm}
\usepackage{epsfig}
\usepackage{graphics}
\usepackage{xspace}
\usepackage{hyperref}
\usepackage{slashed}
\usepackage{xcolor}
\usepackage{longtable}
\usepackage[small]{subfigure}
\usepackage[normalem]{ulem}

\newcommand{\nl}{n_{\ell}}
\newcommand{\mbar}{\overline{m}}
\newcommand{\MSR}{\mathrm{MSR}}
\newcommand{\pole}{\mathrm{pole}}
\newcommand{\MSb}{\overline{\mathrm{MS}}}
\newcommand{\Ord}{\mathcal{O}}
\newcommand{\LQCD}{\Lambda_\mathrm{QCD}}
\newcommand{\leri}[1]{\left(#1\right)}
\newcommand{\leris}[1]{\left[#1\right]}
\newcommand{\dd}{\mathrm{d}}
\newcommand{\PS}{\mathrm{PS}}
\newcommand{\OS}{\mathrm{1S}}

\title{\boldmath The MSR Mass  \\ and the \texorpdfstring{$\Ord(\LQCD)$}{Order Lambda QCD} Renormalon Sum Rule}

\preprint{\begin{flushright} UWThPh-2017-6\\ MIT-CTP 4896\\IFT-UAM/CSIC-17-034\end{flushright}\vspace*{-1cm}}

\author[a,b]{Andr\'e H. Hoang}
\author[c]{Ambar Jain}
\author[a]{Christopher Lepenik}
\author[d,e]{Vicent Mateu}
\author[a,f]{\mbox{Moritz Preisser}}
\author[g]{Ignazio Scimemi}
\author[f]{Iain W. Stewart}

\affiliation[a]{University of Vienna, Faculty of Physics,\\Boltzmanngasse 5, A-1090 Wien, Austria}
\affiliation[b]{Erwin Schr\"odinger International Institute for Mathematical Physics,\\
University of Vienna, Boltzmanngasse 9, A-1090 Wien, Austria}
\affiliation[c]{Indian Institute of Science Education and Research Bhopal,\\Bhopal Bypass Road, Bhopal 462066, India}
\affiliation[d]{Departamento de F\'isica Fundamental e IUFFyM,\\Universidad de Salamanca, E-37008 Salamanca, Spain}
\affiliation[e]{Instituto de F\'isica Te\'orica UAM-CSIC,\\E-28049 Madrid, Spain}
\affiliation[f]{Center for Theoretical Physics, Massachusetts Institute of Technology,\\Cambridge, MA 02139, USA}
\affiliation[g]{Departamento de F\'isica Te\'orica II, Universidad Complutense de Madrid (UCM),\\E-28040 Madrid, Spain}

\emailAdd{andre.hoang@univie.ac.at}
\emailAdd{ambarj@iiserb.ac.in}
\emailAdd{christopher.lepenik@univie.ac.at}
\emailAdd{vmateu@usal.es}
\emailAdd{moritz.preisser@univie.ac.at}
\emailAdd{iains@mit.edu}
\emailAdd{ignazios@fis.ucm.es}

\abstract{
We provide a detailed description and analysis of a low-scale short-distance mass scheme, called the MSR mass, that is useful for high-precision top quark mass determinations, but can be applied for any heavy quark $Q$. In contrast to earlier low-scale short-distance mass schemes, the MSR scheme has a direct connection to the well known $\MSb$ mass commonly used for high-energy applications, and is determined by heavy quark on-shell self-energy Feynman diagrams. Indeed, the MSR mass scheme can be viewed as the simplest extension of the $\MSb$ mass concept to renormalization scales $\ll m_Q$.  The MSR mass depends on a scale $R$ that can be chosen freely, and its renormalization group evolution has a linear dependence on $R$, which is known as R-evolution.  Using R-evolution for the MSR mass we provide details of the derivation of an analytic expression for the normalization of the ${\cal O}(\Lambda_{\rm QCD})$ renormalon asymptotic behavior of the pole mass in perturbation theory. This is referred to as the ${\cal O}(\Lambda_{\rm QCD})$ renormalon sum rule, and can be applied to any perturbative series.  The relations of the MSR mass scheme to other low-scale short-distance masses are analyzed as well.
}

\begin{document} 
\maketitle
\flushbottom

\section{Introduction}

Achieving higher precision in theoretical predictions in the framework of quantum chromo dynamics (QCD) is one of the main goals in high-energy physics and an essential ingredient in the indirect search for physics beyond the Standard Model. In this endeavor accurate determinations of the masses of the heavy charm, bottom and top quarks play an important role since they enter the description of many observables that are employed in consistency tests of the Standard Model and in the exploration of models of new physics. Because quark masses are formally-defined renormalized quantities and not physical observables, the quantities from which the heavy quark masses are extracted need to be computed in perturbative QCD to high order. Among the most precise recent high-order analyses to determine the heavy quark masses are QCD sum rules and the analysis of quarkonium energies for the charm and bottom quark masses~\cite{ 
Dehnadi:2011gc,Bodenstein:2011ma,Bodenstein:2011fv,Hoang:2012us,Chakraborty:2014aca, 
Colquhoun:2014ica,Beneke:2014pta,Ayala:2014yxa,Dehnadi:2015fra,Erler:2016atg}
and the top pair production threshold cross section at a future lepton collider for the top quark mass~\cite{Hoang:2000yr,Hoang:2013uda,Beneke:2015kwa}.
Over time all of these analyses have been continuously updated and improved by computations of new QCD corrections, and more are being designed and studied currently to also allow for more precise determinations of the top quark mass from available LHC data~\cite{Czakon:2013goa,Khachatryan:2016mqs,Aaboud:2016pbd,Czakon:2016ckf,Alioli:2013mxa,Frixione:2014ala,Chatrchyan:2013boa,
Kharchilava:1999yj}.

In all the analyses of Refs.~\cite{Dehnadi:2011gc,Bodenstein:2011ma,Bodenstein:2011fv,Hoang:2012us,Chakraborty:2014aca, 
Colquhoun:2014ica,Beneke:2014pta,Ayala:2014yxa,Dehnadi:2015fra,Erler:2016atg,Hoang:2000yr,Hoang:2013uda,Beneke:2015kwa} the use of short-distance mass schemes was essential to achieve a well-converging perturbative expansion and a precision in the mass determination well below the hadronization scale $\Lambda_{\rm QCD}\sim 200\,-\,300$\,MeV. The heavy quark pole mass $m_Q^{\pole}$, which is the perturbation theory equivalent of the rest mass of an on-shell quark, on the other hand, leads to a substantially worse perturbative behavior due to its linear infrared-sensitivity, also known as the ${\cal O}(\Lambda_{\rm QCD})$ renormalon problem~\cite{Bigi:1994em,Beneke:1994sw}, and was therefore not adopted as a relevant mass scheme for analyses where a precision better than $\Lambda_{\rm QCD}$ could be achieved. Nevertheless, the pole mass still served as an important intermediate mass scheme during computations because it determines the partonic (but unphysical) poles of heavy quark Green functions.  
Typical short-distance quark mass schemes which have been employed were the renormalization-scale dependent $\overline{\rm MS}$ mass $\mbar_Q(\mu)$ and so-called low-scale short-distance masses such as the kinetic mass~\cite{Czarnecki:1997sz}, the potential-subtracted (PS)~mass~\cite{Beneke:1998rk}, the 1S mass~\cite{Hoang:1998ng,Hoang:1998hm,Hoang:1999ye}, the renormalon-subtracted (RS) mass~\cite{Pineda:2001zq} or the jet mass~\cite{Jain:2008gb,Fleming:2007qr}. 
The basic difference between the  $\overline{\rm MS}$ mass to the low-scale short-distance mass schemes is that the perturbative coefficients of its relation to the pole mass scale linearly with the heavy quark mass,
$\mbar_Q(\mu)\,-\,m_Q^{\pole}\sim m_Q(\alpha_s+\ldots)$, while for the low-scale short-distance mass schemes 
the corresponding series scales linearly with a scale $R\ll m_Q$. This feature enables the low-scale short-distance quark mass schemes to be used for predictions of quantities where the heavy quark dynamics is non-relativistic in nature and fluctuations at the scale of $m_Q$ are integrated out. This is because radiative corrections to the mass in such quantities involve physical scales much smaller than $m_Q$. One very prominent example in the context of top quark physics is the non-relativistic heavy quarkonium dynamics inherent to the top-antitop pair production cross section at threshold at a future lepton collider~\cite{Hoang:2000yr,Hoang:2013uda,Beneke:2015kwa}, where the most important dynamical scale is the inverse Bohr radius $m_t\,\alpha_s\sim 25$\,GeV~$\ll m_t$. On the other hand, the  $\overline{\rm MS}$ mass is a good scheme choice for quantities that involve energies much larger than $m_Q$, such as for high-energy total cross sections, or when the massive quark causes virtual and off-shell effects. This is because in such cases the heavy quark mass yields corrections that either scale with positive or negative powers of $m_Q$ such that QCD corrections associated with the mass have a scaling that is linear in $m_Q$ as well.
The difference between the $\overline{\rm MS}$ mass and the low-scale short-distance masses is most important for the case of the top quark because  in this case the difference between $m_t$ and the dynamical low-energy scales can be very large numerically.

For the top quark mass there are excellent prospects for very precise measurements in low-scale short-distance schemes such as the PS mass or the 1S mass from the top-antitop threshold inclusive cross section at a future lepton collider~\cite{Hoang:2000yr,Hoang:2013uda,Beneke:2015kwa}. Current studies indicate that a precision well below $50$\,MeV can be achieved accounting for theoretical as well as experimental uncertainties~\cite{Seidel:2013sqa,Horiguchi:2013wra,Vos:2016til}. Currently, the most precise measurements of the top quark mass come from reconstruction analyses at the LHC~\cite{Khachatryan:2015hba,Aaboud:2016igd} and the Tevatron~\cite{Tevatron:2014cka} and have uncertainties at the level of $500$\,MeV or larger. Moreover, the mass is obtained from multivariate fits involving multipurpose Monte Carlo (MC) event generators and thus represents a determination of the top quark mass parameter $m_t^{\rm MC}$ contained in the particular MC event generator. Recently, a first high-precision analysis on how the MC top quark mass parameter can be related to a field theoretically well-defined short-distance top quark mass was provided in Refs.~\cite{Butenschoen:2016lpz,Hoang:2017kmk} and general considerations on the relation were discussed in Ref.~\cite{Hoang:2008xm,Hoang:2014oea}. For the analysis, hadron level predictions for the 2-jettiness distribution~\cite{Stewart:2010tn} for electron-positron collisions and ${\cal O}(\alpha_s)$ QCD corrections together with the resummation of large logarithms at next-to-next-to leading order~\cite{Fleming:2007qr,Fleming:2007tv,Hoang:2007vb} were employed. Since the 2-jettiness distribution is closely related to the invariant mass distribution of a single reconstructed top quark, the relevant dynamical scales inherent to the problem are governed by the width of the mass distribution which amounts to only about $5$\,GeV in the peak region of the distribution where the sensitivity to the top mass is the highest. Interestingly, as was shown in Ref.~\cite{Butenschoen:2016lpz}, the dynamical scales increase continuously considering the 2-jettiness distribution further away from the peak. In the analysis of~\cite{Butenschoen:2016lpz} the MSR mass scheme $m_Q^{\rm MSR}(R)$ was employed which depends on a scale $R$ and for which the dependence on $R$ is described by a renormalization group flow such that $R$ can be continuously adapted according to which part of the distribution is predicted. Other applications of the MSR mass using a flavor number dependent evolution in $R$ to account for the mass effects of lighter quarks were given in Ref.~\cite{Hoang:2017btd,Mateu:2017hlz}. In contrast to the $\mu$-dependent $\MSb$ mass $\mbar_Q(\mu)$, which evolves only logarithmically in $\mu$, the MSR mass has logarithmic as well as linear dependence on $R$. 

The MSR mass scheme was succinctly introduced in Ref.~\cite{Hoang:2008yj} and discussed conceptually in Ref.~\cite{Hoang:2014oea}, but a detailed discussion has so far not been provided. A key purpose of this paper is to provide sufficient details such that phenomenological MSR mass analyses, such as the results of Ref.~\cite{Butenschoen:2016lpz}, can be easily related to other common short-distance mass schemes that are being used in the literature.

The definition of the MSR mass given by the perturbative series for the MSR-pole mass difference $m_Q^{\rm MSR}(R)-m_Q^{\pole}$ is obtained directly from the $\overline{\rm MS}$-pole mass relation $\mbar_Q(\mbar_Q)-m_Q^{\pole}$ and is therefore the only low-scale short-distance mass suggested in the literature that is derived directly from on-shell heavy quark self-energy diagrams just like the $\MSb$ mass.\footnote{The name `MSR mass' arises from a combination of
the letters `MS' standing for the close relation to the $\MSb$ mass and the letter `R' standing for R-evolution.} The MSR mass thus automatically
inherits the clean and good infrared properties of the $\overline{\rm MS}$ mass. Furthermore, by construction, the MSR mass matches to the $\overline{\rm MS}$ mass for $R=\mbar_Q(\mbar_Q)$ and is known to the same order as the series of $\mbar_Q(\mbar_Q)\,-\,m_Q^{\pole}$ without any further effort, which is currently ${\cal O}(\alpha_s^4)$ from the results of Refs.~\cite{Tarrach:1980up,Gray:1990yh,Melnikov:2000qh,Chetyrkin:1999ys,Chetyrkin:1999qi,Marquard:2007uj,Marquard:2015qpa,Marquard:2016dcn}. As already argued in Refs.~\cite{Hoang:2008yj,Hoang:2008xm}, the MSR mass can therefore be considered as the natural modification of the ``running'' $\overline{\rm MS}$ mass scheme concept for renormalization scales below $m_Q$, where the logarithmic evolution of the regular $\overline{\rm MS}$ mass is known to be unphysical. 

Since the MSR mass is designed to be employed for scales $R< m_Q$, it can be useful -- for applications where a clean treatment of virtual massive-flavor effects is important -- to integrate out the virtual effects of the massive quark $Q$ from the MSR mass definition. We therefore introduce two types of MSR masses, one where the virtual effects of the massive quark $Q$ are integrated out, called the {\it natural MSR mass}, and one where these effects are not integrated out, called the {\it practical MSR mass}. The 
difference between these two versions of the MSR mass is quite small and very well behaved for all $R$ values in the perturbative region, and the practical definition should be perfectly fine for most phenomenological applications. But the natural definition has conceptual advantages as its evolution for scales $R<m_Q$ does not include the virtual effects of the massive quark $Q$, which is conceptually cleaner since these belong physically to the scale $m_Q$.

We note that the R-evolution concept of a running heavy quark mass scheme for scales $R<m_Q$ elaborated in Ref.~\cite{Hoang:2008yj} has already been suggested a long time ago in Refs.~\cite{Voloshin:1992wg,Bigi:1997fj}. The R-evolution equation we discuss for the MSR mass was already quoted explicitly for the renormalization group evolution of the kinetic mass~\cite{Czarnecki:1997sz} at ${\cal O}(\alpha_s)$ in these references, but the conceptual implications of R-evolution and its connection to the   
${\cal O}(\Lambda_{\rm QCD})$ renormalon problem in the perturbative relations between short-distance masses and the pole mass were first studied systematically in Ref.~\cite{Hoang:2008yj}.
The second main purpose of this paper is to give further details on R-evolution and also to discuss its relation to the Borel transformation focusing mainly on the case of the MSR mass.
We note that the concept of R-evolution is quite general and can in principle be applied to any short-distance mass which depends on a variable infrared cutoff scale (such as the PS and the RS masses) or to cutoff-dependent QCD matrix elements with arbitrary dimensions. In fact, R-evolution has already been examined and applied in a number of other applications which include the factorization-scale dependence in the context of the operator product expansion~\cite{Hoang:2009yr}, the scale dependence of the non-perturbative soft radiation matrix element in  high-precision determinations of the strong coupling from $e^+e^-$ event-shape distributions~\cite{Abbate:2010xh,Abbate:2012jh,Hoang:2014wka,Hoang:2015hka}, even accounting for the finite mass effects of light quarks~\cite{Gritschacher:2013tza,Pietrulewicz:2014qza} and hadrons~\cite{Mateu:2012nk,Hoang:2014wka}.

The basic feature of the R-evolution concept is that for the difference of MSR masses at two scales, $m_Q^{\rm MSR}(R)-m_Q^{\rm MSR}(R^\prime)$, its linear dependence on the renormalization scale provides, completely within perturbation theory, a resummation of the terms in the asymptotic series associated to the pole-mass renormalon ambiguity to all orders. The R-evolution then resums the factorially growing terms in a systematic way that is ${\cal O}(\Lambda_{\rm QCD})$-renormalon free and, at the same time also sums all large logarithms that arise if $R$ and $R^\prime$ are widely separated. This cannot be achieved by more common purely logarithmic renormalization group equations, but is fully compatible with a Wilsonian renormalization group setup. We note that the summations carried out by
the R-evolution was achieved prior to Ref.~\cite{Hoang:2008yj} for the RS mass in~\cite{Bali:2003jq} (see also Ref.~\cite{Campanario:2005np}). Their method (and the RS mass) is based on using an approximate expression for the Borel transform function. The summation for a difference
of RS masses (for scales $R$ and $R^\prime$) is obtained by computing the inverse Borel integral over the difference of the two respective Borel functions. This method and R-evolution lead to consistent results, but the R-evolution does not rely on the knowledge of the Borel functions. 

The essential and probably most interesting conceptual feature of the perturbative series of the \mbox{R-evolution} equations is that it provides a systematic reordering of the terms in the asymptotic series associated to the ${\cal O}(\Lambda_{\rm QCD})$ renormalon ambiguity in leading, subleading, subsubleading, etc.\ contributions. So using the analytic solution of the \mbox{R-evolution} equations allows one to derive analytically (i.e.\ without any numerical procedure or modeling) the Borel-transform of a given perturbative series from the perspective that it carries an ${\cal O}(\Lambda_{\rm QCD})$ renormalon ambiguity. As a result one can rigorously derive an analytic expression for the normalization of the non-analytic terms in the Borel transform that are characteristic for the ${\cal O}(\Lambda_{\rm QCD})$ renormalon. The analytic result for this normalization factor was already given and discussed in Ref.~\cite{Hoang:2008yj}, but no details on the derivation were provided. We take the opportunity to show the details of the derivation here. We call the analytic result for the normalization of the ${\cal O}(\Lambda_{\rm QCD})$ renormalon ambiguity the {\it ${\cal O}(\Lambda_{\rm QCD})$ sum rule}, because it can be quickly applied to any given perturbative series.
To demonstrate the use and the high sensitivity of the ${\cal O}(\Lambda_{\rm QCD})$ renormalon sum rule we apply it also to a number of other cases, pointing out subtleties in its application to avoid inconsistencies and misinterpretations of the results. 

We note that also other methods to determine the normalization factor have been used.
In Ref.~\cite{Pineda:2001zq} it was determined from a computation of the residue of the Borel transform of the series following a proposal in Ref.~\cite{Lee:1996yk}. This approach, which we call {\it Borel method} can also be carried out analytically and provides the correct result, but has been observed to converge very slowly. We can identify the reason for this analytically from the solutions for the R-evolution equations, and we also discuss the connection of this method to our ${\cal O}(\Lambda_{\rm QCD})$ sum rule based on explicit analytic expressions.
In Ref.~\cite{Bali:2013pla} the normalization factor was computed taking the
ratio of the $n$-th term of the series to the asymptotic behavior. This {\it ratio method} converges very fast and provides results very similar to the ${\cal O}(\Lambda_{\rm QCD})$ sum rule. Recently, the ratio method was applied 
in Ref.~\cite{Beneke:2016cbu}, accounting for the ${\cal O}(\alpha_s^4)$ corrections to the pole-$\overline{\rm MS}$ mass relation~\cite{Marquard:2015qpa,Marquard:2016dcn}. We show that our ${\cal O}(\Lambda_{\rm QCD})$ sum rule provides results that are in full agreement with the ones obtained in Ref.~\cite{Beneke:2016cbu} and also leads to very similar uncertainties.

The paper is organized as follows: In Sec.~\ref{sec:msrsetup}
we provide the definition of the natural and practical MSR masses, $m_Q^{\rm MSRn}$ and $m_Q^{\rm MSRp}$, based on the perturbative series of the $\MSb$-pole mass relation $\mbar_Q(\mbar_Q)-m_Q^{\pole}$, and we also analyze the difference between these two MSR masses. This section provides the conventions we use for the coefficients of perturbative series, but it can otherwise be skipped by the reader not interested in the MSR masses.
In Sec.~\ref{sec:Revolution} we present the R-evolution equations which describe the scale dependence of the MSR masses and we also show explicitly how the solutions of the \mbox{R-evolution} equations sum large logarithms together with the high-order asymptotic series terms related to the ${\cal O}(\Lambda_{\rm QCD})$ renormalon. We in particular show for the top quark mass under which conditions the use of the R-evolution equations and its resummation is essential and superior to renormalon-free fixed-order perturbation theory, which does not sum any large logarithms. To our knowledge, such an analysis has not been provided in the literature before.
We also point out that the solution of the R-evolution equations is intrinsically related to carrying out an inverse Borel transform over differences of functions in the Borel plane such that the singularities  
related to the ${\cal O}(\Lambda_{\rm QCD})$ renormalon cancel. In Sec.~\ref{sec:sumrule} we present the analytic derivation of the ${\cal O}(\Lambda_{\rm QCD})$ renormalon sum rule and demonstrate its utility by a detailed analysis concerning the normalization of the ${\cal O}(\Lambda_{\rm QCD})$ renormalon ambiguity in the series for the difference of the pole mass and the MSR masses. 
The derivation of the sum rule allows to derive a new alternative expression for the high-order asymptotic behavior of a series that contains an ${\cal O}(\Lambda_{\rm QCD})$ renormalon which we discuss as well.
To demonstrate the high sensitivity of the sum rule and to explain  its consistent (and inconsistent) application we discuss its strong flavor number dependence and apply it to the massive quark vacuum polarization function, the series for the PS mass-pole mass difference, the QCD $\beta$-function, and the hadronic R-ratio. This section can be bypassed by the reader not interested in applications of the ${\cal O}(\Lambda_{\rm QCD})$ sum rule, but we note that Sec.~\ref{sec:PSmassIR} discusses implications for the PS mass that are relevant for Sec.~\ref{sec:othermasses} and may be important for high-precision top quark mass determinations.
Some subtle issues in the relation of the MSR masses to the PS, 1S and $\MSb$ masses are discussed in Sec.~\ref{sec:othermasses}. Finally, we conclude in Sec.~\ref{sec:conclusions}.
The paper also contains two appendices. In App.~\ref{app:coefficients} we specify our convention for the QCD $\beta$-function coefficients and present a number of expressions and formulae for coefficients, quantities and matching relations that arise in the discussion of R-evolution, the ${\cal O}(\Lambda_{\rm QCD})$ renormalon and on various mass definitions throughout this paper. In App.~\ref{sec:N12alternative} we provide details on the relation of the Borel method and our sum rule method to determine the normalization of the ${\cal O}(\Lambda_{\rm QCD})$ renormalon ambiguity of the pole mass.
Finally, in App.~\ref{sec:othermasscoeff} we quote the coefficients that define the PS and the 1S masses for the convenience of the reader and also show how the MSR masses can be obtained from a given value of the 1S mass in the non-relativistic and  $\Upsilon$-expansion counting scheme~\cite{Hoang:1998hm,Hoang:1998ng}.

\section{MSR Mass Setup}
\label{sec:msrsetup}

\subsection{Basic Idea of the MSR Mass}

The $\overline{\rm MS}$ mass $\mbar_Q(\mu)$ serves as the standard short-distance mass scheme for many high-energy applications with physical scales of the order or larger than the mass of the quark $Q$. It relies on the subtraction of the $1/\epsilon$ divergences in the common  $\overline{\rm MS}$ scheme in the on-shell self-energy corrections calculated in dimensional regularization. Despite the fact that it is an unphysical (i.e.\ theoretically designed) mass definition, it is infrared-safe and gauge invariant to all orders~\cite{Tarrach:1980up,Kronfeld:1998di} and its series relation to the pole mass $m_Q^{\pole}$ thus serves as the cleanest way to precisely quantify the renormalon ambiguity of the pole mass. The relation of $\mbar_Q \equiv \mbar_Q^{(n_\ell+1)}(\mbar_Q^{(n_\ell+1)})$
to the pole mass in the approximation that the masses of all quarks lighter than $Q$ are zero reads
\begin{equation}\label{eqn:msbarpoleseries}
 m_Q^{\pole} - \mbar_Q = \mbar_Q\,\sum_{n=1}^\infty\,a_n^{\overline{\rm MS}} (\nl,n_h)\,\bigg(\frac{\alpha_s^{(n_\ell+1)}(\mbar_Q)}{4\pi}\bigg)^{\!\!n} \,,
\end{equation}
with
\begin{align}\label{eqn:coeffanmsbar}
 a_1^{\overline{\rm MS}}(\nl,n_h) &= {\textstyle \frac{16}{3}}\,,\\
 a_2^{\overline{\rm MS}}(\nl,n_h) &= 213.437 + 1.65707\, n_h - 16.6619\, \nl\,,\nonumber\\
 a_3^{\overline{\rm MS}}(\nl,n_h) &= 12075. + 118.986\, n_h + 4.10115\, n_h^2 - 1707.35\, \nl + 1.42358\, n_h\, \nl + 41.7722\, \nl^2\,,\nonumber\\
 a_4^{\overline{\rm MS}}(\nl,n_h) &= (911588.\pm 417.) + (1781.61\pm 30.72)\,n_h - (60.1637\pm 0.6912)\,n_h^2 \nonumber\\ &\quad- (231.201\pm 0.102)\,n_h\,\nl - (190683.\pm10.)\,\nl + 9.25995\,n_h^2\,\nl \nonumber\\&\quad + 6.35819\,n_h^3 + 4.40363\,n_h\,\nl^2   + 11105.\,\nl^2 - 
 173.604\,\nl^3\,,\nonumber
\end{align}
where $\alpha_s^{(n_f)}$ stands for the strong coupling that renormalization-group (RG) evolves with $n_f$ active flavors, see Eq.~\eqref{eqn:betafct}. The coefficients $a_n^{\overline{\rm MS}}$ at ${\cal O}(\alpha_s,\alpha_s^2,\alpha_s^3)$ are known analytically from Refs.~\cite{Tarrach:1980up, Gray:1990yh,Chetyrkin:1999ys,Chetyrkin:1999qi,Melnikov:2000qh,Marquard:2007uj}. The ${\cal O}(\alpha_s^4)$ coefficient $a_4^{\overline{\rm MS}}$ was determined numerically in Refs.~\cite{Marquard:2015qpa,Marquard:2016dcn}, and the quoted numerical uncertainties have been taken from Ref.~\cite{Marquard:2016dcn}. Using the method of Ref.~\cite{Kataev:2015gvt} the uncertainties of the $\nl$-dependent terms may be further reduced. Using renormalon calculus~\cite{Bigi:1994em,Beneke:1994sw,Beneke:1998ui} one can show that the high-order asymptotic behavior series of Eq.~\eqref{eqn:msbarpoleseries} has an ambiguity of order $\Lambda_{\rm QCD}^{(n_\ell)}$, which depends on the number of massless quarks (indicated by the superscript) but is {\it independent} of the actual value of  $\mbar_Q$. 

A coherent treatment of the mass effects of lighter quarks is beyond the scope of this paper, and we therefore use the approximation that all flavors lighter than $Q$ are massless. These mass corrections come from the insertion of massive virtual quark loops in the self-energy Feynman diagrams and start at ${\cal O}(\alpha_s^2)$. At this order and at ${\cal O}(\alpha_s^3)$ the mass corrections from the virtual massive quark loops have been calculated analytically for all mass values in Ref.~\cite{Gray:1990yh} and \cite{Bekavac:2007tk}, respectively.  The dominant linear mass corrections at ${\cal O}(\alpha_s^3)$ were determined in Ref.~\cite{Hoang:2000fm}.
At ${\cal O}(\alpha_s^4)$ and the mass corrections are not yet known, but the corrections in the limit of large virtual quark masses are encoded in the ultraheavy flavor threshold matching relations of the RG-evolution $\mbar_Q(\mu)$ at scales above $m_Q$~\cite{Chetyrkin:1997un}.

The idea of the MSR mass is based on the fact that the ${\cal O}(\Lambda_{\rm QCD})$ ambiguity of the perturbative series on the RHS of Eq.~\eqref{eqn:msbarpoleseries} does not depend on the value $\mbar_Q$, as already mentioned above. This is an exact mathematical statement within the context of the calculus for asymptotic series and means that we can replace the term $\mbar_Q$ by the arbitrary scale $R$ on the RHS of Eq.~\eqref{eqn:msbarpoleseries} and use the resulting perturbative series as the definition of the $R$-dependent MSR mass scheme. It was pointed out in Ref.~\cite{Hoang:2014oea} that, for a given value of $R$, one can also interpret the MSR mass field theoretically as having a mass renormalization constant that contains the on-shell self-energy corrections of the pole mass only for scales larger than $R$. In other words, the pole mass and the MSR mass at the scale $R$ differ by self-energy corrections from scales below $R$: while the pole mass absorbs all self-energy corrections for quantum fluctuations up to scales $m_Q$, the MSR mass at the scale $R$ absorbs only self-energy corrections between $R$ and $m_Q$. Since the pole mass renormalon problem is related to the self-energy corrections from the scale $\Lambda_{\rm QCD}< R$, this explains why the MSR mass is a short-distance mass. In this illustrative context the $\MSb$ mass absorbs no self-energy corrections up to the scale $m_Q$. 
Since the scale $R$ is variable, the MSR mass can serve as a short-distance mass definition for applications governed by different physical scales and thus can also interpolate between them. Since the MSR mass is expected to have applications primarily for $R<m_Q$, it is further suitable to change the scheme from $n_\ell+1$ dynamical flavors, which includes the UV effects of the quark $Q$, to a scheme with $n_\ell$ dynamical flavors. This can be achieved in two ways, either by simply rewriting $\alpha_s^{(n_\ell+1)}$ in terms of $\alpha_s^{(n_\ell)}$, or by integrating out the virtual loop corrections of the quark $Q$. This results in two different ways to define the MSR mass, where we call the former the {\it practical MSR mass} and the latter the {\it natural MSR mass}, either one having advantages depending on the  application.  

We note that the notion of a scale-dependent short-distance mass which was first suggested in Refs.~\cite{Voloshin:1992wg,Bigi:1997fj}
has also been adopted for the kinetic~\cite{Czarnecki:1997sz},
the PS~\cite{Beneke:1998rk}, RS~\cite{Pineda:2001zq} and jet masses~\cite{Jain:2008gb,Fleming:2007tv}. However, none of these short-distance masses is defined directly from the on-shell self-energy diagrams of the massive quark $Q$ such as the MSR mass. This has a number of advantages, for example when discussing heavy flavor symmetry properties in the pole-$\MSb$ mass relation of different heavy quarks.

\subsection{Natural MSR Mass}
\label{sec:MSRn}

The \emph{natural MSR mass} definition is obtained by integrating out the corrections from the heavy quark $Q$ virtual loops in the self-energy diagrams of the massive quark $Q$, such that its relation to the pole mass reads
\begin{equation}\label{eqn:msrpolenat}
 m_Q^{\pole} - m_Q^{\rm MSRn}(R) = R\,\sum_{n=1}^\infty\,a_n^{\overline{\rm MS}} (\nl,0)\,\bigg(\frac{\alpha_s^{(\nl)}(R)}{4\pi}\bigg)^{\!\!n} \;,
\end{equation}
where the coefficients are given in Eq.~\eqref{eqn:coeffanmsbar}.
The natural MSR mass only accounts for gluonic and massless quark corrections, and has a non-trivial matching relation to the $\overline{\rm MS}$ mass.
The matching between the natural MSR mass and the $\overline{\rm MS}$ mass can be derived from the relation [\,\mbox{$\mbar_Q \equiv\mbar_Q^{(n_\ell+1)}(\mbar_Q^{(n_\ell+1)})$}\,]
\begin{equation}\label{eqn:msbmsr1}
  \!\!m_Q^{\rm MSRn}(\mbar_Q) - \mbar_Q
  = \mbar_Q \!\sum_{k=1}^{\infty}\!\bigg[ a_k^{\overline{\rm MS}}(\nl,1)\!\bigg(\frac{\alpha_s^{(\nl+1)}(\mbar_Q)}{4\pi}\bigg)^{\!\!k} \!- a_k^{\overline{\rm MS}}(\nl,0)\!\bigg(\frac{\alpha_s^{(\nl)}(\mbar_Q)}{4\pi}\bigg)^{\!\!k}\,\bigg],   
\end{equation}
and will be discussed in more detail in Sec.~\ref{sec:MSbar}.

We note that, formally, the natural MSR mass (as well as the practical MSR mass discussed in the next subsection) agrees with the pole mass in the limit $R\to 0$. However, taking this limit is ambiguous as it involves evolving through the Landau pole of the strong coupling and dealing with its non-perturbative definition for $|R\,|<\LQCD$. This issue is a manifestation of the renormalon problem of the pole mass.

\subsection{Practical MSR Mass}
\label{sec:MSRp}

The \emph{practical MSR mass} definition is directly related to the $\overline{\rm MS}$-pole perturbative series of Eq.~\eqref{eqn:msbarpoleseries}. To obtain its defining series one rewrites $\alpha_s^{(\nl+1)}(\mbar_Q)$ as a series in $\alpha_s^{(\nl)}(\mbar_Q)$ in Eq.~\eqref{eqn:msbarpoleseries} using the matching relation given in Eq.~\eqref{eqn:asmatchingmsbar} and then replaces $\mbar_Q$ by $R$, obtaining
\begin{equation}\label{eqn:msrpoleprac}
 m_Q^{\pole} - m_Q^{\rm MSRp}(R) =  R\,\sum_{n=1}^\infty\,a_n^{\rm MSRp} (\nl)\,\bigg(\frac{\alpha_s^{(\nl)}(R)}{4\pi}\bigg)^{\!\!n} , \\
\end{equation}
with
\begin{align}\label{eqn:coeffsanmsrp}
 a_1^{\rm MSRp}(\nl) & = {\textstyle \frac{16}{3}} \,,\\
 a_2^{\rm MSRp}(\nl) & = 215.094 - 16.6619\, \nl \,, \nonumber\\
 a_3^{\rm MSRp}(\nl) &= 12185. - 1705.93\, \nl + 41.7722\, \nl^2 \,,\nonumber\\
 a_4^{\rm MSRp}(\nl) &= (911932.\pm 418.) - (190794.\pm 10.)\, \nl + 11109.4\, \nl^2 - 173.604\, \nl^3 \,.\nonumber
\end{align}
The practical MSR mass still accounts for the virtual corrections from the massive quark Q with an evolving mass $R$ and has the convenient feature that it agrees with the $\overline{\rm MS}$ mass at the scale of the mass to all orders in perturbation theory [\,\mbox{$\mbar_Q \equiv\mbar_Q^{(n_\ell+1)}(\mbar_Q^{(n_\ell+1)})$}\,]:
\begin{equation}\label{eqn:msrpmsbarmatch}
 m^{\rm MSRp}_Q(m^{\rm MSRp}_Q) = \mbar_Q(\mbar_Q) \,.
\end{equation}

The formula for the difference of the natural and practical MSR masses at the same scale $R$ up to $\Ord(\alpha_s^4)$ reads
\begin{align}\label{eqn:msrdiff}
	&m_Q^\mathrm{MSRn}(R)-m_Q^\mathrm{MSRp}(R)\,=\,R\,\bigg[\,1.65707\,\bigg(\frac{\alpha_s^{(\nl)}(R)}{4\pi}\bigg)^{\!\!2}
	+ \big(110.050+1.4236\,\nl\big)\bigg(\frac{\alpha_s^{(\nl)}(R)}{4\pi}\bigg)^{\!\!3}
	\nonumber \\
	& \qquad
	+\,\big( (344.\pm 31.) 
    - (111.59\pm 0.10)\,\nl 
    +4.40\,\nl^2\big)\bigg(\frac{\alpha_s^{(\nl)}(R)}{4\pi}\bigg)^{\!\!4}
    + \dots\bigg] .
\end{align}
\begin{figure}
\center
\includegraphics[width=0.53\textwidth]{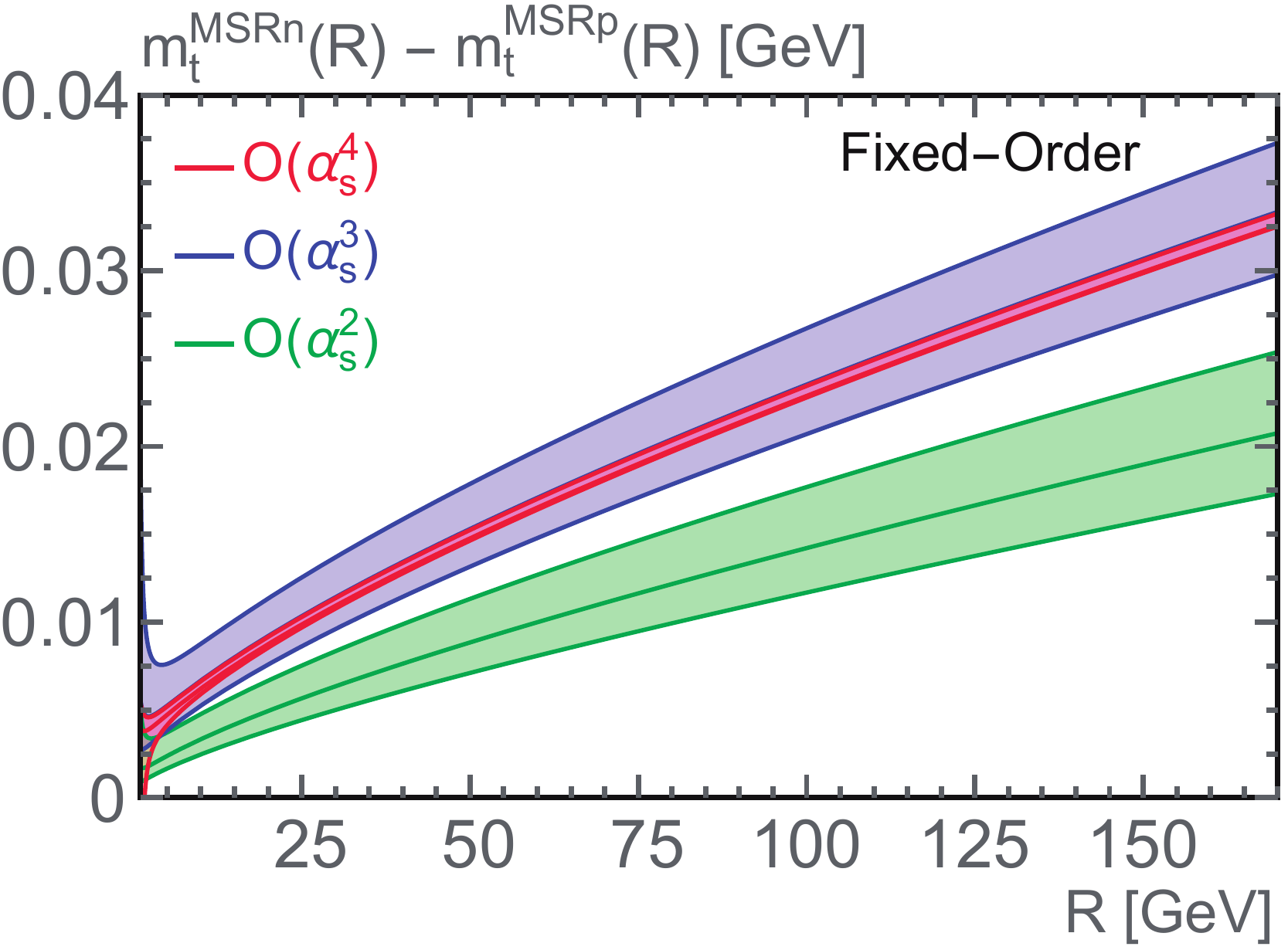}
 \caption{\label{fig:msrnminusmsrp} Difference of the natural and practical MSR top quark masses ($\nl=5$) as a function of $R$ in GeV at two, three and four loop order (the one loop result vanishes). The uncertainty bands are obtained from scale variations in $\alpha_s(\mu)$ with $R/2<\mu<2R$.}
\end{figure}
In Fig.~\ref{fig:msrnminusmsrp} the difference between the natural and the practical MSR top quark masses $m_t^\mathrm{MSRn}(R)-m_t^\mathrm{MSRp}(R)$ is shown for $R$ between $1$ and $170$\,GeV (here $\nl=5$).\footnote{Throughout this article we use $\alpha_s^{(n_f=5)}(m_Z)=0.118$ and $m_Z=91.187\,$GeV.} The numerical difference between these two masses is quite small. The natural MSR mass is larger than the practical MSR mass and the difference increases with $R$ reaching about $30$\,MeV at $R=170$\,GeV. The error bands reflect variations of the renormalization scale $\mu$ in $\alpha_s$ between $R/2$ and $2R$, showing very good convergence, exhibiting a perturbative error of $\pm\, 5$\,MeV for $R\sim 1$\,GeV and below $\pm\, 1$\,MeV for $R\gtrsim 3$\,GeV due to missing terms of ${\cal O}(\alpha_s^5)$ and higher. This indicates that the different way how the natural and practical MSR masses treat the virtual massive quark effects does not reintroduce any infrared sensitivity, as is expected since the mass of the virtual quark provides an infrared cutoff.  The numerical uncertainties in the ${\cal O}(\alpha_s^4)$ correction are below the level of $0.1$\,MeV and negligible. 
Note that the difference between the natural and the practical MSR masses at the common scale $R$ starts at ${\cal O}(\alpha_s^2)$ and that the uncertainty band from scale variation is an underestimate at this lowest order. However, the series results and error bands at ${\cal O}(\alpha_s^{3,4})$ show good behavior and convergence.
In Ref.~\cite{Butenschoen:2016lpz} the practical MSR mass was employed, but the numerical difference to the natural MSR mass is subdominant to the uncertainties obtained in the analysis there.

In the rest of the paper we will simply use the notation of the MSR mass with the definition $m_Q^{\pole}-m_Q^{\MSR}(R)=R\sum_n a_n \big[\alpha_s(R)/(4\pi)\big]^n$ when the difference between the natural and practical definitions and the value of $\nl$ are insignificant but we will specify explicitly our use of the practical or the natural MSR masses (or any other mass scheme) and the massless flavor number $\nl$ for any numerical analysis.

\section{R-Evolution}
\label{sec:Revolution}

The dependence of the MSR mass $m_Q^{\rm MSR}$ on the scale $R$ is described by the R-evolution equation~\cite{Hoang:2008yj}, which is derived from  the logarithmic derivative of the defining equations ~\eqref{eqn:msrpolenat} and \eqref{eqn:msrpoleprac} and using that the pole mass is $R$ independent:
\begin{equation}\label{eqn:revolvdef}
 R\frac{\dd}{\dd R}m_Q^{\MSR}(R)=-\,R\,\gamma^R(\alpha_s(R))=-\,R\sum_{n=0}^\infty\gamma_n^R\bigg(\frac{\alpha_s(R)}{4\pi}\bigg)^{\!\!n+1}\;,
\end{equation}
where
\begin{align}
\label{eqn:gammaintermsofa}
	\gamma_0^R&=a_1\,,\\
	\gamma^R_1&=a_2-2\,\beta_0 \,a_1\,, \nonumber\\
	\gamma^R_2&=a_3-4\,\beta_0\, a_2-2\,\beta_1\, a_1\,, \nonumber\\
	\gamma^R_n&=a_{n+1}-2\sum_{j=0}^{n-1}\, (n-j)\,\beta_j\,a_{n-j}\,. \nonumber
\end{align}
The overall minus sign on the RHS of Eq.~\eqref{eqn:revolvdef} indicates that the MSR mass always decreases with $R$. Note that this equation applies to all MSR schemes and we have therefore suppressed the superscript on the $a_n$'s.
The crucial feature of the R-evolution equation is that it is free from the $\Ord(\LQCD)$ ambiguity contained in the series that relates the MSR mass to the pole mass because the ambiguity is $R$-independent.
This is directly related to the fact that for determining the R-evolution equation also the overall linear factor of $R$ on the RHS of Eqs.~\eqref{eqn:msrpolenat} and \eqref{eqn:msrpoleprac} has to be accounted for. Therefore the R-evolution equation 
does not only have a logarithmic dependence on $R$, as common to usual renormalization group equations (RGEs), but also a linear one. Both of these issues are actually tied together conceptually. The numerical expressions for the coefficients $\gamma_n$ for the natural and practical MSR masses are given explicitly in Eqs.~\eqref{eqn:gammaRn} and \eqref{eqn:gammaRp}.  
We implement renormalization scale variation in the R-evolution equation by simply expanding $\alpha_s(R)$ in Eq.~\eqref{eqn:revolvdef} as a series in $\alpha_s(\lambda R)$ and by varying $\lambda$, typically in the range $0.5<\lambda<2$. In principle one may also consider varying the boundaries of integration, as it is common for usual RGEs, but only the former way of implementing scale variations in the R-evolution leads to variations of the scale solely in logarithms, which is the standard used for the usual logarithmic RGEs. 

By solving the R-evolution equation one sums, at the same time and systematically, the asymptotic renormalon series as well as the large logarithmic terms in $m_Q^{\MSR}(R_0)-m_Q^{\MSR}(R_1)$ to all orders in a manner free from the $\Ord(\LQCD)$ renormalon:
\begin{equation}\label{eqn:rrge}
m_Q^\MSR(R_0)-m_Q^\MSR(R_1)=-\sum_{n=0}^\infty\gamma_n^R
\int_{R_1}^{R_0}\dd R\,\bigg(\frac{\alpha_s(R)}{4\pi}\bigg)^{\!\!n+1} \;.
\end{equation}
It is straightforward to solve the R-evolution equation numerically and it shows very good perturbative stability even for low values of $R$ very close to the Landau pole~\cite{Hoang:2009yr} in the perturbative strong coupling.
Details of how to solve the R-evolution equations analytically have already been given in \cite{Hoang:2008yj} and shall not be repeated here.

It is instructive to briefly discuss what the solution of the R-evolution achieves by considering
the difference of the MSR mass, $m_Q^{\MSR}(R_0)-m_Q^{\MSR}(R_1)$, in the context of fixed-order perturbation theory (FOPT), where it is well-known that the renormalon ambiguity contained in the series for $m_Q^{\pole}-m_Q^\MSR(R_0)$ and the series for $m_Q^{\pole}-m_Q^\MSR(R_1)$ only cancel if one expands in $\alpha_s$ with a common renormalization scale $\mu$. This is nicely illustrated in the $\beta_0$/LL (leading log) approximation where the pole-MSR mass relation has the all order form 
\begin{align}\label{eqn:polemsrbeta0}
\big[m_Q^\pole-m_Q^\MSR(R)\big]_{\beta_0\mathrm{/LL}}&=\frac{a_1}{2\beta_0}R\sum_{n=0}^{\infty}\bigg(\frac{\beta_0 \alpha_s(R)}{2\pi}\bigg)^{\!\!n+1}n! \\
&=\frac{a_1}{2\beta_0}R\sum_{n=0}^\infty\bigg(\frac{\beta_0\alpha_s(\mu)}{2\pi}\bigg)^{\!\!n+1}n!
\sum_{k=0}^n \frac{1}{k!}\log^k\frac{\mu}{R} \,.\nonumber
\end{align}
The series by itself is divergent and not summable, but
\begin{align}\label{eqn:diffmsrbeta0}
\big[m_Q^\MSR(R_0)&-m_Q^\MSR(R_1)\big]_{\beta_0\mathrm{/LL}}=
\\
&=\frac{a_1}{2\beta_0}\sum_{n=0}^{\infty}\bigg(\frac{\beta_0\alpha_s(\mu)}{2\pi}\bigg)^{\!\!n+1}n!\,
\bigg(R_1\sum_{k=0}^n\frac{1}{k!}\log^k\frac{\mu}{R_1}-R_0\sum_{k=0}^n\frac{1}{k!}\log^k\frac{\mu}{R_0}\bigg)\nonumber\\
&=\frac{a_1}{2\beta_0}\sum_{n=0}^\infty\bigg(\frac{\beta_0\alpha_s(R_1)}{2\pi}\bigg)^{\!\!n+1}n!\,\bigg(R_1-R_0\sum_{k=0}^n\frac{1}{k!}\log^k\frac{R_1}{R_0}\bigg),\nonumber
\end{align}
is easily seen to be convergent. In the context of FOPT, when the sum over $n$ is truncated, the unavoidable appearance of large logarithms $\log(R_0/R_1)$ for let's say $R_0\ll R_1$ may degrade the convergence and cause sizable perturbative uncertainties. Due to the additional linear dependence on $R_0$ and $R_1$, as shown in Eq.~\eqref{eqn:diffmsrbeta0}, these logarithms cannot be summed by common logarithmic renormalization group (RG) equations. The same type of logarithms also appear for example in the relation of any other low-scale short-distance mass to the $\mathrm{\overline{MS}}$ mass and their effects can be significant particularly for the top quark.
By solving the R-evolution equation one sums, at the same time and systematically, the asymptotic terms in the renormalon series as well as the large logarithmic terms in $m_Q^{\MSR}(R_0)-m_Q^{\MSR}(R_1)$ to all orders in a manner free from the $\Ord(\LQCD)$ renormalon. It is again instructive to see how this is achieved in the $\beta_0$/LL approximation of Eq.~\eqref{eqn:polemsrbeta0}, which explicitly shows the factorial growth of the perturbative series. When calculating the derivative to get the R-evolution equation, the whole series collapses exactly (i.e. without any truncation!) to
\begin{equation}\label{eqn:revolvbeta0dif}
 \left[R\frac{\dd}{\dd R}m_Q^\MSR(R)\right]_{\beta_0\mathrm{/LL}}=\,-\,a_1\, R\,\bigg(\frac{\alpha_s(R)}{4\pi}\bigg) \;,
\end{equation}
which is the one-loop version of Eq.~\eqref{eqn:revolvdef}. Moreover,
the exact solution of the R-evolution equation at this order
\begin{equation}\label{eqn:rrgebeta0}
 \left[m_Q^\MSR(R_0)-m_Q^\MSR(R_1)\right]_{\beta_0\mathrm{/LL}} =-\,a_1\!\int_{R_1}^{R_0}\!\dd R\,\bigg(\frac{\alpha_s(R)}{4\pi}\bigg) \;,
\end{equation}
can be easily seen to be exactly equal to the RHS of Eq.~\eqref{eqn:diffmsrbeta0} which sums the renormalon series and the large logarithms at the same time into a convergent series.

Conceptually, the solution of the R-evolution equation is directly related to the Borel space integral over the Borel transform for the series for $m_Q^\MSR(R_0)-m_Q^\MSR(R_1)$. Since this has not been shown in \cite{Hoang:2008yj} we briefly outline this calculation here at the $\beta_0$/LL level. Starting from Eq.~\eqref{eqn:rrgebeta0} one can shuffle the integration over $R$ into an integral over $\alpha_s(R)$ by using the QCD $\beta$-function and the relation $\LQCD^{\rm LL}=R\exp\,(-\,2\pi/\beta_0\alpha_s(R))$. Using the variable $t= -\,2\pi/(\beta_0\alpha_s(R))$ one can then rewrite the integral as [\,$t_i= -\,2\pi/(\beta_0\alpha_s(R_i))$\,]
\begin{align}\label{eqn:rrgebeta0t}
 \left[m_Q^\MSR(R_0)\,-\,m_Q^\MSR(R_1)\right]_{\beta_0\mathrm{/LL}}&
=-\,\frac{a_1}{2\beta_0}\LQCD^\mathrm{LL}\int_{t_1}^{t_0}\frac{\dd t}{t}\,\mathrm{e}^{-\,t}\\
& =-\,\frac{a_1}{2\beta_0}\LQCD^\mathrm{LL}\left[\int_{t_1}^\infty\frac{\dd t}{t}\,
\mathrm{e}^{-\,t}-\int_{t_0}^\infty\frac{\dd t}{t}\,\mathrm{e}^{-\,t}\right] ,\nonumber
\end{align}
where the two integrals in the last line are just the difference of the MSR masses at $R_{0,1}$ to the pole mass, and the pole mass ambiguity is encoded in the singularity at $t=0$, which arises because $t_{0,1}<0$,
\begin{equation}
\label{eq:priortoturelation}
 \Big[m_Q^\MSR(R_i)\,-\,m_Q^\pole\Big]_{\beta_0\mathrm{/LL}}\,=\,\frac{a_1}{2\beta_0}\LQCD^\mathrm{LL}\int_{t_i}^\infty\frac{\dd t}{t}\,\mathrm e^{-\,t} \;.
\end{equation}
Upon changing variables to the Borel plane parameter $u=-(t/t_i-1)/2$ and writing $\LQCD$ in terms of $R_i$ and $\alpha_s(R_i)$ in both integrals, this gives
\begin{equation}\label{eqn:rrgebeta0u}
 \left[m_Q^\MSR(R_0)-m_Q^\MSR(R_1)\right]_{\beta_0\mathrm{/LL}}=\int_0^\infty\dd u\,
\left[\,B(R_0,\mu,u)-B(R_1,\mu,u)\,\right]
\mathrm e^{-\frac{4\pi u}{\beta_0\alpha_s(\mu)}} \;.
\end{equation}
Here
\begin{equation}
 B(R,\mu,u)=\frac{a_1}{2\beta_0}\,R\left(\frac{\mu}{R}\right)^{2u}\frac{1}{u-\frac{1}{2}} \;,
\end{equation}
is the well-known Borel transform with respect to $\alpha_s(\mu)$ of the $\beta_0$/LL series in Eq.~\eqref{eqn:polemsrbeta0}.
In Eq.~\eqref{eqn:rrgebeta0u} the singular and non-analytic contributions contained in the individual Borel functions cancel and the integral becomes ambiguity-free.

\begin{figure*}[t]
\center
\subfigure[]
{\label{fig:msrdiffa}\includegraphics[width=0.48\textwidth]{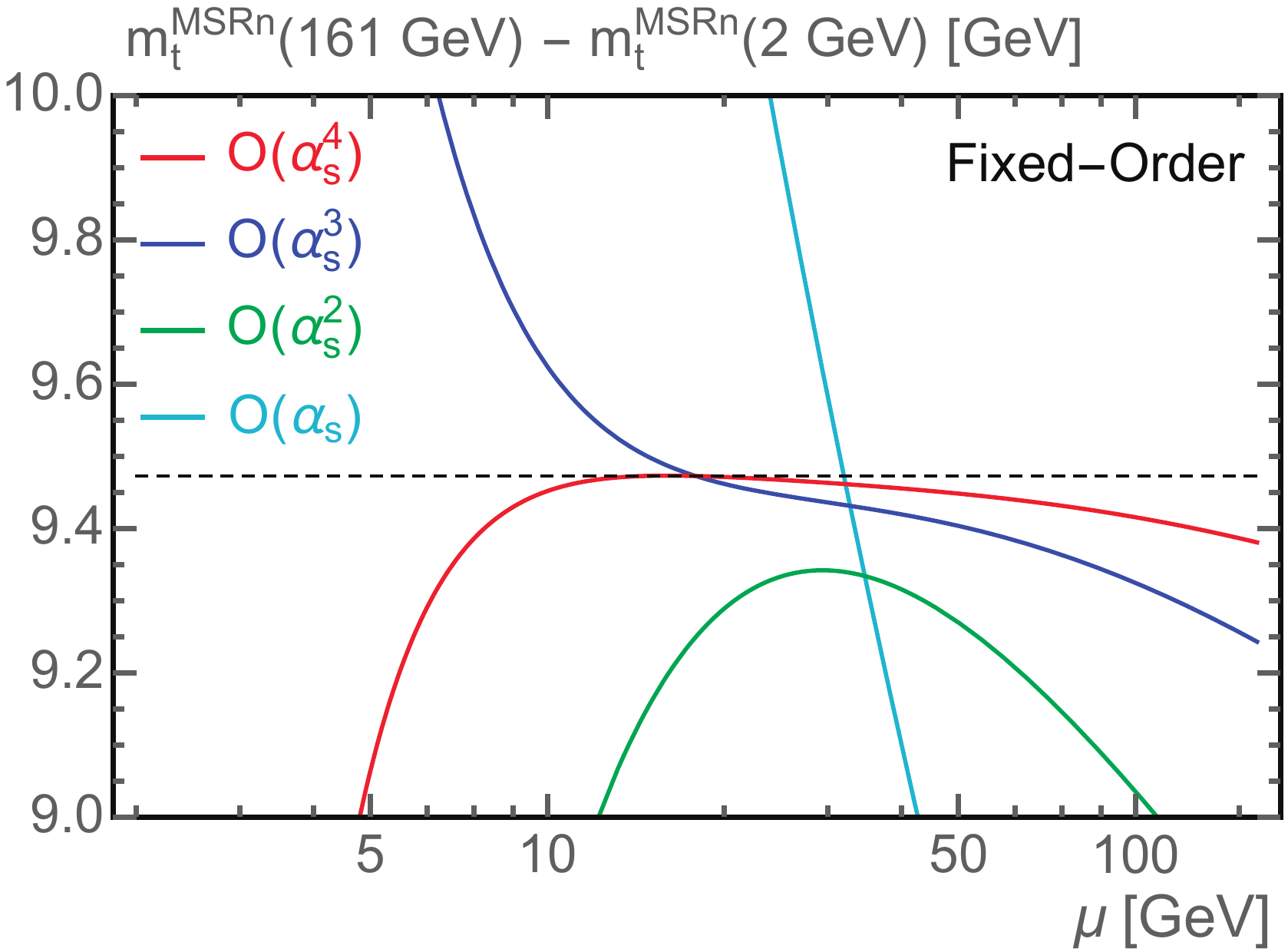}~~~}
 \subfigure[]
{\label{fig:msrdiffb}\includegraphics[width=0.48\textwidth]{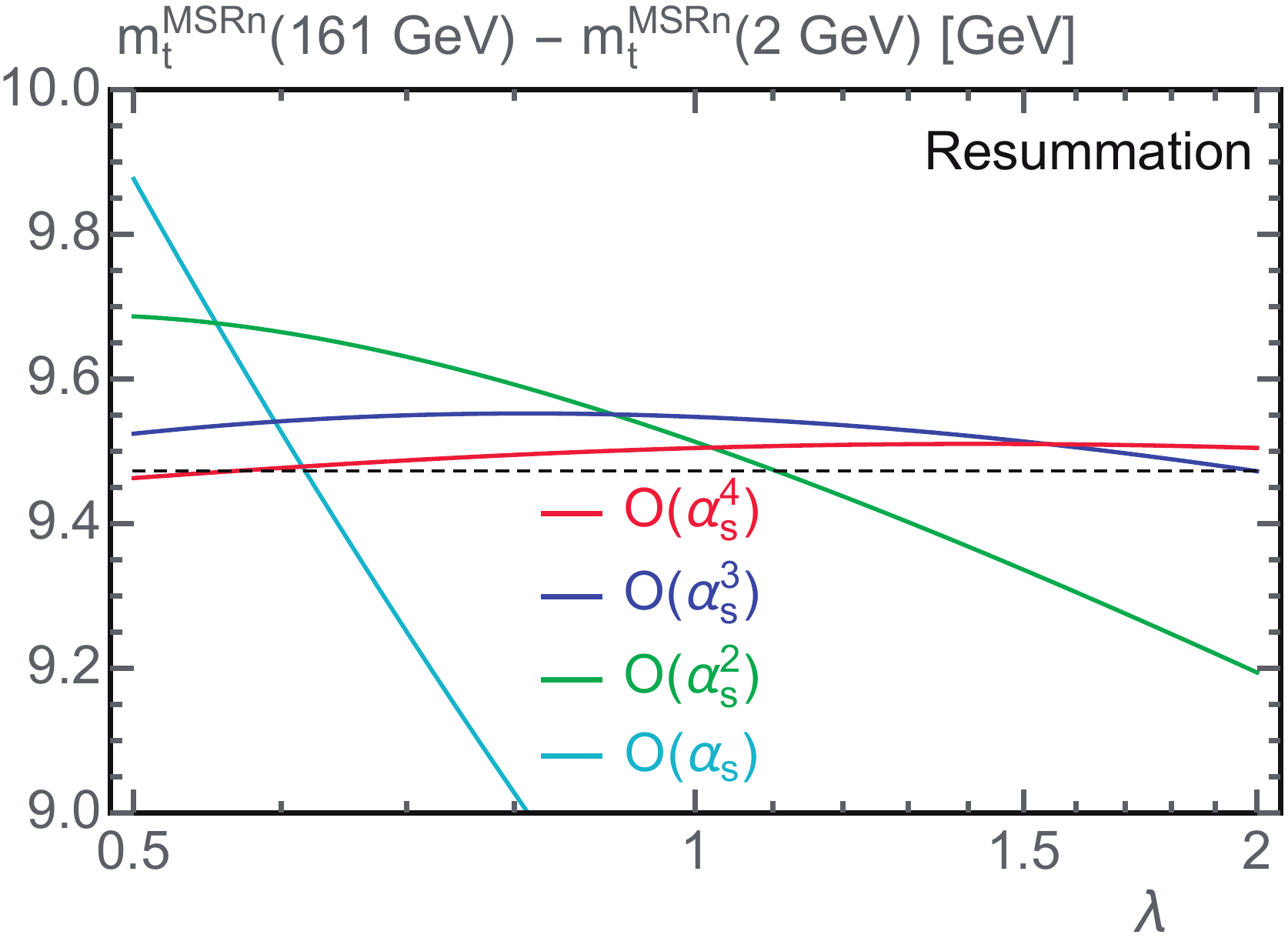}}
\subfigure[]
{\label{fig:msrdiffc}\includegraphics[width=0.47\textwidth]{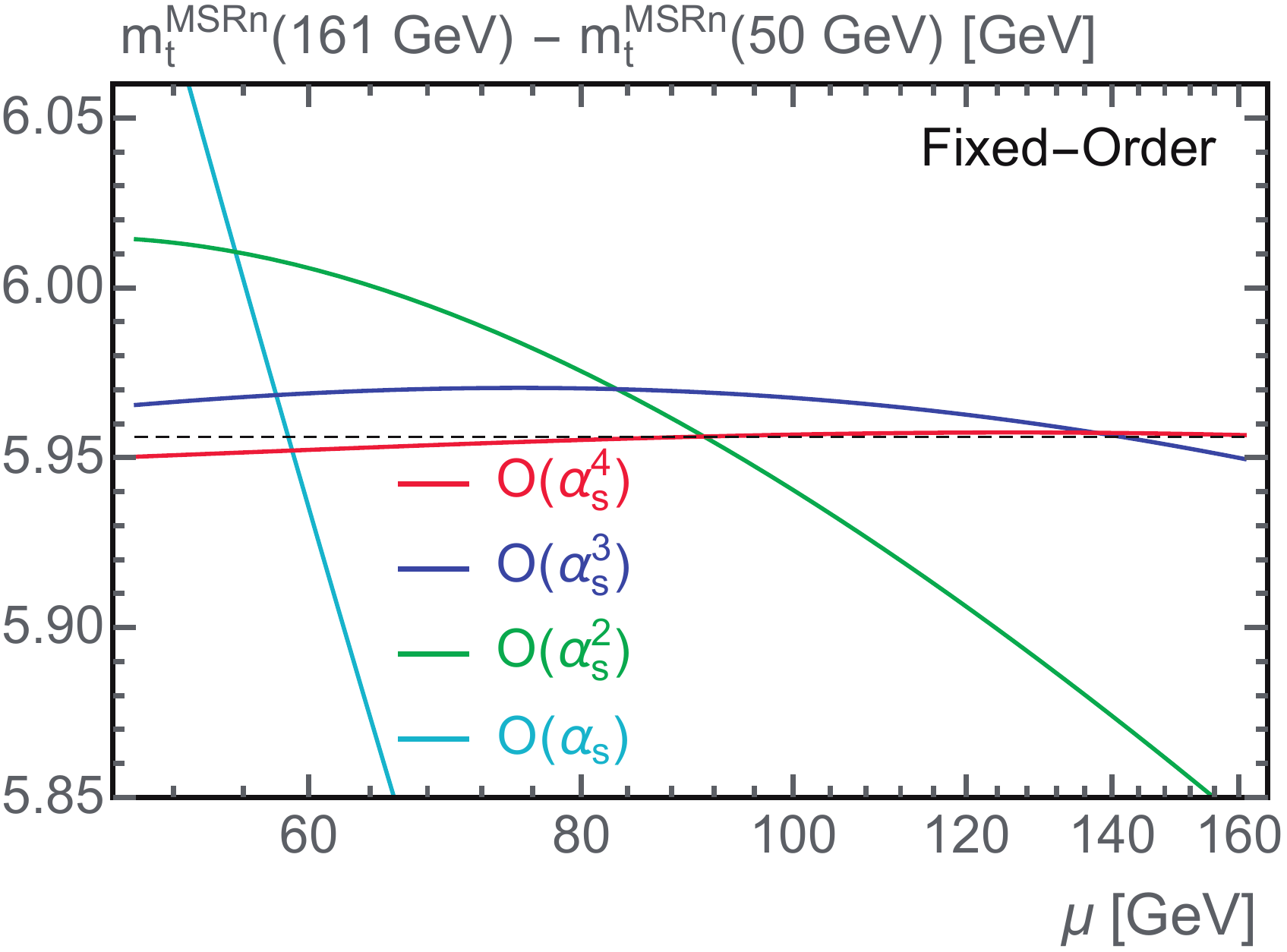}~~~}
\subfigure[]
{\label{fig:msrdiffd}\includegraphics[width=0.48\textwidth]{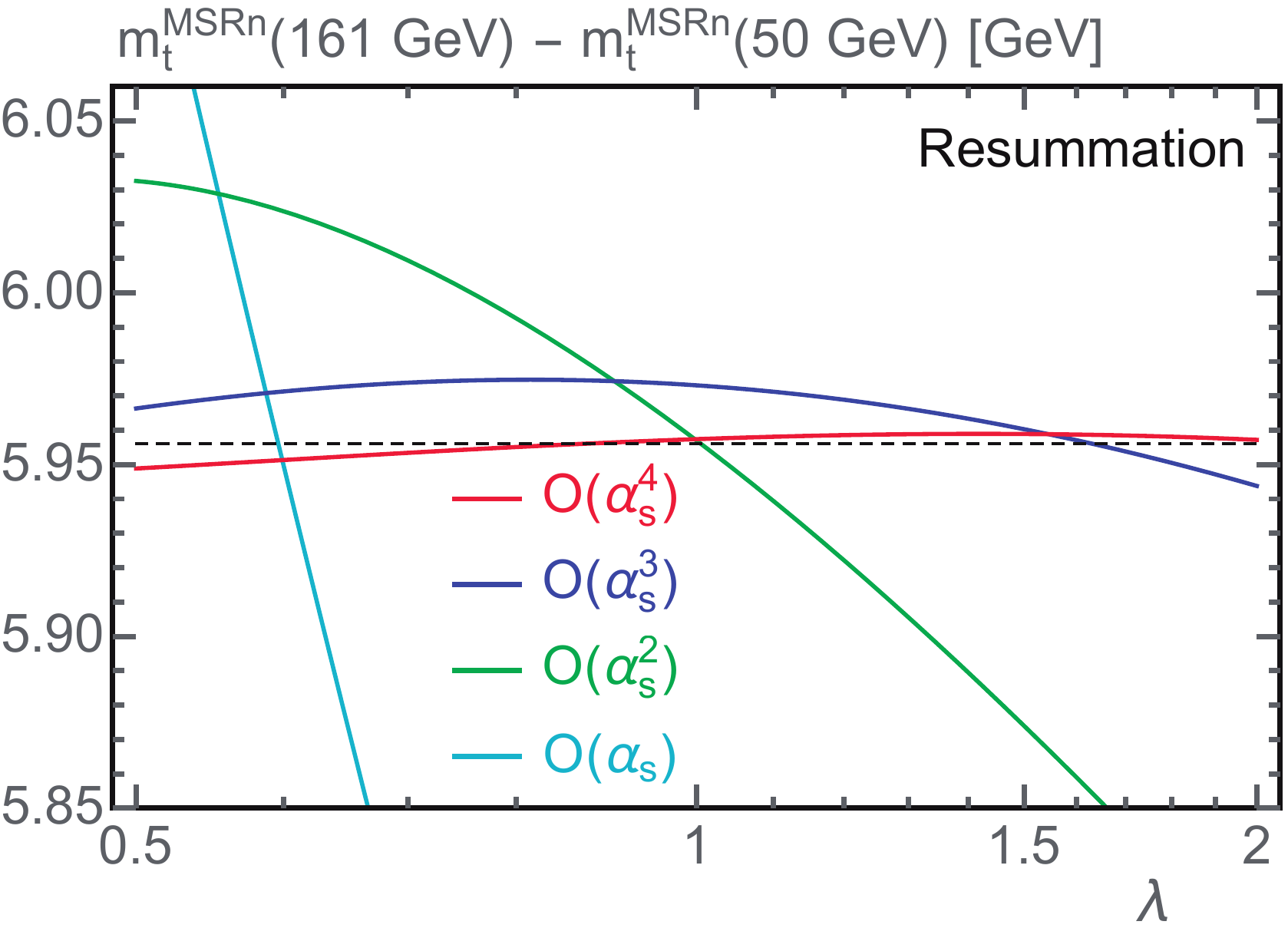}}
 \caption{\label{fig:msrdiff} Difference of the natural top quark MSR mass ($\nl=5$) at two different scales $R$ including contributions from one to four loops. Results are shown for the difference between a high scale $R_1 = 161$\,GeV and two lower scales $R_2 = 2$\,GeV (top two panels) and $R_2 = 50$\,GeV (lower two panels). The high and low scales are connected by a fixed-order perturbation theory conversion [\,left two panels, as a function of the scale $\mu$ in $\alpha_s(\mu)$\,] or via R-evolution [\,right two panels, as a function of the $\lambda$ renormalization parameter\,].}
\end{figure*}

To illustrate the impact of using R-evolution compared to using FOPT we show in Fig.~\ref{fig:msrdiff} the difference of natural MSR masses $\Delta m_t^\mathrm{MSRn}(R_0,R_1)\equiv m_t^\mathrm{MSRn}(R_0)-m_t^\mathrm{MSRn}(R_1)$  for \mbox{$\nl=5$} in fixed-order perturbation theory (FOPT) and with R-evolution. The curves in Fig.~\ref{fig:msrdiffa} show $\Delta m_t^\mathrm{MSRn}$ for $(R_0,R_1)=(2,161)$\,GeV in FOPT for the common renormalization scale $\mu$ between $R_0$ and $R_1$ at 1 loop (cyan), 2 loop (green), 3 loop (blue) and 4 loops (red). We see a good convergence for $\mu$ around $\sqrt{R_0 R_1}$, but a deterioration of the series when $\mu$ gets closer to either $R_0$ or $R_1$. For $\mu\lesssim 1/2 \sqrt{R_0 R_1}$ the series even gets out of bounds and breaks down completely. If one uses scale variation as an estimate of the remaining perturbative error, one therefore obtains a significant dependence on the choice of the lower bound of the variation, and one has no other choice than to abandon in an ad hoc manner scales closer to $R_0$ to estimate the scale variation error. The curves in Fig.~\ref{fig:msrdiffb} show $\Delta m_t^\mathrm{MSRn}$ for $(R_0,R_1)=(2,161)$\,GeV from numerically solving the \mbox{R-evolution} equation as a function of the renormalization scale parameter $\lambda$ between $0.5$ and $2$. The color coding for the order of the \mbox{R-evolution} equation used for the evaluation is the same as for Fig.~\ref{fig:msrdiffa}. As explained below Eq.~\eqref{eqn:revolvdef}, the parameter $\lambda$ is the renormalization scaling parameter in the \mbox{R-evolution} equation which determines by how much the scale in $\alpha_s$ differs from the scale $R$. Thus a variation between $0.5$ and $2$ means that in the solution of the \mbox{R-evolution} equations scales between $R/2$ and $2R$ are covered at each value of $R$ along the evolution, which in this case includes scales between $1$ and $322$\,GeV. Comparing the curves in Fig.~\ref{fig:msrdiffa} and \ref{fig:msrdiffb} we see that the renormalization scale variation in the R-evolved results is much smaller than the one of FOPT. For the FOPT result with scale variation between $\sqrt{R_0 R_1}/2$ -- which we pick by hand -- and $R_1$ we obtain 
$\Delta m_t=(9.838\,\pm\, 2.504,\,8.981\,\pm\,0.361,\,9.465\,\pm\,0.222,\,9.427\,\pm\,0.047)$\,GeV  at (1,\,2,\,3,\,4) loops. Using R-evolution with $\lambda$ variation between $0.5$ and $2$ we obtain
$\Delta m_t=(8.817\,\pm\, 1.059,\,9.440\,\pm\,0.246,\,9.512\,\pm\,0.040,\,9.486\,\pm\,0.025)$\,GeV which is fully compatible with the FOPT result, but shows more stability and smaller errors. It is also quite instructive to see that using R-evolution the 3-loop result is significantly closer to the 4-loop result than the corresponding 3-loop FOPT result. The results show that for $R_0\ll R_1$ employing R-evolution to calculate MSR mass differences is clearly superior to FO perturbation theory.

To compare to a situation where the scales $R_0$ and $R_1$ are of similar size we have also shown in Figs.~\ref{fig:msrdiffc} and \ref{fig:msrdiffd} the results for $\Delta m_t$ in FOPT and from R-evolution for $(R_0,R_1)=(50,161)$\,GeV. Here the results from both approaches are completely equivalent showing that the logarithm $\log(R_0/R_1)$ is not large and the summation of the renormalon contributions from higher orders only constitutes very small effects. Furthermore using renormalization scales close to $R_0$ or $R_1$ in FOPT is not problematic. 
Numerically, using FOPT with scale variations between $R_0$ and $R_1$ we obtain 
$\Delta m_t=(5.618\,\pm\,0.498,\,5.928\,\pm\,0.086,\,5.961\,\pm\,0.010,\,5.954\,\pm\,0.004)$\,GeV at $(1,2,3,4)$ loops, while using R-evolution with $\lambda$ variations between $0.5$ and $2$ we obtain 
$\Delta m_t=(5.555\,\pm\, 0.577,\,5.919\,\pm\, 0.114,\,5.959\,\pm\, 0.015,\,5.954\,\pm\, 0.005)$\,GeV. We find that FOPT and \mbox{R-evolution} give equivalent results even for $(R_0,R_1)=(20,161)$\,GeV, and that the use of R-evolution is essential for $R_0/R_1< 0.1$.
Overall we see that, if $R_0$ and $R_1$ are of similar size, FO perturbation theory and R-evolution lead to equivalent results, but that it is in general safer to use R-evolution. So the situation is very similar to the one we encounter when considering the relation of the strong coupling for two different renormalization scales. 

We note that the possibility to sum the renormalon-type logarithms displayed in Eq.~\eqref{eqn:diffmsrbeta0} by considering the Borel integral over the difference of Borel transforms as shown in Eq.~\eqref{eqn:rrgebeta0u} was pointed out already in Ref.~\cite{Bali:2003jq}
prior to Ref.~\cite{Hoang:2008yj}. However, this exact equivalence [\,via a transformation of variables as given below Eq.~(\ref{eq:priortoturelation})\,] of R-evolution and the method using the integration over Borel transform differences can only be analytically shown at the $\beta_0$/LL approximation. Beyond that, both approaches sum up the same type of logarithms but differ  in subleading terms. Numerically, both approaches converge to the same result and have comparable order-by-order convergence. From a practical point of view, however, the concept of R-evolution may be considered more general.
This is because R-evolution can be applied directly to any series having the form of \eqref{eqn:msrpolenat} or \eqref{eqn:msrpoleprac} while using the Borel integration method requires that the corresponding Borel transforms are known or constructed beforehand. For general series, such as for the difference of MSR masses as discussed above, this is not possible without making additional approximations. In practice, the approach of Ref.~\cite{Bali:2003jq} to sum the renormalon-type logarithms has therefore only been applied for series (referred to as RS-schemes) which were explicitly derived from a given expression for the Borel transform. 

\section{Analytic Borel Transform and Renormalon Sum Rule}
\label{sec:sumrule}

Using the solution of the R-evolution equation it is possible to derive, analytically and rigorously, an expression for the Borel transform of the MSR-pole mass relation. This Borel transform is designed to focus on the singular contributions that quantify the $\Ord(\LQCD)$ renormalon of the pole mass. This result was already quoted in the letter \cite{Hoang:2008yj} where, however, no details on the derivation could be given due to lack of space. In the following we provide these details on how to obtain the analytic result for the normalization of the singular terms. The analytic results for the normalization can be applied to other perturbative series as a probe of $\Ord(\LQCD)$ renormalon ambiguities, and we therefore call it {\it the $\Ord(\LQCD)$ renormalon sum rule}. This sum rule was first given in Ref.~\cite{Hoang:2008yj}, and is very sensitive to even subtle effects if $\Ord(\alpha_s^4)$ corrections are known. We apply the sum rule to obtain an updated determination of the size of the pole mass $\Ord(\LQCD)$ ambiguity, accounting for the $\Ord(\alpha_s^4)$ results of Refs.~\cite{Marquard:2015qpa,Marquard:2016dcn} which became available recently but were unknown when Ref.~\cite{Hoang:2008yj} appeared.  To demonstrate the sum rule's capabilities to probe  $\Ord(\LQCD)$ renormalon ambiguities in perturbative series and to clarify subtleties in how to use it properly, we also apply it to a few other cases.
Interestingly, the analytic manipulations arising in the derivation of the sum rule lead to an alternative expression for the high-order asymptotic behavior of a series that contains an ${\cal O}(\Lambda_{\rm QCD})$ renormalon. This expression differs from the well known asymptotic formula which is known since a long time from~\cite{Beneke:1994rs}, and we therefore discuss it as well.

\subsection{Derivation}
\label{sec:derivation}

The analytic derivation for the Borel transform of the MSR-pole mass relation starts from its expression related to the solution of the R-evolution equation given in Eq.~\eqref{eqn:revolvdef} which was already derived in Ref.~\cite{Hoang:2008yj}.
\begin{align}\label{eqn:msrpolefull}
	 m_Q^\MSR(R)-m_Q^\pole &=-\int_0^R\dd\bar R\,\gamma^R(\alpha_s(\bar R)) \\
	&=-\,\LQCD\int_{t_R}^\infty\dd t\,\gamma^R(t)\,\hat b(t)\,\mathrm e^{-G(t)}\nonumber\\
	&=\LQCD\sum_{k=0}^\infty\mathrm e^{i\pi(\hat b_1+k)}S_k\int_{t_R}^\infty\dd t\, t^{-1-k-\hat b_1}\mathrm e^{-t}\nonumber\\
	&=\LQCD\sum_{k=0}^\infty\mathrm e^{i\pi(\hat b_1+k)}\,S_k\,\Gamma(-\,\hat b_1-k,t_R) \,,\nonumber
\end{align}
where in the second line we changed variable to $t=-\,2\pi/(\beta_0\alpha_s(\bar R))$ and used the identity \eqref{eqn:lambda} to scale out $\LQCD$, and in the third line we employed the coefficients given in Eq.~\eqref{eqn:skcoeff}.
The expression in Eq.~\eqref{eqn:msrpolefull} gives an all-order representation of the original series that is more useful for analyzing $\Ord(\LQCD)$ renormalon issues than Eqs.~\eqref{eqn:msrpolenat} and \eqref{eqn:msrpoleprac}.  This is because using the R-evolution equation of Eq.~(\ref{eqn:revolvdef}) (which is linear in $R$) and its solution, provides, through the sum in $k$, a reordering of the original series in leading and subleading series of terms from the perspective of their numerical importance in the asymptotic high order behavior related to the $\Ord(\LQCD)$ renormalon. This allows to derive rigorously a representation of the Borel transform [\,given in Eq.~(\ref{eqn:borel2})\,] reflecting efficiently the hierarchy of leading and subleading terms with respect to the $\Ord(\LQCD)$ renormalon, which is the information that is not contained in the original series. That such a separation is possible in a systematic way may not be obvious, but it is achieved by the R-evolution equation. We stress that the result of Eq.~(\ref{eqn:borel2}) should not be considered as the exact expression for the Borel transform because it does not encode information on possible poles (or non-analytic cuts) other than at $u=1/2$. We note that these poles and the associated renormalons can be studied by considering solutions of R-evolution equations involving powers of $R$ different from the linear dependence shown in Eq.~(\ref{eqn:revolvdef}), see~\cite{Ambar:thesis}.

We note that the expression in the last line of Eq.~\eqref{eqn:msrpolefull}, which involves the incomplete gamma function $\Gamma(c,t)=\int_t^\infty \dd x\,x^{c-1} e^{-x}$, also arises in the analytic solution of the mass difference~\eqref{eqn:rrge},
\begin{equation}\label{eqn:rrge2}
m_Q^\MSR(R_0)-m_Q^\MSR(R_1)=
\LQCD\sum_{k=0}^\infty\mathrm e^{i\pi(\hat b_1+k)}\,S_k\,\big[\, \Gamma(-\,\hat b_1-k,t_0)-\Gamma(-\,\hat b_1-k,t_1)\, \big].
\end{equation}
Here the cut in the gamma functions $\Gamma(c,t)$ for $t<0$ cancels in the difference for each $k$ in the sum, and the result on the RHS is real. We mention that the first term ($k=0$) in the sum over $k$ provides the summation of the leading terms in the $\beta_0\mathrm{/LL}$ approximation shown in Eqs.~\eqref{eqn:diffmsrbeta0} and \eqref{eqn:rrgebeta0}.
In Eq.~\eqref{eqn:msrpolefull} the cut still remains and arises from the integration of the Landau pole in the strong coupling located at $t=0$ in the integral in the next-to-last line. The resulting imaginary part in the numerical expression corresponds to the imaginary part that arises in the inverse Borel integral for $m_Q^\MSR(R)\,-\,m_Q^\pole$, see Eq.~\eqref{eqn:rrgebeta0u}, and simply reflects the ambiguity of the pole mass. From the point of view of the analytic solution of Eq.~\eqref{eqn:msrpolefull} based on a perturbative expansion, the imaginary part is well-defined and analytically unique. 

To proceed we asymptotically expand the incomplete gamma function in inverse powers of $t$ (i.e. powers of $\alpha_s$) 
\begin{align}\label{eqn:gammaexpand}
\LQCD\mathrm e^{i\pi(\hat b_1+k)}\Gamma(-\hat b_1-k,t)&=-R\left[\mathrm e^{G(t)}\mathrm e^{-t}(-t)^{-\hat b_1}\right]\sum_{m=0}^\infty\frac{\Gamma(1+\hat b_1+k+m)}{\Gamma(1+\hat b_1+k)}\,(-t)^{-1-k-m}\nonumber\\
&=-R\sum_{\ell=0}^\infty g_\ell\sum_{m=0}^\infty\frac{\Gamma(1+\hat b_1+k+m)}{\Gamma(1+\hat b_1+k)}\,(-t)^{-1-\ell-k-m} \,,
\end{align}
where the coefficients $g_\ell$ are given in Eq.~\eqref{eqn:glcoeff}, and coincide
with the $s_k$ coefficients defined in Ref.~\cite{Beneke:1994rs}. We stress that the equality in Eq.~\eqref{eqn:gammaexpand} is the asymptotic expansion and is not an identity, so that the imaginary part due to the cut in the incomplete gamma function does not arise on the RHS. Inserting Eq.~\eqref{eqn:gammaexpand} in Eq.~\eqref{eqn:msrpolefull} gives 
\begin{align}\label{eqn:msrpolefull2}
	m_Q^\MSR(R)-m_Q^\pole&=-R\,\sum_{k=0}^\infty
	S_k	\sum_{\ell=0}^\infty g_\ell\sum_{m=0}^\infty\frac{\Gamma(1+\hat b_1+k+m)}{\Gamma(1+\hat b_1+k)}\,(-t)^{-1-\ell-k-m}	 \,.
\end{align}
We then perform the Borel transform with respect to powers of $\alpha_s(R)$ according to the rule 
\mbox{$(-t)^{-1-n}\to 2\,(2u)^n/\Gamma(n+1)$} giving
\begin{align}\label{eqn:borel1}
	B_{\alpha_s(R)}\Big[ & m_Q^\MSR(R)-m_Q^\pole \Big](u)= \\
	& =
	-\,2R\sum_{\ell=0}^{\infty}g_\ell\sum_{k=0}^\infty S_k\sum_{m=0}^\infty\frac{\Gamma(1+\hat b_1+k+m)}{\Gamma(1+\hat 
 b_1+k)\Gamma(1+k+\ell+m)}\,(2u)^{\ell+k+m}\nonumber\\
	&=-\,2R\sum_{\ell=0}^{\infty}g_\ell\sum_{k=0}^\infty S_k\frac{(2u)^{\ell+k}}{\Gamma(1+k+\ell)}\,
{}_2F_1(1,1+\hat b_1+k,1+k+\ell,2u) \;.\nonumber
\end{align}
Using identities for the hypergeometric function we can rewrite
\begin{align}
	\frac{(2u)^{\ell+k}}{\Gamma(1+k+\ell)}\,&{}_2F_1(1,1+\hat b_1+k,1+k+\ell,2u)=\frac{\Gamma(1+\hat b_1-\ell)}{\Gamma(1+\hat b_1+k)}(1-2u)^{-1-\hat b_1+\ell}\\
	&-\frac{1}{(1+\hat b_1-\ell)\Gamma(k+\ell)}\,{}_2F_1(1+\hat b_1-\ell,1-k-\ell,2+\hat b_1-\ell,1-2u) \,,\nonumber
\end{align}
and the Borel transform can then be cast into the form~\cite{Hoang:2008yj}
\begin{align}\label{eqn:borel2}
  B_{\alpha_s(R)}\left[m_Q^\MSR(R)-m_Q^\pole\right]\!(u)\,=\, &- N_{1/2}\!\left[R\,\frac{4\pi}{\beta_0}
  \sum_{\ell=0}^{\infty}g_\ell\frac{\Gamma(1+\hat b_1-\ell)}{\Gamma(1+\hat b_1)}(1-2u)^{-1-\hat b_1+\ell}\right]\nonumber\\
&+\,2R\sum_{\ell=0}^{\infty}g_\ell \,Q_\ell(u) \,,
\end{align}
where
\begin{align}\label{eqn:P12def}
 N_{1/2}\,=\,& \frac{\beta_0\,\Gamma(1+\hat b_1)}{2 \pi}\,P_{1/2} \;, \\
 P_{1/2}\,=\,& \sum_{k=0}^{\infty}\frac{S_k}{\Gamma(1+\hat b_1+k)} \;,\nonumber
\end{align}
and $N_{1/2}$ and $P_{1/2}$ are two conventions for the normalization. Here
\begin{align}\label{eqn:Qldef}
Q_\ell(u)&=\sum_{k=0}^\infty\frac{S_k\,(2u)^{k+\ell}}{(1+\hat b_1-\ell)\,\Gamma(k+\ell)}
\,{}_2F_1(1,1+\hat b_1+k,2+\hat b_1-\ell,1-2u)\\
&=\sum_{k=0}^\infty S_k\!\sum_{i=0}^{k+\ell-1}
\frac{2^i\,\Gamma(1+ \hat b_1 +i - \ell)}{\Gamma(1+ \hat b_1 + k)\,\Gamma(i+1)}\,u^i \,.\nonumber
\end{align}
Setting $u=1/2$ in Eq.~\eqref{eqn:Qldef} one gets
$Q_\ell(1/2) = 1/(1+\hat b_1-\ell)\sum_{k=0}^\infty S_k/\Gamma(k+\ell)$. Since the $S_k$ coefficients
are renormalon-free and further damped by the factorial in the denominator, this sum is finite.
Furthermore, the sum on the second line of Eq.~\eqref{eqn:borel2} is also finite for $u=1/2$. Therefore
one concludes that the sum of $Q_\ell$ coefficients is regular at $u=1/2$, implying that the first line
of Eq.~\eqref{eqn:borel2} fully contains the leading-renormalon singular behavior.
In Ref.~\cite{Hoang:2008yj} the expression for the Borel transform in Eq.~\eqref{eqn:borel2} was given using $P_{1/2}$, but here we have shown an alternate convention with $N_{1/2}$ which agrees with the terms $N_m$ and $N$ discussed in Refs.~\cite{Ayala:2014yxa,Beneke:2016cbu}, and hence eases comparison of our numerical results with theirs. For the phenomenological relevant values $\nl=(3,4,5)$ we have $N_{1/2}/P_{1/2}=(1.27,1.18,1.09)$. The analytic difference between these normalizations is that
$P_{1/2}$ vanishes in the limit $\nl\to-\infty$ while $N_{1/2}$ is finite in this limit. We will predominantly use $N_{1/2}$ for the numerical examinations in the following subsections. 

The manipulations that lead to the expressions for $P_{1/2}$ and $N_{1/2}$  involve the rearrangement of the infinite sums over $\ell$ and $k$ in Eq.~\eqref{eqn:borel1}. These can be seen to be identities if one assumes that the QCD $\beta$-function and its inverse have some region of convergence. In practice, because only the first few terms in perturbation theory are known and one truncates the sums over $\ell$ and $k$, no formal convergence issue arises. 
We note that the analytic manipulations involving the R-evolution equation and the derivation of Eq.~\eqref{eqn:borel2} are also valid in schemes for the strong coupling other than $\MSb$, and to apply them to such schemes one simply needs to account for the perturbative rearrangement for the coefficients $a_n$ and the QCD $\beta$-function due to the scheme change. As an example, all manipulations and the results simplify considerably in a strong coupling scheme $\bar{\alpha}$ where the coefficients 
$\hat b_n$ vanish for $n>1$ and which also implies $g_{\ell}=0$ for $\ell>0$ and that the coefficients of the QCD $\beta$-function have the exact form $\beta_{n}=\beta_0(\beta_1/\beta_0)^n$. 
Since such a scheme change can be achieved via a relation of the form $\alpha_s(\mu) =\bar\alpha(\mu) + [\,\beta_2/\beta_0-(\beta_1/\beta_0)^2\,]\,\bar\alpha^3(\mu)+\ldots$\,, which does not contain any ${\cal O}(\bar\alpha_s^2)$ term, the overall normalization of $N_{1/2}$ (or $P_{1/2}$) remains unchanged~\cite{Beneke:1998ui}. In this scheme we have $S_{k>0}=\tilde\gamma_k^R\,-\,\hat b_1\tilde\gamma_{k-1}^R$, and 
Eq.~\eqref{eqn:P12def} can be rewritten in the equivalent form
$N_{1/2}=(\beta_0/2\pi) \Gamma(1+\hat b_1) \sum_{k=0}^\infty\,\tilde\gamma_k^R (1+k)/\Gamma(2+\hat b_1+k)$ and was derived recently in Ref.~\cite{Komijani:2017vep}. 
There is, however, no advantage in using this form, because the coefficients $\tilde\gamma_k^R$ in the $\bar{\alpha}$ scheme still have to account for the reordering of the series due to the scheme change from $\alpha_s$ to $\bar\alpha$. 
Other schemes, such as the '\!\! t Hooft scheme, where all coefficients of the QCD $\beta$-function beyond $\beta_0$ and $\beta_1$ vanish,
have been studied in Ref.~\cite{Ambar:thesis}.

We discuss the structure of the non-analytic terms multiplied by $N_{1/2}$ in Eq.~\eqref{eqn:borel2} in Sec.~\ref{sec:asymtotic} below.
The second term in Eq.~\eqref{eqn:borel2} is purely polynomial and represents contributions in the Borel transform $B(u)$ that account for the portions in the original series of Eqs.~\eqref{eqn:msrpolenat} and \eqref{eqn:msrpoleprac} that go beyond the pure $\Ord(\LQCD)$ renormalon corrections that numerically dominate the series. These terms may include renormalon contributions of a different kind [\,such as $\Ord(\LQCD)^{k>1}$\,], which are however not probed by an R-evolution equation that is linear in $R$ \cite{Hoang:2009yr}. Moreover, they account for the difference of the pure $\Ord(\LQCD)$ renormalon asymptotic form of the series (encoded in the value of $N_{1/2}$) and the actual coefficients of the original series given in Eqs.~\eqref{eqn:msrpolenat} and \eqref{eqn:msrpoleprac}. The latter are recovered in the asymptotic limit were the sums over
$k$ and $\ell$ are carried out up to infinity.
Note that in practice, for a finite order determination of the Borel transform for a given value of $N_{1/2}$ or $P_{1/2}$, one truncates the sum over $k$ and $\ell$ in Eq.~\eqref{eqn:Qldef}, and in this case the terms coming from the $Q_\ell$ represent finite polynomials. For the construction of a Borel transform that reproduces the known coefficients exactly, it may then be more suitable to simply fit the coefficients of the remaining polynomial terms such that the known coefficients in the original series are reproduced exactly.

\subsection{Renormalon Sum Rule}
\label{sec:renormalonsumrule}

The analytic expression for $N_{1/2}$ is quite useful as it can be applied to any perturbative series as a probe for $\Ord(\LQCD)$ renormalons, given the information on the available coefficients of a perturbative series. We therefore call the formula for $N_{1/2}$ (or equivalently $P_{1/2}$) in Eq.~(\ref{eqn:P12def}) \textit{the $\Ord(\LQCD)$ renormalon sum rule}~\cite{Hoang:2008yj}. Formally to any given order in $k$, $N_{1/2}$ is a linear functional acting on perturbative series in powers of $\alpha_s$ since the coefficients $S_k$ in Eq.~\eqref{eqn:P12def} are linear in the coefficients $a_n$ of the perturbative series, see Eq.~\eqref{eqn:skcoeff}. So given two series defined by the sequence $\{c_n\}=(c_1,c_2,\dots)$ and $\{d_n\}=(d_1,d_2,\dots)$, where $c_n/d_n$ are the coefficients of order $[\,\alpha_s/(4\pi)\,]^{n}$ in the series, one has
\begin{equation}\label{eqn:N12linear}
N_{1/2}[\{\alpha \,c_n+\beta\, d_n\}]\,=\,\alpha\, N_{1/2}[\{c_n\}]+\beta\, N_{1/2}[\{d_n\}] \;.
\end{equation}

As a word of caution, we emphasize that applying the $N_{1/2}$ sum rule to a truncated series does (like any other type of renormalon calculus in the context of perturbative QCD) not rigorously and mathematically prove or disprove the existence of an $\Ord(\LQCD)$ renormalon, since the existence of renormalons is by definition related to the asymptotic high-order behavior and mathematically strict proofs, if they exist, are related to elaborate all-order studies of Feynman diagrams. 
So using the sum rule should be better thought of as an analytic projection of the known terms of a perturbative series onto the known pattern of a pure $\Ord(\LQCD)$ renormalon series, which is generated from the singular terms in the Borel transform in Eq.~\eqref{eqn:borel2} that are multiplied by $N_{1/2}$ or $P_{1/2}$ and known to all orders. This projection becomes more accurate the more terms of a series are known and mathematically converges (only) if the yet unknown high order terms keep following the renormalon pattern expected from the low order terms.\footnote{For example, applying the sum rule to a series that follows an $\Ord(\LQCD)$ renormalon pattern up to order $m$, but then changes to a convergent series beyond, the value of $N_{1/2}$ approaches a finite value up to order $m$, but then decreases and approaches zero when more terms beyond order $m$ are included. Note however that there is no reason to expect a perturbative series in QCD to behave in such a manner.}  

Although the series in $k$ for $N_{1/2}$ in Eq.~\eqref{eqn:P12def} is not ordered in powers of the strong coupling, it is possible to implement renormalization scale variation by rescaling $R\to \lambda R$ in the original series of Eqs.~\eqref{eqn:msrpolenat} and \eqref{eqn:msrpoleprac} and subsequently expanding again in $\alpha_s(R)$. This leads to
\begin{align}\label{eqn:skcoefflambda}
S'_0\,=&\,\lambda\, S_0 \,,\nonumber\\
S'_1\,=&\,\lambda\big[S_1-S_0\log\lambda\big] ,\nonumber\\
S'_2\,=&\,\lambda\big[S_2-2\,S_1\log\lambda+
S_0\big(\log^2\lambda- (\hat b_2+2\,\hat b_1)\log\lambda\big)\big] ,\nonumber\\
S'_3\,=&\,\lambda\Big[S_3-3\,S_2\log\lambda+S_1\big(3\log^2\lambda-(\hat b_2+3\,\hat b_1)\log\lambda\big)\nonumber\\
&+S_0\Big(\!-\log^3\lambda+\Big(2\,\hat b_2+\frac{9}{2}\,\hat b_1\Big)\log^2\lambda+
\big(3\,\hat b_2+\hat b_3-\hat b_1(\hat b_2+3\,\hat b_1)\big)\log\lambda\Big)\Big] ,
\end{align}
and one can show that in the asymptotic limit, i.e.\ to all orders in $k$, the sum rule expression for $N_{1/2}$ or $P_{1/2}$ is invariant under variations of $\lambda$. Thus for a finite order determination of $N_{1/2}$ the $\lambda$-dependence decreases with order, and the remaining variation with $\lambda$ can be taken as an estimate for the uncertainty due to the missing higher order terms in the same way as renormalization scale variation in RG-invariant power series in $\alpha_s$ is commonly used to estimate perturbative uncertainties. The invariance under changes of $\lambda$ is directly related to the facts that the $\Ord(\LQCD)$ renormalon ambiguity of the series in Eqs.~\eqref{eqn:msrpolenat} and \eqref{eqn:msrpoleprac} is $R$-independent and that carrying out the Borel transform of Eq.~\eqref{eqn:borel1} in the previous section with respect to $\alpha_s(\mu)$ instead of $\alpha_s(R)$ leads to the simple rescaling factor $\mu/R$ of all the non-analytic terms proportional to $N_{1/2}$.

\begin{figure*}[t]
	\center
	\subfigure[]
	{\label{fig:N12band}\includegraphics[width=0.468\textwidth]{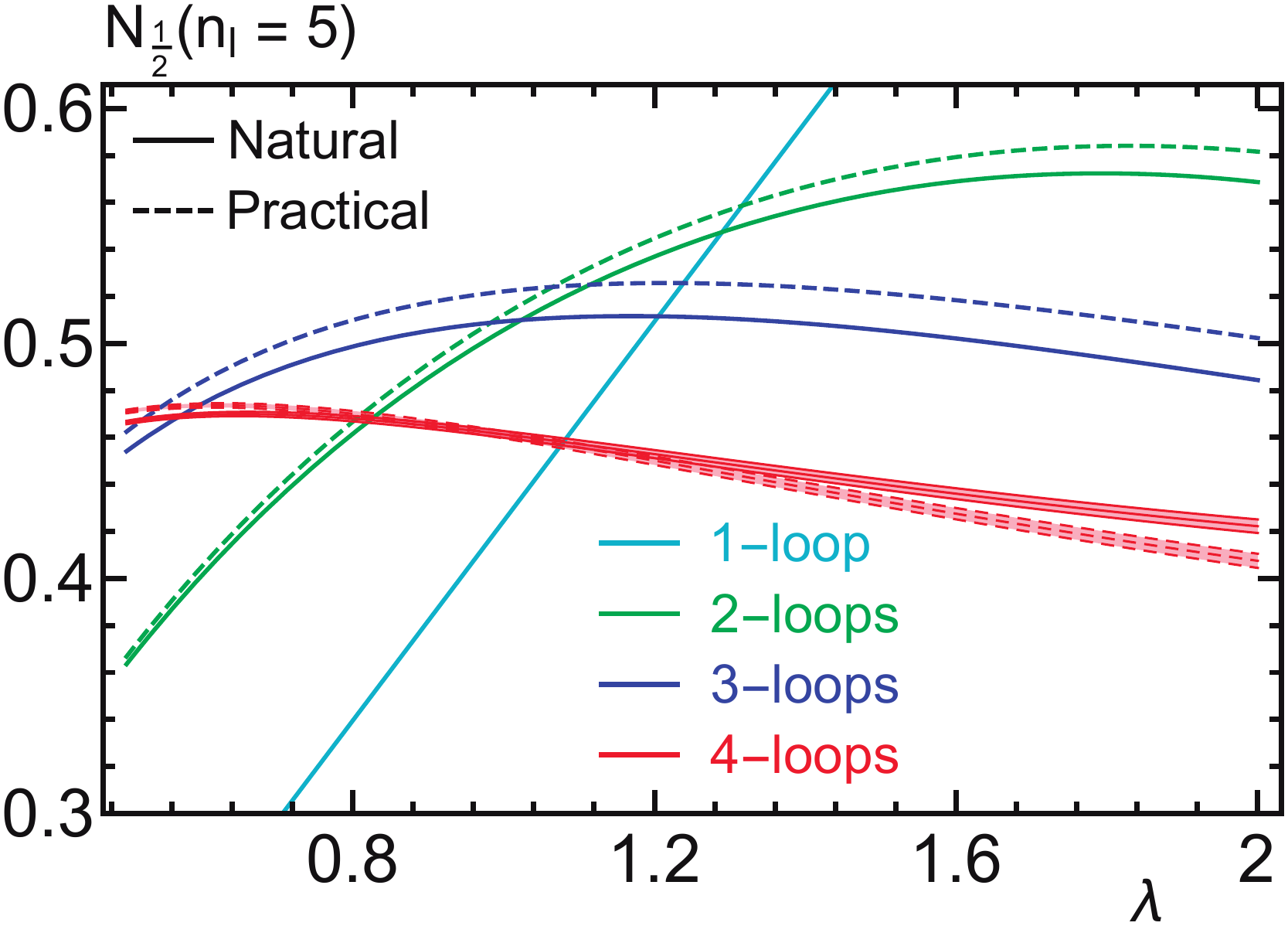}~~~}
	\subfigure[]
	{\label{fig:N12ord}\includegraphics[width=0.468\textwidth]{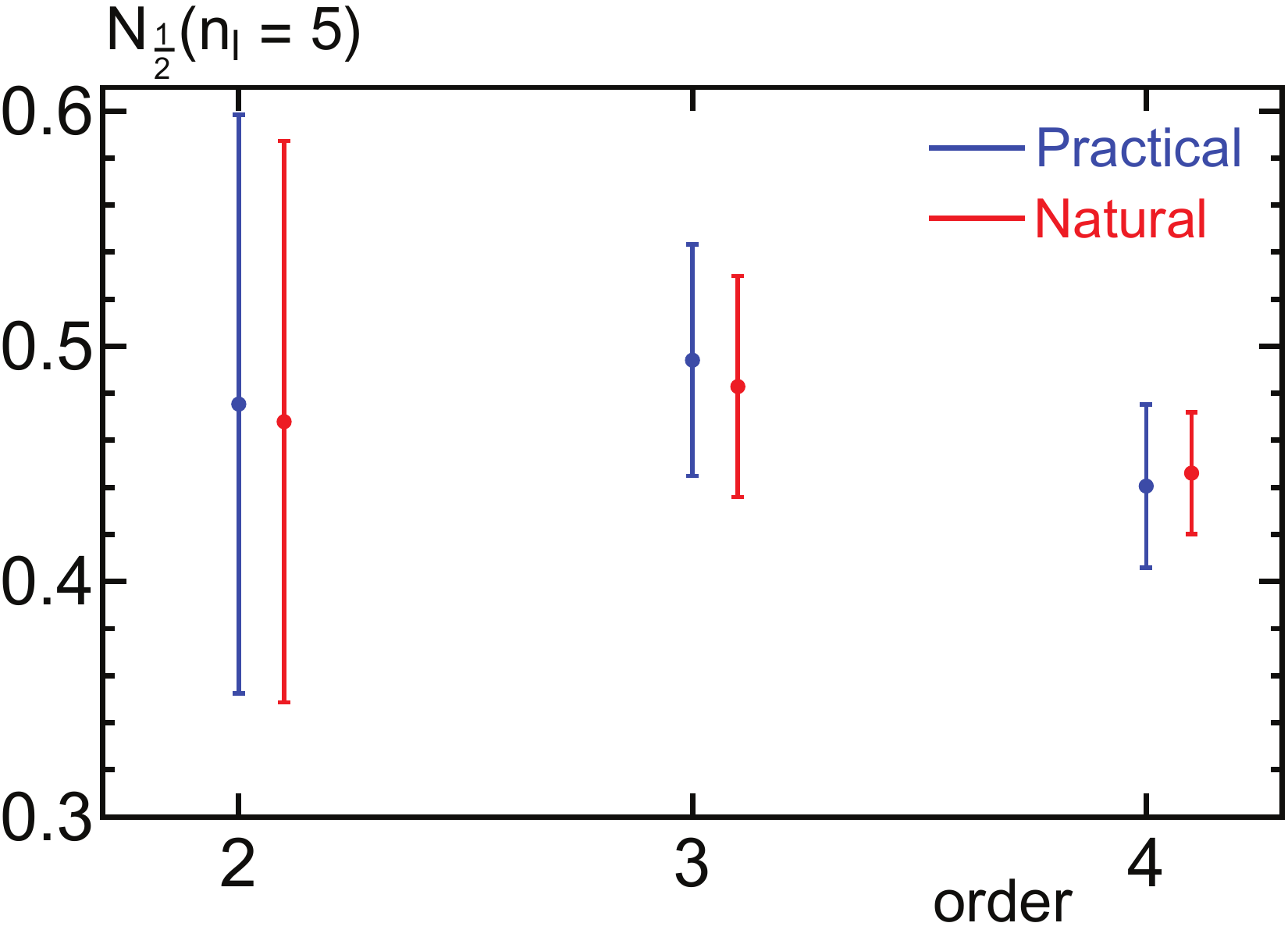} }
 \caption{\label{fig:N12msr} $N_{1/2}(\nl=5)$ for the natural and practical top quark MSR masses.
On panel (a)~results are shown as a function of $\lambda$ including contributions from one to four
loops. The size of the bands at four loops reflects the error introduced by the numerical uncertainty
in the $\Ord(\alpha_s^4)$ coefficient for the $\MSb$-pole conversion series. On panel (b)~results are
shown as error bars in blue (red) for the practical (natural) MSR masses at $k$-loops accounting also for the $\eta$ parameter variation as described after Eq.(\ref{eqn:N12msrfinal5}).}
\end{figure*}

\subsection{Sum Rule for the Pole Mass Renormalon}

We now apply the sum rule to the series of the MSR-pole mass relations to quantify the $\Ord(\LQCD)$ renormalon of the pole mass. Note, that to fully determine the order $k$ result, the $\Ord(\alpha_s^{k+1})$ ($k+1$)-loop corrections from Eq.~\eqref{eqn:msrpolenat} and Eq.~\eqref{eqn:msrpoleprac} and the $\Ord(\alpha_s^{k+3})$ ($k+2$)-loop correction to the QCD $\beta$-function, $\beta_{k+1}$ need to be known. So at $k=3$, both the recently determined $\Ord(\alpha_s^4)$ 4-loop correction from Eqs.~\eqref{eqn:msrpolenat} and \eqref{eqn:msrpoleprac}~\cite{Marquard:2015qpa,Marquard:2016dcn} and the $\Ord(\alpha_s^6)$ 5-loop correction to the QCD $\beta$-function \cite{Baikov:2016tgj} are required.
To simplify terminology we call the result that truncates the series for $N_{1/2}$ after the $k$-th term the ``($k+1$)-loop'' or ``$\Ord(\alpha_s^{k+1})$ result'', referring to the order to which the series is being probed with the sum rule.

In Fig.~\ref{fig:N12msr}a the numerical results for  $N_{1/2}(\nl=5)$ are shown for the natural (solid lines) and practical (dashed lines) MSR masses for $0.5<\lambda<2$ using terms in the series for $N_{1/2}$ up to $k=0$ (cyan), $k=1$ (blue), $k=2$ (green) and $k=3$ (red).  The thickness of the $\Ord(\alpha_s^{4})$ curves correspond to the numerical error of the coefficients quoted in \cite{Marquard:2016dcn} and shown in 
Eqs.~\eqref{eqn:coeffsanmsrp} and \eqref{eqn:coeffanmsbar}
and indicates that this error is more than an order of magnitude smaller than the uncertainty due to missing higher order terms and therefore negligible. We therefore do not account for this uncertainty any further and adopt the central values given in Eqs.~\eqref{eqn:coeffsanmsrp} and \eqref{eqn:coeffanmsbar}.
Using the $\lambda$ dependence in the range $0.5<\lambda<2$ as an error estimate due to the missing higher orders we obtain for $N_{1/2}(\nl=5)$ at ${\cal O}(\alpha_s^k)$, $k=(1,2,3,4)$ the numerical results $N_{1/2}^\mathrm{nat}(\nl=5)=(0.531\pm 0.318,0.468\pm 0.104,0.483\pm 0.029,0.446\pm 0.024)$ for the natural MSR mass and $N_{1/2}^\mathrm{prac}(\nl=5)=(0.531\pm 0.318,0.475\pm 0.109,0.494\pm 0.032,0.441\pm 0.033)$ for the practical MSR mass. The central values are the mean of the respective maximal and minimal value obtained in the range $0.5<\lambda<2$.  Both results are fully compatible, as is expected since the difference of the natural and practical MSR masses is free from an $\Ord(\LQCD)$ renormalon as already discussed in Sec.~\ref{sec:MSRp}.
We see that the $\lambda$-dependence of $N_{1/2}$ nicely decreases when including more higher-order terms and that there is excellent convergence. The convergence and the reducing $\lambda$-dependence both indicate that the numerical size of the recently calculated 4-loop correction in the $\MSb$-pole mass relation \cite{Marquard:2015qpa,Marquard:2016dcn} is fully compatible with the expectations based on the knowledge of the corrections up to 3~loops and the proposition that the $\MSb$-pole mass is dominated by an $\Ord(\LQCD)$ renormalon behavior already at the known low orders.

It is quite instructive that one can invert this line of arguments and use the sum rule as a tool to determine a prediction for higher order terms in the perturbative series under the assumption that the $\Ord(\LQCD)$ renormalon-type behavior observed at lower orders persists also at higher orders.
Indeed, using for example the $\Ord(\alpha_s^3)$ result for the practical MSR mass 
$N_{1/2}^{\rm prac}(\nl=5)=0.494\pm 0.032$ and the coefficients $a^{\rm MSRp}_{1,2,3}$ of the relation between practical MSR and pole masses [\,see Eqs.~\eqref{eqn:coeffsanmsrp}\,] and the $\beta$-function coefficients up to $\beta_4$ as an input, one can fit for the ${\cal O}(\alpha_s^4)$ coefficient 
giving $a^{\rm MSRp}_4(n_\ell=5)=224620\pm 18656$.  
Converting to the $(\nl+1)$ flavor scheme we obtain for the ${\cal O}(\alpha_s^4)$ coefficient  in the $\MSb$-pole mass relation $a_4^{\MSb}(\nl=5,1)=230192\pm 14747$ compared to the result $a_4^{\MSb}(\nl=5,1)=211807\pm5504$ from \cite{Marquard:2015qpa} and $a_4^{\MSb}(\nl=5,1)=214828\pm422$ from Ref.~\cite{Marquard:2016dcn}. The prediction for the ${\cal O}(\alpha_s^4)$ coefficient based on the sum rules has a larger error but is fully compatible with the results from the explicit loop calculations. This is remarkable given that the sum rule result is obtained with essentially no additional computational effort.  
We note that estimates for the coefficient $a^{\MSb}_4$ were given before for example in Refs.~\cite{Beneke:1994qe,Chetyrkin:1997wm,Kataev:2010zh,Sumino:2013qqa,Ayala:2014yxa}. These were not based on the renormalon sum rule but used available information on the high-order asymptotics of the perturbative series (see Sec.~\ref{sec:asymtotic}).
The analyses of Refs.~\cite{Ayala:2014yxa} and \cite{Sumino:2013qqa} were quoting an uncertainty for the estimate using the known corrections up to ${\cal O}(\alpha_s^3)$ and obtained the results
$a_4^{\MSb}(\nl=5,1)=241920\pm 23552$ and $a_4^{\MSb}(\nl=5,1)=229632\,{}^{+~\,7936}_{-\,44800}$, respectively, which are fully compatible with the sum rule estimate we showed above at the same order.

The results for $N_{1/2}(\nl=5)$ represent the $\Ord(\LQCD)$ renormalon ambiguity for the top quark pole mass assuming that the other quark flavors including {\it the charm and bottom quarks are massless}. The other cases of phenomenological interest are $\nl=3$ and $\nl=4$ and the corresponding results for the natural and practical MSR masses are given in Tab.~\ref{tab:N12msr}.
As our final results for the $N_{1/2}$ values for the {\it number of massless flavors} $\nl=3,4,5$ we quote the 4-loop results for the natural MSR mass
\begin{align} 
N_{1/2}(\nl=3)&=0.526\pm0.016\label{eqn:N12msrfinal3} \,,\\
N_{1/2}(\nl=4)&=0.492\pm0.020\label{eqn:N12msrfinal4} \,,\\
N_{1/2}(\nl=5)&=0.446\pm0.026\label{eqn:N12msrfinal5} \,.
\end{align}
Note that the uncertainties are slightly larger than the ones quoted in Tab.~\ref{tab:N12msr}. Following Ref.~\cite{Beneke:2016cbu} we have also included an additional uncertainty coming from varying the defining coefficients $a_n^{\overline{\rm MS}}=a_n^{\overline{\rm MS}}(\nl,0)$ of the natural MSR mass based on the idea that using the association of $R$ with the $\MSb$ mass at the scale of the $\MSb$ mass is in principle not mandatory. Since one may as well consider different renormalization scales for the $\MSb$ mass and the ${\cal O}(\Lambda_{\rm QCD})$ renormalon ambiguity is not affected by this choice, we have determined modified coefficients $a_n$ from Eq.~(\ref{eqn:msrpolenat}) by setting 
$R=\mbar_Q^{(n_\ell)}(\mbar_Q^{(n_\ell)})$ and completely reexpanding the series in terms of $R^\prime=\mbar_Q^{(n_\ell)}(\eta\, \mbar_Q^{(n_\ell)})$ using the RG equation for the $\MSb$ mass for $n_\ell$ dynamic flavors. Using the resulting series coefficients we have reevaluated the sum rule using variations in $\eta$ between $0.5$ and $2$ and added the resulting uncertainty (while keeping $\lambda=1$)
quadratically to the ones shown in Tab.~\ref{tab:N12msr} (which relate to the choice $\eta=1$). The results including the $\eta$ variation are shown in Fig.~\ref{fig:N12ord} exemplarily for $n_\ell=5$.

The results of Eqs.~(\ref{eqn:N12msrfinal3})\,-\,(\ref{eqn:N12msrfinal5}) are compatible with those given in Refs.~\cite{Ayala:2014yxa,Beneke:2016cbu}. For example for $\nl=5$~\cite{Beneke:2016cbu} obtained
$0.4616^{+0.027}_{-0.070} \pm 0.002$, where the first uncertainty is from a double scale variation similar to ours and the second uncertainty is from the numerical determination of the four loop coefficient. In Refs.~\cite{Ayala:2014yxa,Beneke:2016cbu} the determination of the normalization $N_{1/2}$ was based on the ratio method, which arises from a comparison of the perturbative coefficients $a_n$ from explicit QCD loop calculations to the coefficients $a_n^\mathrm{asy}$ of the series generated by a pure $\Ord(\LQCD)$ renormalon in Eq.~\eqref{eqn:anasy} based on the relation that $\lim_{n\to\infty}a_n/a_n^\mathrm{asy}=1$. In Ref.~\cite{Ayala:2014yxa} the static QCD potential and the $\MSb$-pole mass relation were studied, and in Ref.~\cite{Beneke:2016cbu} the $\MSb$-pole mass was examined. (In Ref.~\cite{Ayala:2014yxa} the static potential based numbers are roughly $1.4\sigma$ higher than those in Eqs.~(\ref{eqn:N12msrfinal3})-(\ref{eqn:N12msrfinal5}), which may be related to the points discussed below in Sec.~\ref{sec:PSmass} for the PS mass.) The agreement of our sum rule results and those obtained from the ratio method in Ref.~\cite{Beneke:2016cbu} underlines the capabilities of R-evolution and the renormalon sum rule concept. 

\renewcommand{\arraystretch}{1.2}\setlength{\LTcapwidth}{\textwidth}

\begin{table}[h]
	\center
	\begin{tabular}{|r|rrrr|}
		\hline
		$n_\ell$~~~~~ & $\Ord(\alpha_s)$~~~~~~& $\Ord(\alpha_s^2)$~~~~~~&
		$\Ord(\alpha_s^3)$~~~~~~&  $\Ord(\alpha_s^4)$~~~~~~~~\\
		\hline
		\multicolumn{5}{|c|}{$N_{1/2}(\nl)$ from $m_t^{\rm MSRn}$ }\\
		\hline
		$-1000000$~~ & $ 0.531 \pm  0.318$ & $ 1.022 \pm  0.378$ & $0.817 \pm  0.121$ & $ 1.009 \pm  0.068$~~  \\
		$-10$~~ & $ 0.531 \pm  0.318$ & $ 0.654 \pm  0.220$ & $ 0.640 \pm  0.062$ & $ 0.684 \pm  0.030$~~  \\
		$0$~~ & $ 0.531 \pm  0.318$ & $ 0.558 \pm  0.169$ & $ 0.567 \pm  0.058$ & $ 0.582 \pm  0.017$~~  \\
		$3$~~ & $ 0.531 \pm  0.318$ & $ 0.514 \pm  0.140$ & $ 0.527 \pm  0.046$ & $ 0.526 \pm  0.012$~~  \\
		$4$~~ & $ 0.531 \pm  0.318$ & $ 0.494 \pm  0.124$ & $ 0.508 \pm  0.039$ & $ 0.492 \pm  0.016$~~  \\
		$5$~~ & $ 0.531 \pm  0.318$ & $ 0.468 \pm  0.104$ & $ 0.483 \pm  0.029$ & $ 0.446 \pm  0.024$~~  \\
		$6$~~ & $ 0.531 \pm  0.318$ & $ 0.434 \pm  0.079$ & $ 0.437 \pm  0.027$ & $ 0.381 \pm  0.038$~~  \\
		$7$~~ & $ 0.531 \pm  0.318$ & $ 0.387 \pm  0.047$ & $ 0.340 \pm  0.059$ & $ 0.271 \pm  0.063$~~  \\
		$8$~~ & $ 0.531 \pm  0.318$ & $ 0.184 \pm  0.141$ & $ 0.165 \pm  0.142$ & $ 0.053 \pm  0.097$~~  \\
		$10$~~ & $ 0.531 \pm  0.318$ & $-\,3.381 \pm  2.714$ & $-\,1.811 \pm 0.492$ & $-\,2.434 \pm  1.041$~~ \\
		\hline
		\multicolumn{5}{|c|}{$N_{1/2}(\nl)$ from $m_t^{\rm MSRp}$  }\\
		\hline
		$-1000000$~~ & $ 0.531 \pm  0.318$ & $ 1.022 \pm  0.378$ & $0.817 \pm  0.121$ & $ 1.009 \pm  0.068$~~  \\
		$-10$~~ & $ 0.531 \pm  0.318$ & $ 0.658 \pm  0.222$ & $ 0.641 \pm  0.062$ & $ 0.684 \pm  0.028$~~  \\
		$0$~~ & $ 0.531 \pm  0.318$ & $ 0.563 \pm  0.172$ & $ 0.572 \pm  0.059$ & $ 0.583 \pm  0.016$~~  \\
		$3$~~ & $ 0.531 \pm  0.318$ & $ 0.520 \pm  0.144$ & $ 0.535 \pm  0.048$ & $ 0.522 \pm  0.017$~~  \\
		$4$~~ & $ 0.531 \pm  0.318$ & $ 0.501 \pm  0.129$ & $ 0.517 \pm  0.041$ & $ 0.487 \pm  0.023$~~  \\
		$5$~~ & $ 0.531 \pm  0.318$ & $ 0.475 \pm  0.109$ & $ 0.494 \pm  0.032$ & $ 0.441 \pm  0.033$~~  \\
		$6$~~ & $ 0.531 \pm  0.318$ & $ 0.442 \pm  0.083$ & $ 0.457 \pm  0.023$ & $ 0.373 \pm  0.052$~~  \\
		$7$~~ & $ 0.531 \pm  0.318$ & $ 0.394 \pm  0.050$ & $ 0.366 \pm  0.051$ & $ 0.259 \pm  0.083$~~  \\
		$8$~~ & $ 0.531 \pm  0.318$ & $ 0.200 \pm  0.134$ & $ 0.201 \pm  0.127$ & $ 0.027 \pm  0.132$~~  \\
		$10$~~ & $ 0.531 \pm  0.318$ & $-\,3.325 \pm  2.681$ & $-\,1.638 \pm 0.439$ & $-\,3.057 \pm  0.649$~~ \\
		\hline
	\end{tabular}
	\caption{\label{tab:N12msr} $N_{1/2}(\nl)$ for the natural and practical heavy quark MSR masses. The results are given for different theoretically interesting values of $\nl$ including contributions from one to four loops. The errors shown are obtained from $\lambda$ variations in the interval $[\,0.5,2\,]$ and the central values are the mean value of the respective maximal and minimal values obtained in that interval.}
\end{table}

In Tab.~\ref{tab:N12msr} we have also shown the results for a number of other $\nl$ values as these results are also of theoretical interest. Our results are in full agreement with and have compatible uncertainties to the results given in Tab.~1 of Ref.~\cite{Beneke:2016cbu} and in particular confirm that $N_{1/2}\to 1$ for $\nl\to-\infty$, which is the classic large-$\nl$ limit where the perturbative series are fully dominated by the massless quark bubble chain and the non-Abelian QCD effects are diluted away.  Our result for $n_\ell=0$ is also in agreement with Ref.~\cite{Ayala:2014yxa} and the lattice determinations of Refs.~\cite{Bali:2013pla,Bali:2013qla}, which found $N_{1/2}(n_\ell=0)=0.600\pm 0.029$, $N_{1/2}(n_\ell=0)=0.660\pm 0.056$ and $N_{1/2}(n_\ell=0)=0.620\pm 0.035$, respectively. We note that our analytic expression for $N_{1/2}$ gets unstable and non-conclusive for $10\lesssim\nl\lesssim 30$ which is the so-called conformal region where the coefficient $\beta_0$ of the QCD $\beta$-function becomes small and in particular $\hat b_1=\beta_1/(2\beta_0^2) $ becomes large. In this region the analytic formula for $N_{1/2}$ has singularities and does not approach any stable value. This is connected to the fact that in this region no definite statement on the asymptotic large order behavior of the perturbative series and in particular on the $\Ord(\LQCD)$ renormalon can be made because the infrared and ultraviolet structure of the QCD $\beta$-function strongly depend on a complicated numerical interplay of the coefficients $\beta_{i>0}$, which can become quite large and have different signs. The unstable behavior of our analytical formula for $10\lesssim\nl\lesssim 30$ differs from the results obtained in Refs.~\cite{Ayala:2014yxa,Beneke:2016cbu}, where the normalization $N_{1/2}$ was observed being tiny.
However, as emphasized in Ref.~\cite{Beneke:2016cbu}, this feature was an artifact of the ratio method used in Refs.~\cite{Ayala:2014yxa,Beneke:2016cbu}, and again indicates that in this $n_\ell$ region the canonical renormalon calculus cannot be applied.

In Ref.~\cite{Pineda:2001zq} the Borel method to compute $N_{1/2}$ was suggested based on the idea that the Borel function $(1-2u)^{1+\hat b_1}B_{\alpha_s}(u)$ eliminates all non-analytic contributions in the first term on the RHS of Eq.~\eqref{eqn:borel2} and thus isolates the term $N_{1/2}$ in the limit $u\to 1/2$~\cite{Lee:1996yk}. This approach entails that after the low-order terms
in the expansion of the Borel transform $B_{\alpha_s}(u)$ around $u=0$ are determined from the original series, one expands $(1-2u)^{1+\hat b_1}B_{\alpha_s}(u)$ in powers of $u$ and subsequently evaluates the resulting series for $u=1/2$.
The results of Refs.~\cite{Pineda:2001zq,Lee:1996yk} were based on the assumption that the analytic contributions [\,involving the functions $Q_\ell(u)$\,] on the RHS of Eq.~\eqref{eqn:borel2} quickly tend to zero
when multiplied by $(1-2u)^{1+\hat b_1}$ and are unimportant. This is not the case, as the Taylor expansion
$(1-2u)^{1+\hat b_1}$ around $u=0$ converges very slowly to zero if one sets $u=1/2$. This can
be traced to the fact that $\hat b_1$ is non-integer and in general the convergence radius of the
binomial series is $1$. Here $u=1/2$ corresponds exactly to the border of this radius.
These terms are therefore numerically sizable at any truncation order. As we show in App.~\ref{sec:N12alternative}, neglecting them leads to a much larger dependence on the renormalization parameter $\lambda$ at a given truncation order. This is because the $\lambda$ dependence of these terms is multiplied by a factor converging to zero, but the convergence is rather slow. When many orders are included, as shown in Ref.~\cite{Bali:2013pla} which accounted for terms up to $\mathcal{O}(\alpha_s^{20})$,
the dependence vanishes and the
method converges to $N_{1/2}$, which we have confirmed through a reanalysis.  This observation is consistent with the large scale uncertainties found in the detailed numerical analysis of Ref.~\cite{Ayala:2014yxa}. The Borel method to determine $N_{1/2}$ is therefore not very precise if only the first few terms of the series are known. Interestingly, accounting for the analytic terms on the RHS of Eq.~\eqref{eqn:borel2}, which are contained in the polynomials $Q_\ell$ and are computed systematically from R-evolution as shown in Sec.~\ref{sec:derivation}, one can derive an improved version of the Borel approach which agrees exactly with our sum rule formula of Eq.~\eqref{eqn:P12def}. The corresponding analytic calculation and a brief numerical analysis are given in App.~\ref{sec:N12alternative}.

\subsection{Asymptotic Higher Order Behavior} 
\label{sec:asymtotic}

In this section we use the analytic manipulations that arise in the derivation of the sum rule to derive an alternative expression for the high-order asymptotic form of a series containing an ${\cal O}(\Lambda_{\rm QCD})$ renormalon that differs from the well known formula derived in ~\cite{Beneke:1994rs}.  	
The latter formula is related to the sum of the non-analytic terms, which are multiplied by $N_{1/2}$ or $P_{1/2}$ in the Borel function of Eq.~\eqref{eqn:borel2}, and reads
\begin{align}\label{eqn:N12asymptotic}
	\Big[ m_Q^\pole & - m_Q^\MSR(R)\Big]_{\rm asy} = 
		N_{\rm 1/2}\, R\,\sum_{n=0}^\infty\,a_{n+1}^{\rm asy} \left(\frac{\alpha_s(R)}{4\pi}\right)^{\!\!n+1}\\
	&=
	N_{\rm 1/2}\, R\,\sum_{n=0}^\infty
	4\pi\,(2\beta_0)^{n}
	\left(\frac{\alpha_s(R)}{4\pi}\right)^{\!\!n+1}
	\sum_{\ell=0}^\infty 
	g_\ell\,\frac{\Gamma(1+\hat b_1+n-\ell)}{\Gamma(1+\hat b_1)}\nonumber\\
	&=
	P_{\rm 1/2}\, R\,\sum_{n=0}^\infty
	\,(2\beta_0)^{n+1}
	\left(\frac{\alpha_s(R)}{4\pi}\right)^{\!\!n+1}
	\sum_{\ell=0}^\infty 
	g_\ell\,\Gamma(1+\hat b_1+n-\ell)\,,\nonumber
\end{align}
giving the asymptotic form of the coefficients
\begin{equation}
\label{eqn:anasy}
a_n^{\rm asy} = 4\pi\,N_{1/2} (2\beta_0)^{n-1}\, \sum_{\ell=0}^{\infty} g_\ell\,
(1+\hat b_1)_{n-1-\ell}\,,
\end{equation}
where $(b)_n=b\, (b+1)\cdots(b+n-1)=\Gamma(b+n)/\Gamma(b)$ is the Pochhammer symbol.
Given the value for $P_{1/2}$ or $N_{\rm 1/2}$ the structure of the perturbative coefficients of Eq.~\eqref{eqn:N12asymptotic} is completely fixed by the properties of the QCD $\beta$-function and does not depend any more on the coefficients of the original series of Eqs.~\eqref{eqn:msrpolenat} and \eqref{eqn:msrpoleprac}. Thus Eq.~\eqref{eqn:N12asymptotic} has been frequently used as the standard form for the asymptotic high-order behavior of perturbative series dominated by an ${\cal O}(\Lambda_{\rm QCD})$ renormalon.
This is also reflected by the fact that the imaginary part of the inverse Borel integration over the non-analytic terms in Eq.~\eqref{eqn:borel2} is exactly proportional to $\LQCD$
\begin{align}\label{eqn:borelambiguity}
  \mathrm{Im}&\int_{0}^{\infty}\dd u\bigg[-N_{1/2}\,R\,\frac{4\pi}{\beta_0}\sum_{\ell=0}^{\infty}\,g_\ell\,
  \frac{\Gamma(1+\hat b_1-\ell)}{\Gamma(1+\hat b_1)}\,(1-2u)^{-1-\hat b_1+\ell}\,\bigg]
   \mathrm e^{-\frac{4\pi u}{\beta_0\alpha_s(R)}}\\
	&=P_{1/2}\,\pi\,\LQCD=N_{1/2}\,\frac{2\pi^2}{\beta_0\,\Gamma(1+\hat b_1)}\,\LQCD \,,\nonumber
\end{align}
with $\LQCD$ given in Eq.~\eqref{eqn:lambda}. 
As a side remark, we note that inserting the series in Eq.~\eqref{eqn:N12asymptotic}, with a given value for $N_{1/2}$, into the sum rule expression of Eq.~\eqref{eqn:P12def} one recovers $N_{1/2}$ in the limit of carrying out the sums over $k$, $n$ and $\ell$ to infinity.

Interestingly, Eq.~\eqref{eqn:msrpolefull2} provides a remarkable alternative expression for the high-order asymptotic of the MSR-pole mass series as it can be rewritten in the form
\begin{align}\label{eqn:fullasymptotic}
m_Q^\pole-m_Q^\MSR(R) &=
R\,\sum_{n=0}^\infty \left(\frac{\alpha_s(R)}{4\pi}\right)^{\!\! n+1}\sum_{k=0}^n \sum_{\ell=0}^{n-k} (2\beta_0)^{n+1} S_k \,g_\ell\, \frac{\Gamma(1+\hat b_1+n-\ell)}{\Gamma(1+\hat b_1+k)
}	 \,.
\end{align}
In contrast to Eq.~\eqref{eqn:N12asymptotic} this expression still depends on the $S_k$ coefficients non-trivially and thus carries all the information contained in the original series due to the identity
\begin{equation}
\label{eqn:anintermsofsk}
a_n = (2\beta_0)^n\, \sum_{k=0}^{n-1} \, S_k\! \sum_{\ell=0}^{n - 1-k} g_\ell\, (1+\hat b_1 + k)_{n - 1 -\ell- k}\,.
\end{equation}
This relation is interesting because it provides a separation of the coefficients of the original series into leading and subleading terms with respect to the asymptotic high-order behavior.
So truncating the sums over $k$ and $\ell$ in Eq.~\eqref{eqn:anintermsofsk} (e.g. accounting for the coefficients $S_k$ and $g_\ell$ up to the order they are known) provides the correct high-order asymptotic behavior for $n$ beyond the truncation order and, at the same time, reproduces exactly the coefficients of the original series up to the truncation order. 

Currently the coefficients $a_n$ for the MSR-pole and the $\MSb$-pole mass relations are known to order ${\cal O}(\alpha_s^4)$ and the QCD $\beta$-function is known to order ${\cal O}(\alpha_s^6)$ so that the coefficients $S_k$ and $g_\ell$ are known up to $k_{\rm max}=\ell_{\rm max}=3$. We may therefore write down estimates for the still uncalculated coefficients $a_{n>4}$ using the expression
\begin{equation}
\label{eqn:anestimated}
a_{n>4}^{\rm asy} = 
4 \pi \,N_{\rm 1/2}\,(2\beta_0)^{n-1}
	\sum_{\ell=0}^3 
	g_\ell\,
	(1+\hat b_1)_{n-1-\ell}\,,
\end{equation}
which is the established formula from~\cite{Beneke:1994rs} shown in Eq.~\eqref{eqn:N12asymptotic}, and
\begin{equation}
\label{eqn:anestimatedprime}
a_{n>4}^{\rm asy\,\prime} = (2\beta_0)^n
\sum_{k=0}^{3}
S_k \!\!\!\!  
 \sum_{\ell=0}^{\min(n-k-1,3)}
\!\!\!\!\!\!g_\ell\, (1+\hat b_1 + k)_{n - 1 -\ell- k}\,,
\end{equation}
based on Eq.~\eqref{eqn:anintermsofsk},  which encodes information on both the regular
and asymptotic behavior of the series.\footnote{One can easily write Eq.~\eqref{eqn:anestimatedprime} as
the sum of Eq.~\eqref{eqn:anestimated} and a term build from the inverse Borel transform of the $Q_\ell$
polynomials defined in Eq.~\eqref{eqn:Qldef}.}
In Tab.~\ref{tab:anestimates} we show estimates for the yet uncalculated coefficients $a_{5\le n\le 9}$ for the relations of the natural MSR mass and the $\MSb$ mass $\mbar_Q \equiv \mbar_Q^{(n_\ell+1)}(\mbar_Q^{(n_\ell+1)})$
to the pole mass 
using Eqs.~\eqref{eqn:anestimated} and \eqref{eqn:anestimatedprime} for $\nl=3,4,5$ and the results of Eqs.~\eqref{eqn:N12msrfinal3}--\eqref{eqn:N12msrfinal5} for $N_{1/2}$. The uncertainties for the coefficients $a_{n}^{\rm asy}$ are based on the uncertainties shown in Eqs.~\eqref{eqn:N12msrfinal3}\,--\,\eqref{eqn:N12msrfinal5} and those for the coefficients $a_{n}^{\rm asy\,\prime}$ are determined from $\lambda$ variations $1/2<\lambda<2$, as explained in Sec.~\ref{sec:renormalonsumrule} and $\eta$ variations $1/2<\eta<2$, as explained below Eq.~(\ref{eqn:N12msrfinal5}). The coefficient estimates for the $\MSb$ mass have been obtained by using the second equality of~\eqref{eqn:msbmsr2} and Eq.~\eqref{eqn:asmatchingmsbar} to the order shown.
We see that both estimates are completely equivalent and have the same uncertainties.
Our estimates for the  $\MSb$ mass coefficients for $\nl=5$ also agree perfectly with those given in Ref.~\cite{Beneke:2016cbu} which used the approach of Eq.~\eqref{eqn:anestimated}.
We note that the relation~\eqref{eqn:anintermsofsk} can also be inverted to provide closed iterative expressions for the $S_k$ coefficients to all orders, which are given in App.~\ref{app:coefficients} and in particular in Eq.~\eqref{eqn:skcoeff3}.

We note that the asymptotic series coefficients $a_n^{\rm asy}$ in Eq.~\eqref{eqn:anasy} 
and the expression for the coefficients $a_n$ in Eq.~\eqref{eqn:anintermsofsk} allow for an alternative derivation of the renormalon sum rule formula since the ratio $a_n/a_n^{\rm asy}$ approaches unity for $n\to\infty$. Taking that ratio  
one arrives at
\begin{align}
\label{eqn:anratio}
N_{1/2}\frac{a_n}{a_n^{\rm asy}}& \,=\,\frac{ (2\beta_0)^n\, \sum\limits_{k=0}^{n-1} \, S_k\! \sum\limits_{\ell=0}^{n - 1-k} g_\ell\, (1+\hat b_1 + k)_{n - 1 -\ell- k}}{4\pi\,(2\beta_0)^{n-1}\, \sum\limits_{\ell=0}^{\infty} g_\ell\,
	(1+\hat b_1)_{n-1-\ell}} \\ 
&\,=\,  
 \frac{\beta_0\,\Gamma(1+\hat b_1)}{2 \pi\,}\,
 \sum_{k=0}^{n-1}\frac{S_k}{\Gamma(1+\hat b_1+k)}\,
 \frac{\sum\limits_{\ell=0}^{n-1-k} \, g_\ell\, \Gamma(\hat b_1+n-\ell)}{\sum\limits_{\ell=0}^{\infty} \ g_\ell \,\Gamma(\hat b_1+n-\ell)}\,. \nonumber
\end{align}
To the extent that the sums over $k$ in the sum rule formula of Eq.~\eqref{eqn:P12def} 
and in Eq.~\eqref{eqn:anratio} for $n\to\infty$ 
are convergent, one can use the Cauchy convergence  
criterion to show that the expression of Eq.~\eqref{eqn:anratio} is equivalent to Eq.~\eqref{eqn:P12def} for $n\to\infty$. This shows analytically the equivalence of the ratio method and the sum rule.

		\begin{table}[t]
		\center
		\begin{tabular}{|c|ccccc|}
			\hline
			\!$\nl$\! & \!$a^\mathrm{MSRn}_5\times 10^{-7}$\! & \!$a^\mathrm{MSRn}_6\times 10^{-9}$\! & \!$a^\mathrm{MSRn}_7\times 10^{-11}$\! & \!$a^\mathrm{MSRn}_8\times 10^{-13}$\! & \!$a^\mathrm{MSRn}_9\times 10^{-15}$\!\\\hline
			$3$ & \!$ 3.394\pm 0.105$\! & \!$ 3.309 \pm 0.102$\! & \!$3.819 \pm 0.118$\! & \!$ 5.093 \pm 0.157$\! & \!$7.706 \pm 0.238$\!\\
			$4$ & \!$ 2.249\pm 0.090$\! & \!$ 2.019 \pm 0.081$\! & \!$2.147 \pm 0.086$\! & \!$ 2.641 \pm 0.106$\! & \!$3.687 \pm 0.148$\!\\
			$5$ & \!$ 1.379\pm 0.080$\! & \!$ 1.128 \pm 0.066$\! & \!$1.095 \pm 0.064$\! & \!$1.231 \pm 0.072$\! & \!$1.572 \pm 0.091$\!\\\hline
			& \!$a^\mathrm{MSRn\,\prime}_5\times 10^{-7}$\! & \!$a^\mathrm{MSRn\,\prime}_6\times 10^{-9}$\! & \!$a^\mathrm{MSRn\,\prime}_7\times 10^{-11}$\! & \!$a^\mathrm{MSRn\,\prime}_8\times 10^{-13}$\! & \!$a^\mathrm{MSRn\,\prime}_9\times 10^{-15}$\!\\\hline
			$3$ & $ 3.393\pm 0.105$ & $ 3.309 \pm 0.102$ & $3.819 \pm 0.118$ & $ 5.093 \pm 0.157$ & $7.706 \pm 0.238$\\
			$4$ & $ 2.248\pm 0.090$ & $ 2.019 \pm 0.081$ & $2.147 \pm 0.086$ & $ 2.641 \pm 0.106$ & $3.687 \pm 0.148$\\
			$5$ & $ 1.378\pm 0.080$ & $ 1.128 \pm 0.066$ & $1.095 \pm 0.063$ & $1.231 \pm 0.072$ & $1.572 \pm 0.091$\\\hline
			& \!$a^{\MSb}_5\times 10^{-7}$\! & \!$a^{\MSb}_6\times 10^{-9}$\! & \!$a^{\MSb}_7\times 10^{-11}$\! & \!$a^{\MSb}_8\times 10^{-13}$\! & \!$a^{\MSb}_9\times 10^{-15}$\!\\\hline
			$3$ & $ 3.401\pm 0.105$ & $ 3.315 \pm 0.102$ & $3.824 \pm 0.118$ & $ 5.099 \pm 0.158$ & $7.714 \pm 0.239$\\
			$4$ & $ 2.255\pm 0.090$ & $ 2.023 \pm 0.081$ & $2.151 \pm 0.086$ & $ 2.644 \pm 0.106$ & $3.692 \pm 0.148$\\
			$5$ & $ 1.383\pm 0.080$ & $ 1.130 \pm 0.066$ & $1.097 \pm 0.064$ & $1.233 \pm 0.072$ & $1.575 \pm 0.091$\\\hline
			& \!$a^{\MSb\,\prime}_5\times 10^{-7}$\! & \!$a^{\MSb\,\prime}_6\times 10^{-9}$\! & \!$a^{\MSb\,\prime}_7\times 10^{-11}$\! & \!$a^{\MSb\,\prime}_8\times 10^{-13}$\! & \!$a^{\MSb\,\prime}_9\times 10^{-15}$\!\\\hline
			$3$ & $ 3.400\pm 0.106$ & $ 3.315 \pm 0.103$ & $3.824 \pm 0.118$ & $ 5.099 \pm 0.158$ & $7.714 \pm 0.239$\\
			$4$ & $ 2.254\pm 0.091$ & $ 2.023 \pm 0.081$ & $2.151 \pm 0.086$ & $ 2.644 \pm 0.106$ & $3.692 \pm 0.148$\\
			$5$ & $ 1.382\pm 0.081$ & $ 1.130 \pm 0.066$ & $1.097 \pm 0.064$ & $1.233 \pm 0.072$ & $1.575 \pm 0.091$\\\hline
		\end{tabular}
 \caption{\label{tab:anestimates} Numerical estimates for the perturbative coefficients $a^\mathrm{MSRn}_n$ (MSRn-pole mass relation in Eq.~\eqref{eqn:msrpolenat}) and $a^{\MSb}_n$ [\,$\MSb$-pole mass relation in Eq.~\eqref{eqn:msbarpoleseries}\,] for $5\le n \le 9$ and $\nl=3,4,5$ using formulae \eqref{eqn:anestimated} and \eqref{eqn:anestimatedprime} for their asymptotic high-order behavior. The quoted errors arise from $\lambda$ and $\eta$ variations in the interval $[0.5,2]$ and the central values are the mean of the maximum and minimum values in that interval.}
\end{table}

\subsection{Other Applications of the Sum Rule}\label{sec:sumruleapplication}

To conclude our considerations concerning the $\Ord(\LQCD)$ renormalon sum rule we discuss in this section a number of subtleties in its proper use and a few interesting applications. As it is sufficient for the purpose of the examinations, we use for simplicity only $\lambda$ variations, as explained in Sec.~\ref{sec:renormalonsumrule}, when quoting uncertainties of the sum rule evaluated here.

\subsubsection{Number of Massless Flavors}

An important feature of the $\Ord(\LQCD)$ renormalon sum rule is that it probes the infrared sensitivity of the perturbative series, which physically depends on the number of massless quarks, $\nl$, one employs in the computation of the series. In a computation in QCD, however, $\nl$ might not be equal to the number of active flavors, $n_f$, which governs the ultraviolet behavior and the renormalization group evolution of the strong coupling $\alpha_s^{(n_f)}$ and other renormalized quantities, and a naive application of the sum rule may lead to inconsistent results. In such a case, the series in $\alpha_s^{(n_f)}$ should be better converted to the $\nl$-flavor scheme for the strong coupling, $\alpha_s^{(\nl)}$, before its coefficients are inserted in the sum rule expression. This can be either realized by simply rewriting $\alpha_s^{(n_f)}$ as a series in $\alpha_s^{(\nl)}$, as it is done in the definition of the practical MSR mass, or by integrating out the effects of the $n_f-\nl$ massive quarks, as it is done in the definition of the natural MSR mass. The latter approach is the physically cleaner way (which was the reason for using the name `natural'), but both approaches are consistent as far as the application of the sum rule is concerned. 

In the following we discuss the pitfalls of using the sum in an inconsistent way. To discuss the issue we recall that, since the  $\Ord(\LQCD)$ renormalon sum rule is a functional on the perturbative series, it can also be seen as a function $N_{1/2}[\,n_\ell,\{a_n\}]$ acting on the coefficients $a_n$ of the $[\,\alpha_s/(4\pi)\,]^n$ terms in the series. As indicated, $N_{1/2}$ is a function of the number of massless flavors $\nl$ through its dependence on $\beta_0$ and the coefficients $\hat b_k$, which appear in Eq.~\eqref{eqn:P12def} and a function of the coefficients $a_n$ contained in the expressions for the $S_k$ as shown in Eq.~\eqref{eqn:skcoeff}.   
The function $N_{1/2}[\,n_\ell,\{a_n\}]$ is therefore probing the series defined by the set of coefficients $\{a_n\}$ with respect to an $\Ord(\LQCD)$ renormalon for $n_\ell$ massless flavors, and it is essential for the sum rule to work properly that the value of $n_\ell$ agrees with the number of massless flavors used for the computation of the coefficients $a_n$. Let us now apply the sum rule to the coefficients $\{a_n^{\overline{\rm MS},\nl}\}$ of the series for \mbox{$m_Q^\pole-\mbar_Q(\mbar_Q)^{(\nl+1)}$} in Eq.~\eqref{eqn:msbarpoleseries}, which is a series in $\alpha_s^{(\nl+1)}$, but contains the effects of $n_\ell$ massless flavors.
Here we use the shorthand notation 
\begin{align}
 a_n^{\overline{\rm MS},\nl} \equiv a_n^{\overline{\rm MS}}(\nl,n_h=1) \,.
\end{align}
To be specific we  take $n_\ell=5$. Probing the series with respect to an $\Ord(\LQCD)$ renormalon for $\nl+1=6$ massless flavors, in accordance with the scheme for $\alpha_s$, one obtains 
$N_{1/2}[\,6,\{a_n^{\rm \overline{\rm MS},\nl=5}\}]=(0.531\pm 0.318,0.526\pm 0.1298,0.623\pm0.070,0.6360\pm0.016)$ at order $n=(0,1,2,3)$, 
where the errors are obtained from varying $\lambda$ in the range $0.5<\lambda<2$ 
and the central values are the mean value of the respective maximal and minimal values obtained in the $\lambda$ variation.
We see that the sum rule appears to approach a value that is much larger than the correct result of Eq.~\eqref{eqn:N12msrfinal5}, but this is a consequence of an inconsistent application of the sum rule. Indeed, one can show by simple algebra in the $\beta_0$/LL approximation [\,where $\hat b_{i\geq 1}=\beta_{i\geq 1}=0$, $a_{n+1}^{{\rm asy},n_\ell}=a_1(2\beta_{0,\nl})^{n} n!$ and $\beta_{0,n_\ell}= 11-2/3\,n_\ell$\,] that the order $n$ expression for $N_{1/2}$ that is obtained -- when probing with respect to an $\Ord(\LQCD)$ renormalon for $n_f$ massless flavors -- has the form 
\begin{equation}\label{eqn:N12inconsistent}
	\leris{N_{1/2}^{(n)}[\,n_f,\{a_n^{{\rm asy},n_\ell}\}]}_{\beta_0\mathrm{/LL}}=\frac{\beta_0}{2\pi}\sum_{k=0}^{n}\frac{S_k}{k!}=\frac{a_1}{4\pi}\leri{\frac{\beta_{0,\nl}}{\beta_{0,n_f}}}^{\!\!n} \,.
\end{equation}
As long as $\beta_{0,n}$ is a positive number this expression diverges for $n_f>\nl$ in the limit $n\to\infty$, which explains the behavior of the sum rule results shown above. On the other hand, the expression of Eq.~\eqref{eqn:N12inconsistent} 
converges to zero for $n_f<\nl$. So when probing the 
coefficients $\{a_n^{\overline{\rm MS},\nl}\}$ of the series for $m_Q^\pole-\mbar_Q(\mbar_Q)^{(\nl+1)}$  
with respect to an $\Ord(\LQCD)$ renormalon for $\nl-1=4$ massless flavors we obtain $N_{1/2}[\,4,\{a_n^{\rm \overline{\rm MS},\nl=5}\}]=(0.531\pm0.318,0.433\pm0.089,0.405\pm0.027,0.327\pm0.051)$ at order $n=(0,1,2,3)$ which is a sequence of decreasing terms, as expected from Eq.~\eqref{eqn:N12inconsistent}, which in addition does not behave in a stable way. But, again, the behavior is a consequence of an inconsistent application of the sum rule. 
On the other hand, if we probe the 
coefficients $\{a_n^{\overline{\rm MS},\nl}\}$ of the series for $m_Q^\pole-\mbar_Q(\mbar_Q)^{(\nl+1)}$  
with respect to an $\Ord(\LQCD)$ renormalon for $\nl=5$ massless flavors we obtain
$N_{1/2}[\,5,\{a_n^{\rm \overline{\rm MS},\nl=5}\}]=(0.531\pm0.318,0.475\pm 0.109,0.494\pm 0.032,0.442\pm 0.033)$ at order $n=(0,1,2,3)$, which converges to the correct result of Eq.~\eqref{eqn:N12msrfinal5}. We also learn that adopting for the strong coupling $\alpha_s^{(n_f)}$ a flavor number scheme where $n_f$ agrees with the number of massless flavors is clean conceptually, but not crucial numerically such that the sum rule works reliably. This is related to the fact that the matching relation of the strong coupling in different flavor number schemes does not suffer from an $\Ord(\LQCD)$ renormalon behavior.

This brief examination above underlines the importance that the $\Ord(\LQCD)$ sum rule, which probes the infrared sensitivity of the perturbative series, is applied consistently with respect to the number of massless quarks, which may not agree with the number of active flavors in the normalization group equation that is governed by ultraviolet effects. Of course this feature may as well be used as a tool, as studying the convergence of the sum rule may be employed to determine the number of massless flavors used, let's say, in a numerical computation of a perturbative series.

\subsubsection{Moments of the Vacuum Polarization Function}

The zero-momentum moments $M_i$, $i=1,2,3,\ldots$, of the massive quark $Q$ vector current correlator $\Pi(q^2)$, defined by [\,$j^\mu(x)\equiv\overline\psi_Q(x)\gamma^\mu\psi_Q(x)$\,]
\begin{align}\label{eqn:Mm1}
  M_i &=\left.\frac{12\pi^2 Q_Q^2}{m!}\frac{\dd^i}{\dd q^{2i}}\Pi(q^2)\right|_{q^2=0} \,,\\
 \leri{g_{\mu\nu}q^2-q_\mu q_\nu}\Pi(q^2)&=-\,i\!\int\dd x\,\mathrm e^{iqx}
\left\langle 0\middle|\mathrm T j_\mu(x)j_\nu(0)\middle| 0\right\rangle \,, \nonumber
\end{align}
provide one of the most precise methods to determine the charm and bottom quark $\MSb$ masses~\cite{
Dehnadi:2011gc,Bodenstein:2011ma,Bodenstein:2011fv,Hoang:2012us,Chakraborty:2014aca, 
Colquhoun:2014ica,Beneke:2014pta,Ayala:2014yxa,Dehnadi:2015fra,Erler:2016atg} and are known to utterly fail in precision when expressed in terms of the charm and bottom pole masses.
This mass sensitivity comes from the fact that the perturbative series for the moments $M_i$ is due to dimensional reasons proportional to $m_Q^{-2i}$ in the form $M_i=m_Q^{-2i}\sum_{n=0}^\infty c_{i,n}(m_Q)[\,\alpha_s^{(n_\ell)}(m_Q)/(4\pi)\,]^n$, where $\nl$ is the number of massless flavors and we use the $\nl$-flavor scheme for the strong coupling.\footnote{
In the recent sum-rule analyses~\cite{
Dehnadi:2011gc,Bodenstein:2011ma,Bodenstein:2011fv,Hoang:2012us,Chakraborty:2014aca, 
Colquhoun:2014ica,Beneke:2014pta,Ayala:2014yxa,Dehnadi:2015fra,Erler:2016atg}
for the bottom quark mass $\nl=4$ was used, while for charm mass determinations $\nl=3$ was employed, and the $(\nl+1)$ flavor scheme was employed for the renormalization group evolution.}
The moments $M_i$ are related to weighted integrals over the hadronic R-ratio of $Q\overline Q$ production and thus free from the $\Ord(\LQCD)$ renormalon. They can be rewritten in the form
\begin{align}\label{eqn:Mm2}
 & m_Q-\leri{\frac{M_i}{c_{i,0} }}^{-\frac{1}{2i}}=m'_Q\sum_{n=1}^\infty
 a_{i,n}[\,m_Q, m'_Q\,]\bigg(\frac{\alpha_s^{(n_\ell)}(m'_Q)}{4\pi}\bigg)^{\!\!n}\,,
\end{align}
where $m_Q$ and $m'_Q$ may be in general different quark mass schemes.

The moments $M_i$ are suitable quantities to discuss the parametric aspect of renormalon ambiguities and how they affect the proper application of the $\Ord(\LQCD)$ sum rule. The first three moments $M_{1,2,3}$ are known to $\Ord(\alpha_s^3)$ \cite{Kallen:1955fb,Chetyrkin:1995ii, Chetyrkin:1996cf, Boughezal:2006uu, 
Czakon:2007qi,Maier:2007yn, Chetyrkin:2006xg, Boughezal:2006px, Sturm:2008eb,Maier:2008he, Maier:2009fz} and the corresponding series coefficients $a_{i,n}$  for $\nl=4$
in the $\MSb$ mass scheme $m_Q=m'_Q=\mbar_Q^{(\nl+1)}(\mbar_Q^{(\nl+1)})$ and the pole mass scheme $m_Q=m'_Q=m_Q^\pole$ using the $\nl$-flavor scheme $\alpha_s^{(\nl)}$ for the coupling are quoted in Tab.~\ref{tab:moments}.
Applying the sum rule to the series for the $M_{1,2,3}$ on the RHS of Eq.~\eqref{eqn:Mm2} in the $\MSb$ scheme we obtain for $\nl=4$, relevant for the bottom quark, the results
\begin{align}
 N_{1/2}^{i=1}&=(0.477\pm 0.286,-\,0.178\pm 0.261,~~\,\,0.013\pm 0.036) \,,\\
 N_{1/2}^{i=2}&=(0.241\pm 0.145,-\,0.007\pm 0.083,-\,0.029\pm 0.058)\,,\nonumber\\
 N_{1/2}^{i=3}&=(0.127\pm 0.076,~~\,\,0.031\pm 0.026,-\,0.029\pm 0.048) \,,\nonumber
\end{align}
at order $n=(0,1,2)$, where the errors are obtained by $\lambda$ variations in the range $0.5<\lambda<2$ and the central values are obtained from the mean of the respective maximal and minimal values in the $\lambda$ variation.
\begin{table}[t]
\center
\begin{tabular}{|c|ccc|ccc|}
\hline
~$i$~ & ~~$\Ord(\alpha_s)$~~ & ~~$\Ord(\alpha_s^2)$~~ & ~~$\Ord(\alpha_s^3)$~~
& ~~$\Ord(\alpha_s)$~~ & ~~$\Ord(\alpha_s^2)$~~ & ~~~$\Ord(\alpha_s^3)$~~~\\
\hline
 & \multicolumn{3}{|c|}{$a_{i,n}[\,m_{\pole},m_{\pole}\,]$} &
   \multicolumn{3}{|c|}{$a_{i,n}[\,\overline m(\overline m),\overline m(\overline m)\,]$}\\
\hline
1 & ~$10.1235$~&~$83.7296$~&~$4669.92$~&~$4.79012$~&~$-\,10.7255$~&~$-\,310.275$~\\
2 & ~$7.76049$  & $120.609$ & ~$4589.81$ & ~$2.42716$ & ~~~\,$13.5516$ & $-\,334.42$ \\
3 & ~$6.61153$  & $127.821$ & ~$4754.39$ & ~$1.2782$  & ~~~\,$14.6354$ & $-\,199.81$ \\
\hline
& \multicolumn{3}{|c|}{$a_{i,n}[\,m_{\pole}, \overline m(\overline m)\,]$} &
  \multicolumn{3}{|c|}{$a_{i,n}[\,m_{\pole}, \widetilde M\,]$}\\\hline
1 &~$10.1235$~&~$137.721$~&~$5719.41$~& $10.1235$ & $186.214$ & $ 5831.25$\\
2 & ~~$7.76049$ & ~$161.998$ & ~$5695.26$~& $7.76049$ & $ 180.834$ & $ 6005.71$\\
3 & ~~$6.61153$ & ~$163.082$ & ~$5829.87$~& $6.61153$ & $ 171.533$ & $ 6063.32$\\
\hline
\end{tabular}
\caption{\label{tab:moments} $a_{i,n}(m_Q,m'_Q)$ coefficients of the perturbative expansion for the mass-subtracted linearized 
moments, as displayed in Eq.~\eqref{eqn:Mm2}, at one (left column of each block), two  (middle column of each block), and three
(right column of each block) loops. The numerical values correspond to the case $\nl = 4$, studied in this section. The table is
split into four blocks: the upper left one corresponds to the pole mass expansion in terms of the pole mass, the upper right
one shows the $\overline{\rm MS}$ mass expansion in terms of the $\overline{\rm MS}$ mass, the lower left block displays the pole
mass expansion in terms of the $\overline{\rm MS}$ mass, and the lower right displays the linearized iterative expansion for the
pole mass.}
\end{table}
We see that the results for $N_{1/2}$ are compatible with zero beyond ${\cal O}(\alpha_s)$
and have uncertainties that decrease with order, illustrating the known fact that the series are free from an $\Ord(\LQCD)$ renormalon in the $\MSb$ mass scheme.

Applying the sum rule to the series for the $M_{1,2,3}$ in the pole mass scheme $m_Q=m'_Q=m_Q^\pole$ the corresponding results for $\nl=4$ read
\begin{align}
 N_{1/2}^{i=1}&=(1.007\pm 0.604,\,0.092\pm 0.278,\,0.510\pm 0.113) \,,\\
 N_{1/2}^{i=2}&=(0.772\pm 0.463,\,0.345\pm0.094,\,0.420\pm 0.012) \,,\nonumber\\
 N_{1/2}^{i=3}&=(0.658\pm 0.395,\,0.416\pm 0.053,\,0.424\pm 0.013) \,.\nonumber
\end{align}
Apart from the outcome for $M_1$, which still happens to have a rather large error at order $n=2$ the results converge to the result $0.42\pm 0.01$ which is incompatible with the correct result $0.49\,\pm\, 0.02$  from Eq.~\eqref{eqn:N12msrfinal4}. So the $\Ord(\LQCD)$ renormalon ambiguity inherent to the coefficients in the series of Eq.~\eqref{eqn:Mm2} in the pole mass scheme appears to be about $15\%$ smaller than for the coefficients of the MSR-pole mass series analyzed before. The discrepancy is resolved by the fact that in the pole scheme with both $m_Q=m'_Q=m_Q^\pole$ the RHS of Eq.~\eqref{eqn:Mm2} is expressed using the ambiguous pole mass as a parameter. As a consequence, the perturbative coefficients of the series and factors of $m_Q^{\pole}$ on the RHS share the full $\Ord(\LQCD)$ pole mass renormalon ambiguity contained in the LHS of Eq.~\eqref{eqn:Mm2}. 

To recover the full $\Ord(\LQCD)$ pole mass renormalon ambiguity in the coefficients on the RHS one has to rewrite the series on the RHS in terms of parameters that are free from the $\Ord(\LQCD)$ renormalon ambiguity. This can be achieved by re-expanding the series for $m^\pole_Q-(M_m/c_{i,0})^{-1/(2i)}$ completely in terms of the $\MSb$ mass using $m'_Q=\mbar_Q^{(\nl+1)}(\mbar_Q^{(\nl+1)})$. The resulting coefficients in powers of $\alpha_s^{(\nl)}(m'_Q)$ are given in the lower left column of Tab.~\ref{tab:moments}. Using these coefficients, the renormalon sum rule applied to the series for the $M_{1,2,3}$ and $\nl=4$ gives 
\begin{align}\label{eqn:pole1}
 N_{1/2}^{i=1}&=(1.007\pm 0.604,\,0.350\pm 0.159,\,0.547\pm 0.047) \,,\\
 N_{1/2}^{i=2}&=(0.772\pm 0.463,\,0.525\pm 0.078,\,0.495\pm 0.032) \,,\nonumber\\
 N_{1/2}^{i=3}&=(0.658\pm 0.395,\,0.535\pm 0.110,\,0.501\pm 0.034) \,,\nonumber
\end{align}
at order $n=(0,1,2)$. This is in full agreement with the result $0.49\,\pm\, 0.02$  given in Eq.~\eqref{eqn:N12msrfinal4}, and also shows a substantially better behavior for the moment $M_1$. 

As an alternative to using the series for $m_Q=m_Q^\pole$ and $m'_Q=\mbar_Q^{(\nl+1)}(\mbar_Q^{(\nl+1)})$, one can also define  
$\widetilde M_i\equiv (M_i/c_{i,0})^{-1/2i}$ and re-express the RHS of Eq.~\eqref{eqn:Mm2} perturbatively in terms of $m_Q^\prime=\widetilde M_i$ for the different moments. (We refer to Ref.~\cite{Dehnadi:2011gc} for details on this iterative procedure.)
The resulting coefficients in powers of $\alpha_s^{(\nl)}(\widetilde M_i)$ are given in the lower right column of Tab.~\ref{tab:moments}. Using these coefficients, the renormalon sum rule applied to the series for the $M_{1,2,3}$ and $\nl=4$ gives 
\begin{align}
  N_{1/2}^{i=1}&=(1.007\pm 0.604,\,0.604 \pm 0.075,\,0.493 \pm 0.071) \,,\\
  N_{1/2}^{i=2}&=(0.772\pm 0.463,\,0.589 \pm 0.109,\,0.501 \pm 0.056) \,,\nonumber\\
  N_{1/2}^{i=3}&=(0.658\pm 0.395,\,0.568 \pm 0.129,\,0.516 \pm 0.040) \,.\nonumber
\end{align}
These results behave similarly to those of Eq.~\eqref{eqn:pole1} and are again in full agreement with the result $0.49\,\pm\, 0.02$ given in Eq.~\eqref{eqn:N12msrfinal4}.

This analysis underlines the importance of using renormalon-free parameters for series coefficients that are being probed with the $\Ord(\LQCD)$ renormalon sum rule, but also illustrates the high sensitivity of the sum rule to even subtle high order effects.

\subsubsection{Infrared Sensitivity of the PS Mass Definition}
\label{sec:PSmassIR}

The PS (potential subtracted) mass~\cite{Beneke:1998rk} is based on the concept that the total static potential energy of a color singlet massive quark-antiquark pair with separation $r$, $2m_Q^\pole+V(r)$, is $\Ord(\LQCD)$ renormalon free. It is defined from the integral 
\begin{align}
\label{eqn:PSmassdef}
  m_Q^\pole-m_Q^\PS(\mu_f)=-\,\frac12 \int_{|\vec q\,|<\mu_f} \frac{\dd^3\vec q}{(2\pi)^3}\,\tilde V(\vec q^{\,2}) \,,
\end{align} 
where $\tilde V(\vec q^{\,2})$ is the momentum-space static potential calculated in perturbation theory. To the extent that the total static potential is a well-defined and unambiguous quantity, the PS mass is free from an $\Ord(\LQCD)$ renormalon. The coefficients of the series for $m_Q^\pole-m_Q^\PS(\mu_f)$, expressed as a series in powers of $\alpha_s^{(\nl)}(\mu_f)/(4\pi)$, are given in Eq.~\eqref{eqn:PSmass}. 

We now apply the $\Ord(\LQCD)$ renormalon sum rule to the relation of the pole mass to the potential PS mass. The examination is of interest because the static potential has infrared divergences starting at $\Ord(\alpha_s^4)$ arising from higher Fock $Q\overline Q$-gluon states which lead to retardation effects that invalidate the frame-independent static limit~\cite{Appelquist:1977tw,Appelquist:1977es}. The definition of the PS mass at $\Ord(\alpha_s^4)$ and beyond is therefore  known to depend on the scheme used for the subtraction prescription for these infrared divergences. In Refs.~\cite{Beneke:2005hg} the authors defined the following convention: the infrared divergence in the $\Ord(\alpha_s^4)$ corrections to the momentum-space static potential~\cite{Anzai:2009tm,Smirnov:2009fh} is
regularized dimensionally (with the $\MSb$ convention for the definition of $\mu$), and the $1/\epsilon$ divergence together with the corresponding logarithm $\log(\mu/\mu_f)$ that arises from the integral over the momentum-space static potential in Eq.~\eqref{eqn:PSmassdef} are subtracted.
We call this the standard convention, and it leads to the coefficient $a_4^\PS$ shown in Eq.~\eqref{eqn:PSmasscoeff}, where the term with the logarithm $\log(\mu/\mu_f)$ is dropped. In a minimal subtraction convention, only the $1/\epsilon$ divergence is subtracted and the logarithmic term displayed in $a_4^\PS$ remains. So the convention of Ref.~\cite{Beneke:2005hg} is equivalent to the choice $\mu/\mu_f=1$ for the dimensional scale in the minimal subtraction convention. 

Using the $\Ord(\LQCD)$ renormalon sum rule we can now track quantitatively if and how much the convention for the infrared subtraction may affect the higher-order behavior in the PS-pole mass relation.
Applying the sum rule to the PS mass in the standard convention of Ref.~\cite{Beneke:2005hg} we obtain for $\nl=5$, relevant for the top quark,
\begin{equation}
 N_{1/2}^{\mu/\mu_f=1}=(0.531\pm 0.318,0.376\pm 0.057,0.503\pm 0.078,0.545\pm 0.045) \,,
\end{equation}
at order $n=(0,1,2,3)$, where the errors come from $\lambda$ variations in the interval $[\,0.5,\,2\,]$. The order $n=3$ result that involves the $\Ord(\alpha_s^4)$ coefficient $a_4^\PS$ is $22\%$ higher and within errors only marginally compatible with the result $N_{1/2}(\nl=5)=0.446\pm0.026$ of Eq.~\eqref{eqn:N12msrfinal5}. This indicates that $a_4^\PS$ in the standard convention is somewhat larger than expected assuming that the pole-PS mass series is dominated by the pole mass renormalon. The same observation has also 
been made in Refs.~\cite{Marquard:2015qpa,Kiyo:2015ooa} in the context of relating the PS mass to the $\MSb$ mass.

It is interesting to consider other minimal subtraction scheme choices that differ from the standard scheme by reasonable variations of the subtraction scale $\mu$. For example, for the choice $\mu/\mu_f=1/5$ we obtain $N_{1/2}^{\mu/\mu_f=1/5}=0.455\,\pm\, 0.021$ at order $n=3$ for $\nl=5$, which is fully compatible with Eq.~\eqref{eqn:N12msrfinal5}. That the sum rule result for the PS mass agrees with the correct result of Eq.~\eqref{eqn:N12msrfinal5} much better for a smaller infrared subtraction scale is quite suggestive because the infrared divergence in the static potential is known to be physically regulated by the massive quark kinetic energy, which is of order $\vec{q}^{\,2}/m_Q\sim \mu_f v$ where $v$ is the relative velocity, and hence is parametrically smaller than $|\vec q\,|\sim \mu_f$. 
We stress that our analysis does neither validate nor invalidate the concept of the standard PS mass as a suitable mass scheme to carry out ongoing high-precision threshold studies \cite{Hoang:2000yr,Beneke:2015kwa}, as the sum rule only probes the calculated orders and the effect of the retardation singularity on the perturbative coefficients in the static potential beyond ${\cal O}(\alpha_s^4)$ on the PS mass scheme is unknown. However, the analysis demonstrates that the scheme dependence in the PS mass coming from the infrared divergences in the static potential at $\Ord(\alpha_s^4)$ is not a numerically irrelevant issue and may become even more serious beyond $\Ord(\alpha_s^4)$. As far as the known $\Ord(\alpha_s^4)$ results are concerned the issue already seems to affect the relation of the standard PS mass to the MSR and $\MSb$ masses 
as discussed in Sec.~\ref{sec:PSmass}.

\subsubsection{QCD \texorpdfstring{$\beta$}{beta}-Function and Massless Quark R-ratio}

As the concluding part of the discussion in this section we now apply the $\Ord(\LQCD)$ renormalon sum rule to series that are known not to be plagued by any $\Ord(\LQCD)$ renormalon. As examples we take the series for the QCD $\beta$-function with
\begin{equation}
 a_n^{\beta} = \beta_{n-1} \;,
\end{equation}
as defined in Eq.~\eqref{eqn:betafct} and the hadronic R-ratio for $\nl$ massless quarks
\begin{equation}
\label{eq:Rratioseries}
 R(s) = 3\sum_{f=1}^{\nl} Q_f^2\!\left[1 + \sum_{n=1}^\infty a_n^R\left(\frac{\alpha_s^{(\nl)}(\sqrt{s})}{4\pi}\right)^{\!\!n}\,\right] \;,
\end{equation}
where $\sqrt{s}$ stands for the center-of-mass energy, with~\cite{Chetyrkin:1979bj,Gorishnii:1990vf,Surguladze:1990tg,Surguladze:1990tgE,Baikov:2008jh}
\begin{align}
 a_1^R &= 4\;,\\
 a_2^R &= 31.7712 - 1.8432\,\nl\;,\nonumber\\
 a_3^R &= -\,424.764 - 76.8083\,\nl - 0.33152\,\nl^2\;,\nonumber\\
 a_4^R &= -\,40092.2 + 4805.12\,\nl - 204.134\,\nl^2 + 5.504\,\nl^3\;,\nonumber
\end{align}
and $Q_f$ stands for the quark electric charges.
Applying the sum rule for $\nl=4$ to the series for the QCD $\beta$-function we obtain 
\begin{equation}
\label{eqn:N12beta}
 N_{1/2}^{\beta}=(0.829\pm0.497,-\,0.004\pm0.272,0.065\pm0.092,0.038\pm0.032) \;,
\end{equation}
and applying it to the hadronic R-ratio we obtain
\begin{equation}
\label{eq:N12Rratio}
 N_{1/2}^{R}=(0.398\pm0.239,-\,0.003\pm0.1311,-\,0.071\pm0.105,-\,0.009\pm0.029) \;,
\end{equation}
at order $n=(0,1,2,3)$. The errors are obtained from the variation $0.5<\lambda<2$. In both cases all results for $N_{1/2}$ beyond $\Ord(\alpha_s)$ are compatible with zero as expected. We note that at least for the hadronic R-ratio it is known that its perturbative series given in Eq.~\eqref{eq:Rratioseries} has a renormalon ambiguity that is suppressed and scales with the fourth power of the hadronic scale $\Lambda_{\rm QCD}$. This leads to an ambiguity in the R-ratio of ${\cal O}(\Lambda_{\rm QCD}^4/s^2)$, which is associated to the gluon condensate, and adding the effects of the gluon condensate in the context of an operator product expansion in terms of low-energy QCD matrix elements~\cite{Shifman:1978bx,Shifman:1978by} this ambiguity is compensated in a physical prediction. 
For the QCD $\beta$-function no conclusive statements on a higher-order renormalon ambiguity exist. The results in Eqs.~\eqref{eqn:N12beta} and \eqref{eq:N12Rratio} show that the $\Ord(\LQCD)$ renormalon sum rule is only probing for an ${\cal O}(\Lambda_{\rm QCD})$ renormalon and not sensitive to any higher order renormalon ambiguity. 

It is straightforward to generalize the sum rule discussed here to higher order renormalons, which has already been studied in Ref.~\cite{Ambar:thesis}.

\section{Relation to Other Short-Distance Masses}
\label{sec:othermasses}

From the perturbative series that relate other short-distance masses to the pole mass it is straightforward to determine the perturbative series for the difference of these short-distance masses to the MSR masses by eliminating the pole mass systematically such that the ${\cal O}(\Lambda_{\rm QCD})$ renormalon is canceled exactly. If regular fixed-order perturbation theory can be applied this is achieved by simply using a common renormalization scale $\mu$ and a consistent scheme for the strong coupling throughout the calculation when the pole mass is eliminated order by order. The corresponding formulae and codes for the relation of frequently used short-distance mass schemes (such as the kinetic mass~\cite{Czarnecki:1997sz}, the PS mass~\cite{Beneke:1998rk}, the 1S mass~\cite{Hoang:1998ng,Hoang:1998hm,Hoang:1999ye}, the RS mass~\cite{Pineda:2001zq} and the jet mass~\cite{Jain:2008gb,Fleming:2007tv}) to the MSR masses can be obtained on request, and we therefore do not intend to cover all possible cases in this paper. However, we will cover several of them explicitly since there are a number of non-trivial practical and conceptual aspects that arise in the relation of the MSR masses to a number of other short-distance mass schemes we would like to point out in the following. 

\subsection{Potential Subtracted Mass}
\label{sec:PSmass}

\begin{figure*}[t]
	\center
	\subfigure[]
	{\label{fig:mPSmsra}\includegraphics[width=0.479\textwidth]{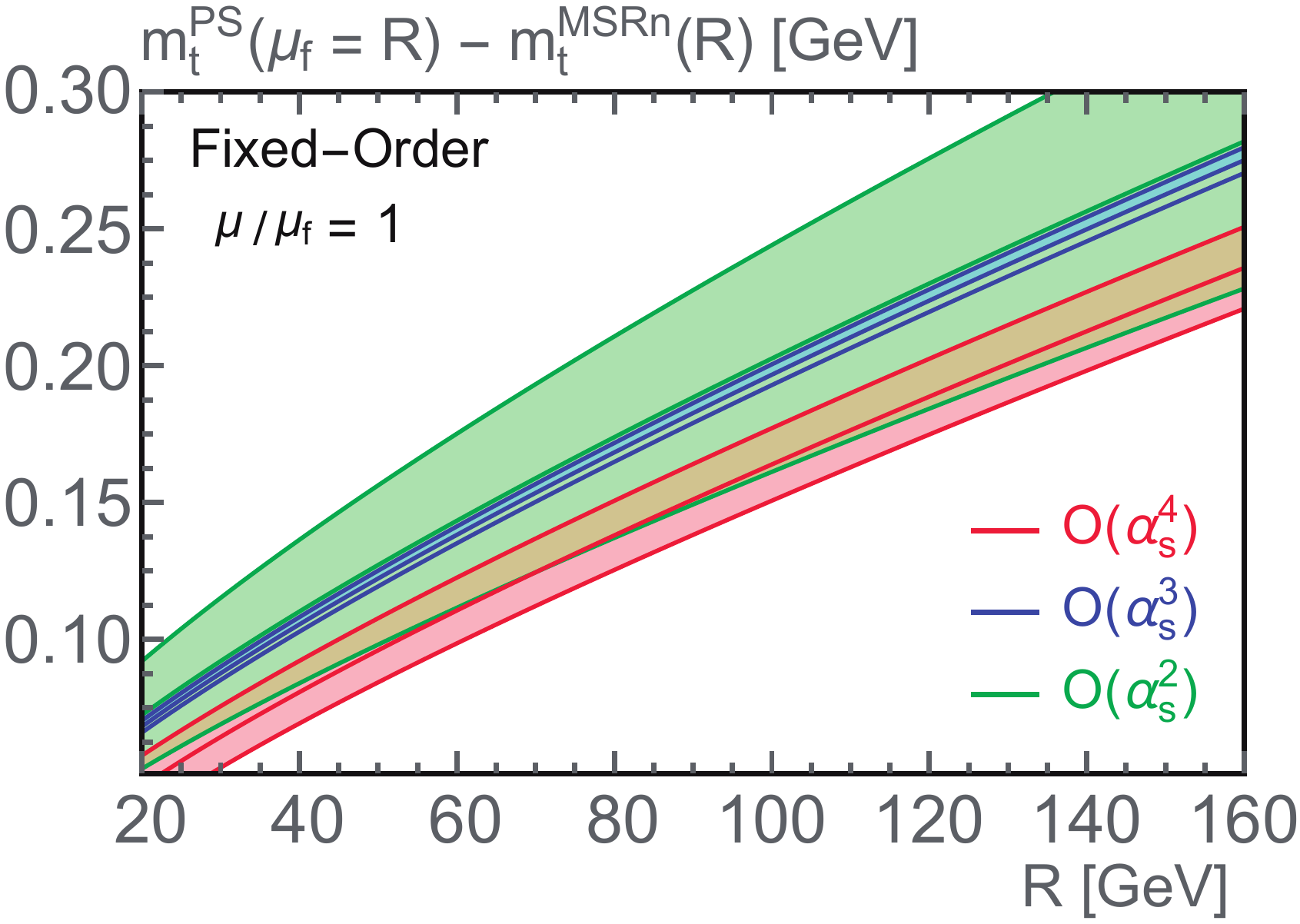}~~~}
	\subfigure[]
	{\label{fig:mPSmsrb}\includegraphics[width=0.479\textwidth]{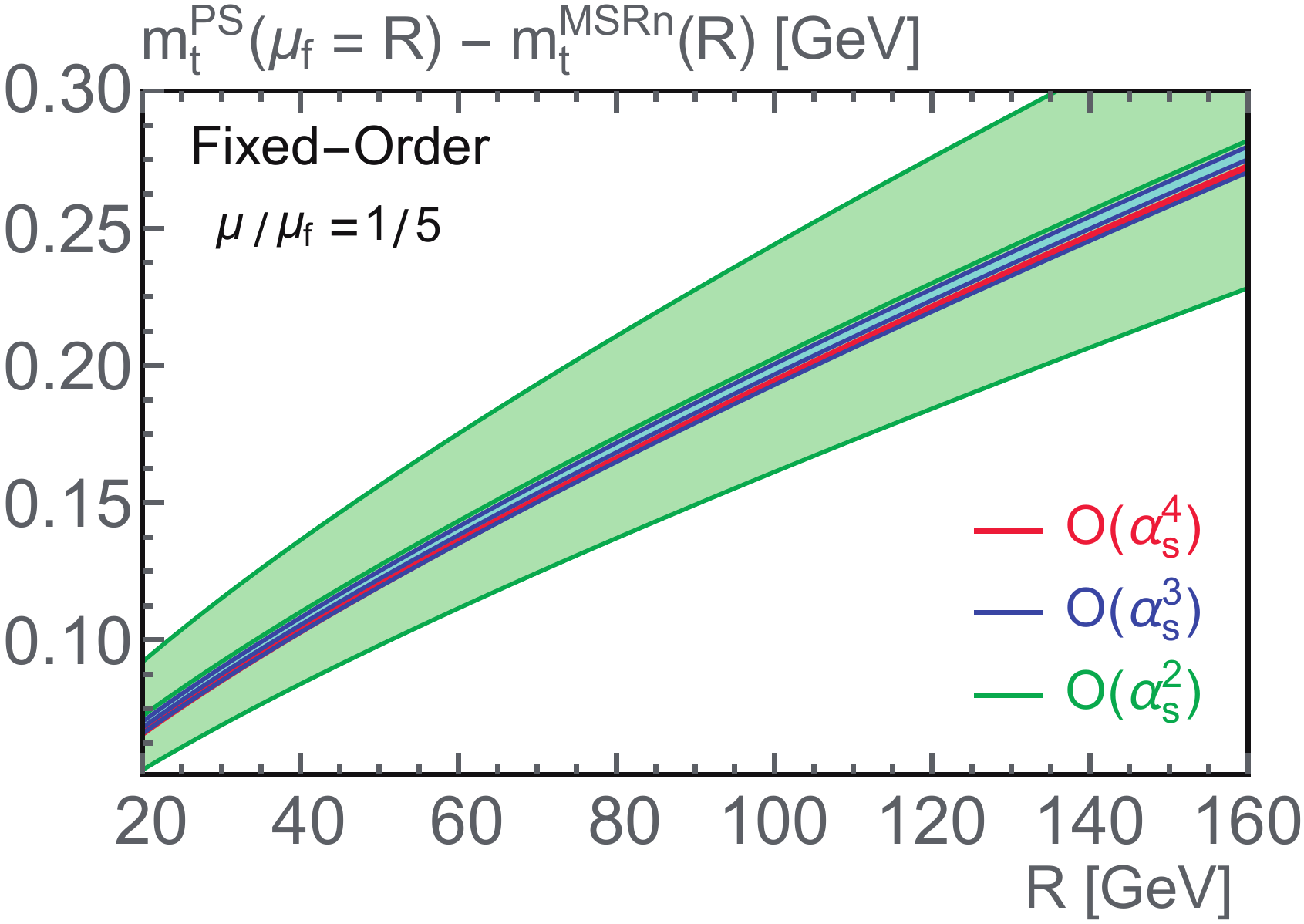} }
 \caption{\label{fig:mPSmsr} Difference between the natural MSR and PS ($\mu_f = R$) top quark mass ($\nl=5$) as a function of $R$ in GeV at two, three and four loops (the one loop result vanishes). Results are shown for two different choices of the IR subtraction scale, $\mu/\mu_f=1$ and $\mu/\mu_f=1/5$.}
\end{figure*}

The relations of the PS mass \cite{Beneke:1998rk} and the natural and practical MSR masses at the common scale $R$ up to $\Ord(\alpha_s^4)$ have the form [\,$a_s\equiv\alpha_s^{(\nl)}(R)/(4\pi)$\,]
\begin{align}
  m_Q^\PS(\mu_f=R)-m_Q^\mathrm{MSRn}(R)=&\,R\,\Big\{[\,40.9928-3.6248\,\nl]\,a_s^2+[\,963.44-184.87\,\nl\\
 &+0.422\,\nl^2\,]\,a_s^3+\Big[-(1749.\pm417.)-(11168.\pm10.)\,\nl\nonumber\\
 &+569.34\,\nl^2-0.89\,\nl^3-22739.57\,\log\leri{\frac{\mu}{R}}\Big]a_s^4+\ldots\Big\} ,\nonumber\\[0.3cm]
 m_Q^\PS(\mu_f=R)-m_Q^\mathrm{MSRp}(R)=& \,R\,\Big\{[\,42.6499-3.6248\,\nl]\,a_s^2+[\,1073.49-183.45\,\nl\\
 &+0.422\,\nl^2\,]\,a_s^3+\Big[-(1405.\pm 418.)-(11279.\pm 10.)\,\nl\nonumber\\
 &+573.74\,\nl^2-0.89\,\nl^3-22739.57\,
\log\leri{\frac{\mu}{R}}\Big]a_s^4+\ldots\Big\}\,.\nonumber
\end{align}
For a conversion at the common scale $\mu_f=R$ the $\Ord(\alpha_s)$ corrections vanish identically indicating that this is the natural way to carry out the conversion. As pointed out already in Sec.~\ref{sec:PSmassIR}, the standard convention for the PS mass~\cite{Beneke:2005hg} corresponds to $\mu/\mu_f=1$, such that the logarithmic term in the $\Ord(\alpha_s^4)$ correction is eliminated.
In Tab.~\ref{tab:mPSmsr} we show numerical results for the PS-MSR mass difference \mbox{$m_Q^\PS(\mu_f=R)-m_Q^\mathrm{MSR}(R)$} for representative $R$ values for $\nl=5$ (relevant to the top quark) and $\nl=4$ (relevant for the bottom quark) at different orders in $\alpha_s$. The  errors come from the variation of the renormalization scale $\mu$ of the strong coupling in the interval $[R/2,2R\,]$, and the central values are the mean of the respective maximal and minimal values obtained in that interval. In Fig.~\ref{fig:mPSmsra} $m_Q^\PS(\mu_f=R)-m_Q^\mathrm{MSRn}(R)$ is shown at $\Ord(\alpha_s^2)$ (green), $\Ord(\alpha_s^3)$ (blue) and $\Ord(\alpha_s^4)$ (red) for $\nl=5$ as a function of $R$ between $20$ and $160$\,GeV. The error bands are again obtained from variations of $\mu$ in the interval  $[R/2,2R\,]$.
For the top quark case ($\nl=5$) the PS and the MSR masses differ by about $20$ to $300$\,MeV for $R$ values between $2$ and $160$\,GeV and for the bottom quark case ($\nl=4$) they differ by about $30$ to $40$\,MeV for $R$ values below $5$\,GeV. So the PS and the MSR masses are quite close numerically.

\begin{table}[t]
	\center
	\begin{tabular}{|c|ccrc|}
		\hline
		$R$ & ~~~~~ $\Ord(\alpha_s^2)$~~~~~  & ~~~~ $\Ord(\alpha_s^3)$ ~~~~ & $\Ord(\alpha_s^4)^{\mu/\mu_f = 1}$ ~ &
		$\Ord(\alpha_s^4)^{\mu/\mu_f = 1/5}$ ~~ \\
		\hline
		& \multicolumn{4}{|c|}{$m_t^{\rm PS}(\mu_f=R) - m_t^{\rm MSRn}(R)$\quad ($\nl=5$) [\,GeV\,]}  \\\hline
		$2$ & $ 0.031 \pm 0.016$ & $ 0.022 \pm 0.004$ & $-\,0.027 \pm 0.042$ & $ 0.017 \pm 0.006$ \\
		$5$ & $ 0.037 \pm 0.014$ & $ 0.032 \pm 0.002$ & $0.007 \pm 0.017$ & $ 0.030 \pm 0.002$ \\
		$10$ & $ 0.050 \pm 0.016$ & $ 0.046 \pm 0.002$ & $ 0.024 \pm 0.013$ & $ 0.044 \pm 0.002$ \\
		$40$ & $ 0.110 \pm 0.026$ & $ 0.105 \pm 0.003$ & $ 0.081 \pm 0.011$ & $ 0.103 \pm 0.001$ \\
		$80$ & $ 0.174 \pm 0.037$ & $ 0.168 \pm 0.003$ & $ 0.138 \pm 0.013$ & $ 0.166 \pm 0.002$ \\
		$160$ & $ 0.282 \pm 0.054$ & $ 0.275 \pm 0.005$ & $ 0.236 \pm 0.015$ & $ 0.272 \pm 0.002$ \\\hline
		& \multicolumn{4}{|c|}{$m_t^{\rm PS}(\mu_f=R) - m_t^{\rm MSRp}(R)$\quad ($\nl=5$) [\,GeV\,] }\\\hline
		$2$ & $ 0.034 \pm 0.018$ & $ 0.028 \pm 0.004$ & $-\,0.024 \pm 0.043$ & $ 0.020 \pm 0.007$ \\
		$5$ & $ 0.040 \pm 0.015$ & $ 0.037 \pm 0.003$ & $ 0.012 \pm 0.017$ & $ 0.034 \pm 0.003$ \\
		$10$ & $ 0.054 \pm 0.017$ & $ 0.052 \pm 0.003$ & $ 0.030 \pm 0.013$ & $ 0.050 \pm 0.002$ \\
		$40$ & $ 0.118 \pm 0.028$ & $ 0.118 \pm 0.004$ & $ 0.094 \pm 0.011$ & $ 0.116 \pm 0.002$ \\
		$80$ & $ 0.186 \pm 0.039$ & $ 0.188 \pm 0.005$ & $ 0.158 \pm 0.013$ & $ 0.186 \pm 0.002$ \\
		$160$ & $ 0.302 \pm 0.058$ & $ 0.306 \pm 0.007$ & $ 0.267 \pm 0.015$ & $ 0.303 \pm 0.002$\\\hline
		& \multicolumn{4}{|c|}{$m_b^{\rm PS}(\mu_f=R) - m_b^{\rm MSRn}(R)$\quad ($\nl=4$) [\,GeV\,] }\\\hline
		$2$ & $ 0.044 \pm 0.027$ & $ 0.034 \pm 0.007$ & $-\,0.041 \pm 0.065$ & $ 0.032 \pm 0.005$ \\
		$3$ & $ 0.041 \pm 0.021$ & $ 0.036 \pm 0.005$ & $-\,0.003 \pm 0.030$ & $ 0.036 \pm 0.002$ \\
		$4$ & $ 0.042 \pm 0.019$ & $ 0.038 \pm 0.004$ & $ 0.010 \pm 0.021$ & $ 0.039 \pm 0.001$\\\hline
		& \multicolumn{4}{|c|}{$m_b^{\rm PS}(\mu_f=R) - m_b^{\rm MSRp}(R)$\quad
        ($\nl=4$) [\,GeV\,] }\\\hline
		$2$&  $ 0.047 \pm 0.029$ & $ 0.040 \pm 0.009$ & $-\,0.039 \pm 0.068$ & $ 0.034 \pm 0.008$ \\
		$3$ & $ 0.044 \pm 0.022$ & $ 0.041 \pm 0.007$ & $ 0.001 \pm 0.031$ & $ 0.039 \pm 0.003$ \\
		$4$ & $ 0.045 \pm 0.020$ & $ 0.043 \pm 0.006$ & $ 0.014 \pm 0.022$ & $ 0.043 \pm 0.002$\\\hline
	\end{tabular}
	\caption{Differences between the top mass in the PS and MSR schemes, showing both the natural and practical MSR mass definitions. Results are given for various scales $\mu_f=R$ and orders in $\alpha_s$. At ${\cal O}(\alpha_s^4)$ results are shown for two choices of the infrared subtraction scale, $\mu/\mu_f=1$ and $\mu/\mu_f=1/5$. \label{tab:mPSmsr}}
\end{table}

The conspicuous property of the relation of the standard PS mass to the MSR masses at the common scale $R$ is that the $\Ord(\alpha_s^4)$ correction is very large and far away from the $\Ord(\alpha_s^3)$ uncertainty band such that the $\Ord(\alpha_s^4)$ error band from scale variation is three to four times larger than the $\Ord(\alpha_s^3)$ one. For the top quark ($\nl=5$) for $R$ around $40$ to $80$\,GeV, the typical range employed in studies of top pair production at threshold~\cite{Hoang:2000yr}, the $\Ord(\alpha_s^3)$ and $\Ord(\alpha_s^4)$ central values differ by $23$\,MeV compared to scale variations of $\pm \,4$\,MeV at $\Ord(\alpha_s^3)$ and $\pm\,12$\,MeV at $\Ord(\alpha_s^4)$. 
For $R=160$~GeV, the $\Ord(\alpha_s^3)$ and $\Ord(\alpha_s^4)$ central values even differ by $40$\,MeV compared to scale variations of about $\pm \,4$\,MeV at $\Ord(\alpha_s^3)$ and $\pm\,15$\,MeV at $\Ord(\alpha_s^4)$. A similar observation was made earlier in Ref.~\cite{Marquard:2015qpa}.
Given this situation it is reasonable to use the difference of the $\Ord(\alpha_s^3)$ and $\Ord(\alpha_s^4)$ central values as the $\Ord(\alpha_s^4)$ uncertainty due to the missing higher order terms rather than the scale variation, leading to uncertainties of about $(20,25,30,40)$\,MeV at $R=(10,40,80,160)$\,GeV. In Ref.~\cite{Marquard:2015qpa} the $\Ord(\alpha_s^4)$ uncertainty in the relation between the $\MSb$ mass $\mbar_Q \equiv \mbar_Q^{(n_\ell+1)}(\mbar_Q^{(n_\ell+1)})$ and the PS mass for $\mu_f=20$~GeV was quoted as $23$~MeV, defined as half the size of the $\Ord(\alpha_s^4)$ correction.  This issue is directly related to our observation made in Sec.~\ref{sec:PSmassIR} that the $\Ord(\alpha_s^4)$ correction in the relation of the pole mass and the PS mass in the standard scheme~\cite{Beneke:2005hg} (with infrared subtraction scale $\mu/\mu_f=1$) is much larger than expected from the $\Ord(\LQCD)$ renormalon of the pole mass. 

In Sec.~\ref{sec:PSmassIR} we also found that for the PS top mass in the infrared subtraction scheme with $\mu/\mu_f=1/5$ there is much better consistency concerning the $\Ord(\Lambda_{\rm QCD})$ sum rule.
Using the PS mass in this modified scheme the $\Ord(\alpha_s^4)$ corrections in this relation to the MSR masses reduce substantially, as can be easily spotted from the corresponding results in Tab.~\ref{tab:mPSmsr} and in Fig.~\ref{fig:mPSmsrb}: for the modified PS mass the $\Ord(\alpha_s^4)$ result for the PS-MSR mass difference is fully compatible with the $\Ord(\alpha_s^3)$ result and leads to scale variations that are about half the ones at $\Ord(\alpha_s^3)$.
In this scheme it is therefore reasonable to quote the scale variations as the remaining perturbative error at $\Ord(\alpha_s^4)$. For all $R$ values above $2$\,GeV and $\nl=4$ and $5$, the error in the $\Ord(\alpha_s^4)$ relation of the natural and the practical MSR masses and the PS mass in the modified scheme with $\mu/\mu_f=1/5$ for the infrared scale is only about $\pm \,2-3$\,MeV. 

One may alternatively make the conversion between the PS mass $m_Q^{\rm PS}(\mu_f)$ and the MSR masses $m_Q^{\rm MSR}(R)$ for $\mu_f\neq R$, where we expand consistently in $\alpha_s(\mu)$ with a common scale $\mu$. For the case  $\mu_f<R$ 
we observe in general that the scale dependence of the $\Ord(\alpha_s^4)$ conversion formula for 
the standard convention for the PS scheme, $m_Q^{{\rm PS},\mu/\mu_f=1}(\mu_f)-m_Q^{\rm MSR}(R)$, decreases compared to the choice  $\mu_f=R$, but the size of the $\Ord(\alpha_s^4)$ correction is still many times larger than the $\Ord(\alpha_s^3)$ scale variation.
This can be seen for example for the case $(\mu_f,R)=(50,100)$\,GeV were we obtain for $\nl=5$ the numerical results
$m_Q^{{\rm PS},\mu/\mu_f=1}(\mu_f)-m_Q^{\rm MSRn}(R) =(2.612\pm 0.143,2.925\pm 0.042,2.946\pm 0.005,2.922\pm 0.005) \,{\rm GeV}$
at $\Ord(\alpha_s^{1,2,3,4})$ for the standard PS mass scheme with renormalization scale variation $\mu_f<\mu<R$. This may be compared to the corresponding results for the modified PS mass scheme, which read
$m_Q^{{\rm PS},\mu/\mu_f=1/5}(\mu_f)-m_Q^{\rm MSRn}(R)=(2.612\pm 0.143,2.925\pm 0.042,2.946\pm 0.005,2.939\pm 0.002)$\,GeV and show again  a fully consistent behavior between the $\Ord(\alpha_s^3)$ and $\Ord(\alpha_s^4)$ results and their scale variations.
On the other hand, for the case $\mu_f>R$ we observe in general that, at each given order, the size of the scale dependence of $m_Q^{{\rm PS},\mu/\mu_f=1}(\mu_f)-m_Q^{\rm MSRn}(R)$ is much smaller than the next correction. This can be seen for example for the case $(\mu_f,R)=(50,25)$\,GeV were we obtain for $\nl=5$ the numerical results
$m_Q^{{\rm PS},\mu/\mu_f=1}(\mu_f)-m_Q^{\rm MSRn}(R)=(-\,1.468\pm 0.091,-\,1.456\pm 0.005,-\,1.478\pm 0.004,-\,1.504\pm 0.007)$\,GeV at $\Ord(\alpha_s^{1,2,3,4})$ for the standard PS mass scheme with the renormalization scale variation $R<\mu<\mu_f$. This may be compared to the corresponding results for the modified PS mass scheme which read
$m_Q^{{\rm PS},\mu/\mu_f=1/5}(\mu_f)-m_Q^{\rm MSRn}(R)=(-\,1.468\pm 0.091,-\,1.456\pm 0.005,-\,1.478\pm 0.004,-\,1.4767\pm 0.0003)$\,GeV, and yet again show a better behavior. So, also when the conversion between the standard PS mass and the MSR masses is carried out for $\mu_f\neq R$, the size of the $\Ord(\alpha_s^4)$ correction and not the usual renormalization scale variation must be taken as an estimate for the remaining perturbative error. Since the $\Ord(\alpha_s^4)$ corrections are typically in the range $20$\,--\,$40$\,MeV, 
making the conversion $\mu_f\neq R$ does not lead to any improvement in the perturbative relation between the standard PS mass and the MSR masses.

We conclude that the conversion of the MSR masses to the PS mass in the standard scheme of Ref.~\cite{Beneke:2005hg} has, even at ${\cal O}(\alpha_s^4)$, perturbative uncertainties due to unknown higher-order terms of about $20$\,--\,$40$\,MeV and that this behavior is related to the fact that the ${\cal O}(\alpha_s^4)$ coefficient in the relation of the PS mass to the pole mass in the standard scheme appears to be unnaturally large in the context of its expected size with respect to the pole mass ${\cal O}(\Lambda_{\rm QCD})$ renormalon ambiguity. On the other hand, using an infrared subtraction scheme for the PS mass, where the subtraction scale is
much lower, leads to a much better perturbative behavior and to much smaller uncertainties in its relation to the MSR masses. 
This observation is fully consistent with the conclusions from the renormalon sum rule analysis we carried out for the PS mass in Sec.~\ref{sec:PSmassIR}. Since the MSR masses for $R=\mbar_Q$ are very close or identical to the $\MSb$ mass $\mbar_Q(\mbar_Q)$ 
the conclusions we draw on the perturbative relation of the standard PS mass to the MSR masses also applies to the perturbative relation of the standard PS mass to the $\MSb$ mass. For $R=\mbar_Q$ the ${\cal O}(\alpha_s^4)$ correction is typically at the level of $40$\,MeV. 
We note that this issue of the standard PS mass scheme becomes problematic once a precision in top quarks mass determinations below $30$\,--\,$40$\,MeV can be reached. Given the projections of top mass determinations of a future lepton collider, see e.g.\ \cite{Simon:2016pwp,dEnterria:2016sca}, this may become a pressing issue, but for current studies of high-precision top quark mass determinations the standard PS mass scheme is adequate for most applications.

\subsection{1S Mass}

The 1S mass \cite{Hoang:1998ng,Hoang:1998hm,Hoang:1999ye} is defined as half of the mass of the heavy quarkonium spin triplet ground state. In terms of the pole mass the 1S mass is defined as 
\begin{equation}\label{eqn:1Smass}
 m_Q^\OS=m_Q^\pole+ \big[C_F\alpha_s^{(\nl)}(\mu)m_Q^\pole\big] \sum_{n=1}^\infty\sum_{k=0}^{n-1}c_{n,k}
   \bigg(\frac{\alpha_s^{(\nl)}(\mu)}{4\pi}\bigg)^{\!\!n}\log^k\! \Bigg(\frac{\mu}{C_F\alpha_s^{(\nl)}(\mu)m_Q^\pole}\Bigg) ,
\end{equation}
where the coefficients $c_{n,k}$ are known up to $n=4$ and given for convenience in Eq.~\eqref{eqn:1Smasscoeff}.
Because the 1S mass originates from a calculation in the non-relativistic context, there are a few subtleties when calculating its relation to the MSR masses so that the $\Ord(\LQCD)$ renormalon cancels properly.

For the case $R\sim m_Q$ it is essential that terms of order $[C_F\alpha_s m_Q] \alpha_s^n$ are formally counted as $\Ord(\alpha_s^n)$ in the conversion. This is because $[C_F\alpha_s m_Q]$ is the inverse Bohr radius, which is the relevant physical mass scale and should not be counted as an ${\cal O}(\alpha_s)$ correction. This counting is called the $\Upsilon$-expansion \cite{Hoang:1998ng,Hoang:1998hm} or the relativistic order counting, and must also be used when relating the 1S mass to the $\MSb$ masses in fixed-order perturbation theory. The resulting formula for the 1S mass as a function of the MSR mass for $\mu=R$ up to $\Ord(\alpha_s^4)$ reads [\,defining parameters \,$M_B=C_F\alpha_s^{(\nl)}(R)\,m_Q^\MSR(R)$, $R_B=C_F\alpha_s^{(\nl)}(R)\,R$, $a_s=\alpha_s^{(\nl)}(R)/(4\pi)$, $L=\log(R/M_B)$\, which are all functions of $R$\,]
\begin{align}\label{eqn:1SinMSRupsilon}
 & m_Q^\OS - m_Q^\MSR(R)\,=\,[\,R \,a_1+M_B\, c_{1,0}\,]\,a_s\\
 &+[\,R\, a_2+R_B\, a_1 \,c_{1,0}+M_B(c_{2,0}+c_{2,1}L)\,]\,a_s^2\nonumber\\
 &+\leris{R \,a_3+R_B\Big(a_2\, c_{1,0}+a_1\,(c_{2,0}-c_{2,1}(1-L))\Big)+M_B\leri{c_{3,0}+c_{3,1}L+c_{3,2}L^2}}\!a_s^3\nonumber\\
 &+\bigg[R\,\bigg(\!a_4-\frac{R_B}{2m_Q^\MSR(R)}\,a_1^2\, c_{2,1}\!\bigg)+R_B\Big(a_3\, c_{1,0}+a_2\,(c_{2,0}-(1-L)\,c_{2,1})\nonumber\\
 &\quad+a_1\,(c_{3,0}-c_{3,1}+(c_{3,1}-2\,c_{3,2})L+c_{3,2}L^2)\Big)+
M_B\leri{c_{4,0}+c_{4,1}L+c_{4,2}L^2+c_{4,3}L^3}\!\bigg]a_s^4 \,.\nonumber
\end{align}
Here $a_n$ are the coefficients in the MSR scheme. 
The inverse of Eq.~\eqref{eqn:1SinMSRupsilon} is given in Eq.~\eqref{eqn:MSRin1Supsilon}. 

For the case $R\sim m_Q\alpha_s$, which is relevant for non-relativistic applications where $\alpha_s$ may scale with the quark velocity $\alpha_s\sim v\ll1$, the non-relativistic counting $R\sim M_B\sim m_Q\alpha_s$ should be used, such that the leading correction in the 1S-MSR mass difference is of order $\alpha_s^2$. In this case the formula for the 1S mass as a function of the MSR mass for $\mu=R$ up to $\Ord(\alpha_s^5)$ reads [\,$M_B=C_F\alpha_s^{(\nl)}(R)\,m_Q^\MSR(R)$,  
$a_s=\alpha_s^{(\nl)}(R)/(4\pi)$, $L=\log(R/M_B)$\,]
\begin{align}\label{eqn:1SinMSRnonrel}
 m_Q^\OS-m_Q^\MSR(R)={}&\big[R\,a_1+M_B\, c_{1,0}\big]\,a_s\\
 +\,&\big[R\, a_2+M_B\,(c_{2,0}+c_{2,1}L)\,\big]\,a_s^2\nonumber\\
 +\,&\big[R\,(a_3+4\pi\, C_F\, a_1\, c_{1,0})+M_B(c_{3,0}+c_{3,1}L+c_{3,2}L^2)\,\big]\,a_s^3\nonumber\\
 +\,&\Big[R\Big(a_4+4\pi\, C_F \,a_2\, c_{1,0}+4\pi\, C_F\, a_1\,\big[\,c_{2,0}-c_{2,1}(1-L)\big]\Big)\nonumber\\
 &+M_B\,\big(c_{4,0}+c_{4,1}L+c_{4,2}L^2+c_{4,3}L^3\big)\Big] a_s^4 \,.\nonumber
\end{align}
The inverse of Eq.~\eqref{eqn:1SinMSRnonrel} is given in Eq.~\eqref{eqn:MSRin1Snonrel}. 
We note that in order to implement a general renormalization scale $\mu$ in Eqs.~\eqref{eqn:1SinMSRupsilon} as well as \eqref{eqn:1SinMSRnonrel}, also the dependence of $M_B$ on $\alpha_s$ needs to be accounted for consistently, which leads to quite involved expressions for the relativistic counting of the $\Upsilon$-expansion.
For the top quark and $R\sim m_t\alpha_s\sim 30$\,GeV the numerical difference between using the relativistic or the non-relativistic counting is below $10$\,MeV at the highest order and may be not significant. However, for all other cases the difference can be more sizable such that a consistent use of the order counting is mandatory in general.

\begin{table}[t]
	\center
	\begin{tabular}{|c|cccc|}
		\hline
		$R$ & \multicolumn{4}{|c|}{$m_t^{1S}$ [GeV]}\\\hline
		& ~~~~~~ $\Ord(\alpha_s)$~~~~~~  & ~~~~~~ $\Ord(\alpha_s^2)$ ~~~~~~ & ~~~~~~ $\Ord(\alpha_s^3)$ ~~~~~~ &
		~~~~~~$\Ord(\alpha_s^4)$~~~~~~\\
		\hline
		$160$ & $ 167.934\pm 0.968$ & $ 168.315 \pm 0.151$ & $\,168.397 \pm 0.019$ & $ 168.368 \pm 0.021$ \\\hline
			& ~~~~~~ $\Ord(\alpha_s^2)$~~~~~~  & ~~~~~~ $\Ord(\alpha_s^3)$ ~~~~~~ & ~~~~~~ $\Ord(\alpha_s^4)$ ~~~~~~ &
			~~~~~~$\Ord(\alpha_s^5)$~~~~~~\\
		\hline
		$40$ & $ 168.156\pm 0.113 $ & $ 168.409 \pm 0.054$ & $168.373 \pm 0.019 $ & $ 168.372 \pm 0.007$
        \\[-5pt]
        &$\phantom{000.000} \hspace{.22em}(\pm\hspace{.22em} 0.113)$ & $\phantom{000.000} \hspace{.22em}(\pm\hspace{.22em} 0.054)$
          & $\phantom{000.000} \hspace{.22em}(\pm\hspace{.22em} 0.021)$ & $\phantom{000.000}\hspace{.22em}(\pm\hspace{.22em} 0.011)$
        \\
		$35$ & $ 168.197\pm 0.077$ & $ 168.421 \pm 0.048$ & $ 168.365 \pm 0.026$ & $ 168.371 \pm 0.006$ 
          \\[-5pt]
        & $\phantom{000.000} \hspace{.22em}(\pm\hspace{.22em} 0.078)$ & $\phantom{000.000} \hspace{.22em}(\pm\hspace{.22em} 0.049)$
         & $\phantom{000.000} \hspace{.22em}(\pm\hspace{.22em} 0.028)$ & $\phantom{000.000} \hspace{.22em}(\pm\hspace{.22em} 0.011)$
        \\
		$30$ & $ 168.232 \pm 0.037$ & $ 168.434 \pm 0.046$ & $ 168.353 \pm 0.036$ & $ 168.372 \pm 0.008$ 
          \\[-5pt]
        & $\phantom{000.000} \hspace{.22em}(\pm\hspace{.22em} 0.038)$& $\phantom{000.000} \hspace{.22em}(\pm\hspace{.22em} 0.047)$
         & $\phantom{000.000} \hspace{.22em}(\pm\hspace{.22em} 0.038)$ & $\phantom{000.000} \hspace{.22em}(\pm\hspace{.22em} 0.012)$
       \\\hline
	\end{tabular}
	\caption{Results for the top mass in the 1S mass scheme at different orders using as input the practical MSR mass  $m_t^\mathrm{MSRp}(m_t^\mathrm{MSRp})=160\,{\rm GeV}$. The results at the top of the table show the 1S mass using FOPT conversion in the relativistic order counting of Eq.~\eqref{eqn:1SinMSRupsilon} with $R=160$~GeV. The conversion still contains large logarithms $\ln(m_Q/M_B)$.  The lower three lines show the 1S mass using R-evolution from $160$~GeV to $R=(30,35,40)$~GeV and then FOPT in the non-relativistic order counting of Eq.~\eqref{eqn:1SinMSRnonrel} with the scale $R$. The logarithms $\ln(m_Q/M_B)$ are then summed to all orders, and the uncertainties are about a factor two smaller at the highest order. The uncertainties shown are explained in detail in the text.
    \label{tab:m1Sinmsr}}
\end{table}

In the top line of Tab.~\ref{tab:m1Sinmsr} the top quark 1S mass is shown for the practical MSR top mass $m_t^\mathrm{MSRp}(m_t^\mathrm{MSRp})=\mbar_t(\mbar_t)=R_0=160$\,GeV using directly the relativistic conversion of Eq.~\eqref{eqn:1SinMSRupsilon} at $\Ord(\alpha_s)$ to $\Ord(\alpha_s^4)$,  where the quoted uncertainties have been obtained by renormalization scale variations $\sqrt{R_0\,M_B}/2<\mu<R_0$ with $M_B=23.2$\,GeV and the central values are the mean of the respective maximal and minimal values obtained in the scale variation.
In the lower three lines the conversion to the 1S mass is achieved by first using $\Ord(\alpha_s^4)$ R-evolution of 
$m_t^\mathrm{MSRp}(160\,\mathrm{GeV})$ to $R=(30,35,40)$\,GeV, which gives $m_t^\mathrm{MSRp}(R)=(167.181\pm 0.010,166.854\pm 0.009,166.535\pm 0.008)$\,GeV, where the uncertainties are obtained by variations of $\lambda$ in the interval $[\,0.5,2\,]$ and central values are the mean of the respective maximal and minimal values. Then the non-relativistic formula of Eq.~\eqref{eqn:1SinMSRnonrel} is used to determine the 1S mass at $\Ord(\alpha_s^2)$ to $\Ord(\alpha_s^5)$. The quoted uncertainties are from renormalization scale variations $R/2<\mu<2R$. To these uncertainties the errors from the R-evolution calculation just shown above still have to be added quadratically to obtain the complete conversion uncertainty, which is shown in the parentheses.
We see that the direct relativistic conversion, which does not account for the resummation of logarithms and renormalon corrections, leads to uncertainties of $\pm\, 20$\,MeV at highest order, compared to $\pm$\,($10$\,--\,$13$)~MeV for the conversion that uses R-evolution from $160$\,GeV down to non-relativistic scales $\sim M_B$. Given the projections of high precision top mass determinations at future lepton colliders \cite{Simon:2016pwp,dEnterria:2016sca,dEnterria:2015kmd}, the increased precision obtained by using the resummation of higher order terms provided by R-evolution could be relevant, but for the conversion of the MSR mass (and also the $\MSb$ mass) to the 1S mass the fixed-order expansion is adequate for most current applications in top quark physics.

\subsection{\texorpdfstring{$\MSb$}{MS-bar} Mass}\label{sec:MSbar}

The relation of the MSR masses to the $\MSb$ mass is conceptually special since the MSR masses are directly derived from the perturbative series of the pole-$\MSb$ mass relation. The concept of the MSR mass addresses the conceptual question of how the $\MSb$ mass evolves for scales much smaller than the quark mass. This question simply expresses the situation that the $\MSb$ mass $\mbar_Q(\mu)$ for $\mu\ll m_Q$ can be readily computed solving its renormalization group equation, but does not have any physical significance, because it breaks the power counting of heavy quark problems involving (non-relativistic) physical scales much smaller than the mass. This power counting breaking comes from the perturbative series of the pole-$\MSb$ mass relation that scales with $m_Q$ even for $\mu\ll m_Q$ and which spoils the perturbative series for non-relativistic problems where smaller dynamical scales govern the size of the perturbative corrections and the scale $m_Q$ is integrated out and hence not a dynamical scale any more.

Since the perturbative series for the pole-MSR mass relations scale with $R$, which is adjustable, but also match to the pole-$\MSb$ mass series for $R=m_Q$, one can consider the concept of the MSR mass $m_Q^\MSR(\mu)$ as the most reasonable answer of how the $\MSb$ mass concept should be extended to scales $\mu\lesssim m_Q$. Thus for $\mu\lesssim m_Q$ R-evolution is the proper concept of the renormalization group running of a heavy quark mass for scales below $m_Q$. Both the natural and the practical MSR masses differ by the way how the virtual massive quark $Q$ effects are treated in their matching relation to the $\MSb$ mass at the scale $\mu\sim m_Q$, and this matching may be considered in analogy to the flavor-number matching of the strong coupling schemes $\alpha_s^{(\nl)}(\mu)$ and $\alpha_s^{(\nl+1)}(\mu)$ when the scale $\mu$ crosses $m_Q$. In this context, the natural MSR mass is conceptually cleaner than the practical MSR mass, since in the natural MSR mass the virtual massive quark loops are integrated out at the scale $\mu=m_Q$, but this issue is irrelevant for practical applications, where the practical MSR mass has an advantage due to its simpler matching relation to the $\MSb$ mass. 

The most efficient way to relate the MSR masses $m_Q^\mathrm{MSRn}(R)$ and $m_Q^\mathrm{MSRp}(R)$ to the $\MSb$ mass $\mbar_Q(\mu)$ is to (i) evolve the MSR masses from $R$ to $m_Q$ using the R-evolution equations Eq.~\eqref{eqn:rrge} with $\nl$ active flavors, (ii) employing the regular renormalization group equation for $\mbar_Q(\mu)$ to evolve it from $\mu$ to $m_Q$ with $(\nl +1)$ active flavors,
\begin{equation}\label{eqn:msbarRGE}
 \mbar_Q^{(\nl+1)}( m_Q) = \mbar_Q^{(\nl+1)}(\mu)\,
\exp\!\Bigg[\!-\sum_{k=0}^\infty\gamma_{m,k}^{(\nl+1)}\!\int_{\log\mu^2}^{\log m_Q^2}
\dd\log\bar\mu^2\leri{\frac{\alpha_s^{(\nl+1)}(\bar\mu)}{4\pi}}^{\!\!k+1}\Bigg] ,
\end{equation}
and then (iii) to apply the simple matching relations based on Eq.~\eqref{eqn:msbmsr1} or Eq.~\eqref{eqn:msrpmsbarmatch}.

The solution of the R-evolution equation is~\cite{Hoang:2008yj} [\,$t_m=-\,2\pi/(\beta_0\alpha_s^{(\nl)}(m_Q))$, $t_R=-\,2\pi/(\beta_0\alpha_s^{(\nl)}(R))$\,]
\begin{align}\label{eqn:msrsolution}
 m_Q^\MSR(m_Q)-m_Q^\MSR(R)&=-\sum_{n=0}^\infty\gamma_n^R\int_{R}^{m_Q}\dd R\leri{\frac{\alpha_s^{(\nl)}(R)}{4\pi}}^{\!\!n+1}\\
 &=\LQCD\sum_{k=0}^\infty\mathrm e^{i\pi(\hat b_1+k)}S_k\,
\big[\,\Gamma(-\,\hat b_1-k,t_m)-\Gamma(-\,\hat b_1-k,t_R)\,\big], \nonumber
\end{align}
where $\LQCD$ and the coefficients $\gamma_n^R$, $S_k$ and $\hat b_1$ are given in Eqs.~\eqref{eqn:lambda}, \eqref{eqn:revolvdef}, \eqref{eqn:skcoeff} and \eqref{eqn:bhatcoeff} and the series may be truncated at the desired order.
The R-evolution equation can be solved numerically or by using the analytic expression in the second line of Eq.~\eqref{eqn:msrsolution}.

The matching relations for the $\MSb$ and the natural MSR mass can be derived from Eq.~\eqref{eqn:msbmsr1} and written in various ways quoted in the following. From $\mbar_Q\equiv\mbar_Q^{(\nl+1)}(\mbar_Q^{(\nl+1)})$ one can determine $m_Q^{\mathrm{MSRn}}(\mbar_Q)$ using the matching relations [\,$A_s\equiv\alpha_s^{(\nl+1)}(\mbar_Q)/(4\pi)$, $a_s\equiv\alpha_s^{(\nl)}(\mbar_Q)/(4\pi)$\,]
\begin{align}\label{eqn:msbmsr2}
 &m_Q^\mathrm{MSRn}(\mbar_Q^{(\nl+1)})-\mbar_Q^{(\nl+1)}(\mbar_Q^{(\nl+1)})\\
 &=\mbar_Q^{(\nl+1)}(\mbar_Q^{(\nl+1)})
  \Bigl\{\,1.65707\, A_s^2+[\,110.05+1.424\,\nl\,]\,A_s^3+[\,(352.\pm31.)
  \nonumber \\ 
 &\qquad\qquad\qquad\qquad\quad
  -(111.59\pm0.10) \,\nl+4.40\nl^2\,]\,A_s^4\,\Bigr\}\nonumber\\
 &=\mbar_Q^{(\nl+1)}(\mbar_Q^{(\nl+1)})\Bigl\{\,1.65707\, a_s^2\,+[\,110.05+1.424\,\nl\,]\,a_s^3\,+[\,(344.\pm31.)
  \nonumber \\ 
 &\qquad\qquad\qquad\qquad\quad
    - (111.59\pm 0.10) \,\nl  +4.40\,\nl^2\,]\,a_s^4\,\,\Bigr\} \,,\nonumber
\end{align}
where the superscript $(\nl+1)$ is a reminder of the active flavors used to run the $\overline{\rm MS}$ mass. 
Given $m_Q^\mathrm{MSRn}\equiv m_Q^{\mathrm{MSRn},(\nl)}(m_Q^{\mathrm{MSRn},(\nl)})$ one can determine $\mbar_Q^{(\nl+1)}(\mbar_Q^{(\nl+1)})$ by the relations \big[\,$\bar A_s\equiv\alpha_s^{(\nl+1)}(m_Q^\mathrm{MSRn})/(4\pi)$, $\bar a_s\equiv\alpha_s^{(\nl)}(m_Q^\mathrm{MSRn})/(4\pi)$\,\big] 
\begin{align}\label{eqn:msbmsr3}
 &\mbar_Q^{(\nl+1)}(\mbar_Q^{(\nl+1)})
   -m_Q^{\mathrm{MSRn},(\nl)}(m_Q^{\mathrm{MSRn},(\nl)})\\
 &=m_Q^{\mathrm{MSRn},(\nl)}(m_Q^{\mathrm{MSRn},(\nl)})
   \Bigl\{\,- 1.65707\,\bar A_s^2-[\,101.21+1.424\,\nl\,]\,\bar A_s^3
  \nonumber \\ 
 &\qquad\qquad\qquad\qquad\qquad\qquad
 +[\,(349.\pm31.)+ (103.35\pm0.10) \,\nl -4.40\,\nl^2\,]\,\bar A_s^4\,\Bigr\}\nonumber\\
 &=m_Q^{\mathrm{MSRn},(\nl)}(m_Q^{\mathrm{MSRn},(\nl)})
   \Bigl\{\,-1.65707\,\bar a_s^2\,-[\,101.21+1.424\,\nl\,]\,\bar a_s^3\,
  \nonumber \\ 
 &\qquad\qquad\qquad\qquad\qquad\qquad
 +\,[\,(357.\pm31.)+
 (103.35\pm0.10)\,\nl-4.40\,\nl^2\,]\,\,\bar a_s^4\,\,\Bigr\},\nonumber
\end{align}
where the superscript $(\nl)$ is a reminder of the active flavors used to run the MSR mass.
We have displayed the matching relations both for the $\nl$ and the $(\nl+1)$-flavor scheme for the strong coupling. The corresponding matching relations for the strong coupling at the scales $\mbar_Q$ and $m_Q^\mathrm{MSRn}$ are shown for convenience in Eqs.~\eqref{eqn:asmatchingmsbar} and \eqref{eqn:asmatchingmsr}, respectively.

Numerically, $m_t^\mathrm{MSRn}(m_t)-\mbar_t(\mbar_t)$ is about $30$\,MeV for $\mbar_t(\mbar_t)$ around $160$\,GeV. The perturbative uncertainties in this matching relations from missing higher orders are $1$\,MeV or lower for all massive quarks. The numerical uncertainties in the $\Ord(\alpha_s^4)$ coefficients given in Eqs.~\eqref{eqn:msbmsr2} and \eqref{eqn:msbmsr3} are quoted from Ref.~\cite{Marquard:2016dcn} and smaller than $0.01$\,MeV. Thus the matching relations can be taken as exact for all foreseeable applications. 

The matching relations for the $\MSb$ and the practical MSR mass simply reads
\begin{equation}\label{eqn:msrpmsbarmatchv2}
 m_Q^{\mathrm{MSRp},(\nl)}(m_Q^{\mathrm{MSRp},(\nl)})
  =\mbar_Q^{(\nl+1)} (\mbar_Q^{(\nl+1)}) \,,
\end{equation}
to all orders of perturbation theory, where in comparison to Eq.~\eqref{eqn:msrpmsbarmatch} we have also explicitly indicated the flavor number of the evolution of the MSR mass and the $\MSb$ mass.

\begin{figure}[h]
\center
\includegraphics[width=0.53\textwidth]{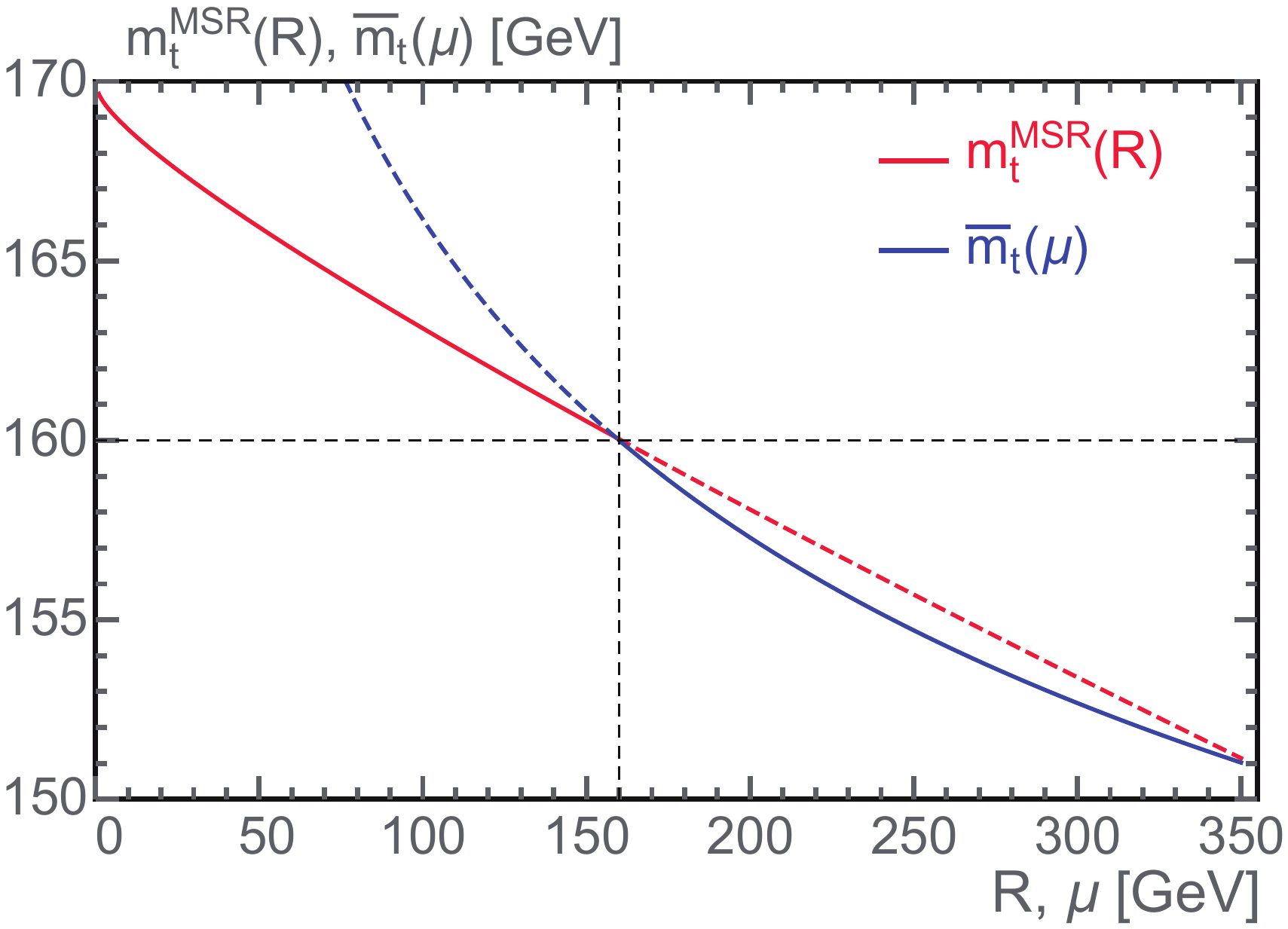}
 \caption{\label{fig:msbarmsr} Comparison of the scale dependence for the $\overline{\rm MS}$ and the  MSR top quark masses ($\nl=5$) as a function of $\mu$ and $R$ in GeV.}
\end{figure}

In Fig.~\ref{fig:msbarmsr} we show the scale dependence of the MSR masses $m_t^\MSR(R)$ (red line) and the $\MSb$ mass $\mbar_t(\mu)$ (blue line) for $\mbar_t^{(\nl+1)}(\mbar_t^{(\nl+1)})=160$\,GeV. The difference between the natural and practical MSR masses is not visible on the scale of this figure. The solid curves represent the evolution of the masses in regions where they should be used for physical applications in close analogy to the flavor-number-dependent scale dependence of the strong coupling, while the dashed lines show their evolution beyond these regions. At the scale $160$\,GeV the two mass schemes are matched via Eq.~\eqref{eqn:msbmsr2}, Eq.~\eqref{eqn:msbmsr3} and Eq.~\eqref{eqn:msrpmsbarmatchv2}. For $R<m_t$ the MSR mass $m_t^\MSR(R)$ is substantially smaller than the $\MSb$ mass $\mbar_t(R)$ and approaches the pole mass for $R\to 0$. The MSR mass remains well defined for all $R\gtrsim \Lambda_{\rm QCD}$, whereas the exact value for the limit $m_t^\MSR(R\to0)$ is ambiguous due to the Landau pole in the evolution of the strong coupling in the R-evolution equation \eqref{eqn:msrsolution}. This illustrates the ambiguity of the pole mass concept.

\section{Conclusions}
\label{sec:conclusions}
This paper had two main aims.
The first aim was to give a detailed presentation of the MSR mass, which is an $R$-dependent short-distance mass designed for high-precision determinations of heavy quark masses from quantities where the physical scales are smaller than the quark mass, $R<m_Q$. Since such scale hierarchies can only be really large for the top quark, the MSR mass concept is most useful in the context of top quark physics, but it may be useful for bottom and charm quark analyses as well. The MSR mass is obtained from the results of heavy quark on-shell self-energy diagrams which is not the case for any earlier low-scale short-distance mass given in the literature. The MSR mass has therefore a very close relation to the well-known $\MSb$ mass $\mbar_Q(\mu)$, and should be viewed as the generalization of the $\MSb$ mass concept for renormalization scales below $m_Q$, where the  $\MSb$ mass is known to be impractical and does not capture the proper physics. The main feature of the MSR mass is that its renormalization group evolution is linear and logarithmic in the scale $R$, compared to the purely logarithmic evolution of the $\MSb$ mass. This linear scale dependence in the renormalization group flow of the MSR mass is called R-evolution and the MSR mass is well defined for any $R\gtrsim\Lambda_{\rm QCD}$. Formally, in the limit $R\to 0$, the MSR mass can be evolved to the pole mass. However, taking this limit is ambiguous as it involves evolving the strong coupling through the Landau pole, which illustrates the ${\cal O}(\Lambda_{\rm QCD})$ ambiguity of the pole mass scheme. Since there are two options to treat the corrections coming from virtual heavy quark loops in the heavy quark self-energy diagrams, we defined two variants of the MSR mass, the {\it natural MSR mass} $m_Q^{\rm MSRn}(R)$, where these effects are integrated out, and the {\it practical MSR mass} $m_Q^{\rm MSRp}(R)$, where they are still included in the mass definition.  
Both MSR masses can be easily related to all other short-distance mass schemes available in the literature. We have provided 
all necessary formulae such that conversions can be carried out to ${\cal O}(\alpha_s^4)$ and we have discussed in detail the cases where there are subtleties in the conversion.

The second aim of the paper was to give a detailed presentation of how R-evolution can be used to derive an analytic expression for the normalization of the high-order asymptotic behavior of the MSR-pole (or $\MSb$-pole) mass perturbative series related to the ${\cal O}(\Lambda_{\rm QCD})$ renormalon ambiguity contained in the pole mass. This analytic result can be applied to any perturbative series and be used to probe the known coefficients for the series pattern related to an ${\cal O}(\Lambda_{\rm QCD})$ renormalon ambiguity. Since using the result does not involve any numerical comparison of the series coefficients, but is a very simple analytic function of the coefficients, we call it the {\it ${\cal O}(\Lambda_{\rm QCD})$ renormalon sum rule}. Using the sum rule we reanalyzed the ${\cal O}(\Lambda_{\rm QCD})$ renormalon in the  MSR-pole (and $\MSb$-pole) perturbative series and showed that the sum rule results are fully compatible with previous available methods. We examined the relation between these methods to our sum rule analytically and explained the reason why one of them has very slow convergence. We also applied the sum rule to a number of other quantities known to high order and demonstrated its high sensitivity. These examples included the PS-pole mass relation, the moments of the massive quark vacuum polarization, the hadronic R-ratio and the QCD $\beta$-function.

\section*{Acknowledgments}
We  acknowledge partial support by the FWF Austrian Science Fund under the Doctoral Program No. W1252-N27 and the Project No.  P28535-N27, the U.S. Department of Energy under the Grant No. DE-SC0011090, the Simons Foundation through the Grant 327942, the  Spanish  MINECO  ``Ram\'on y  Cajal'' program (RYC-2014-16022), MECD grants FPA2016-78645-P, FPA2014-53375-C2-2-P and FPA2016-75654-C2-2-P, the group UPARCOS, the IFT ``Centro de Excelencia Severo Ochoa'' Program under Grant SEV-2012-0249 and by the Ramanujan Fellowship of SERB, DST. We also thank the Erwin-Schr\"odinger International Institute for Mathematics and Physics, the University of Vienna and Cultural  Section  of  the  City  of  Vienna  (MA7)  for  partial support.
\vspace*{0.3cm}

{\it Note Added}: After this paper was originally posted the comments in Ref.~\cite{Pineda:2017uby} appeared. We have added Appendix~\ref{sec:N12alternative} to make a comparison of our sum rule with the method and formulas discussed there. 

\begin{appendix}
 \section{QCD \texorpdfstring{$\beta$}{beta}-Function and Coefficients} 
 \label{app:coefficients}
 
 For the QCD $\beta$-function in the $\overline{\rm MS}$ scheme we use the convention
 \begin{equation}\label{eqn:betafct}
  \frac{\dd\alpha_s(R)}{\dd\log R}=\beta(\alpha_s(R))\,=\,-\,2\,\alpha_s(R)\sum_{n=0}^\infty\beta_n\leri{\frac{\alpha_s(R)}{4\pi}}^{\!\!n+1} ,
 \end{equation}
 where $\beta_0=11-2/3\,\nl$ with $\nl$ being the number of dynamical flavors. The coefficients are known up to $\beta_4$ from Refs.~\cite{Tarasov:1980au, Larin:1993tp, vanRitbergen:1997va, Korchemsky:1987wg, Moch:2004pa, Czakon:2004bu, Baikov:2016tgj}. The equation can be used to write [\,$\alpha_i\equiv\alpha_s(R_i)$, $t=-\,2\pi/(\beta_0\alpha_s(R))$\,]
 \begin{equation}\label{eqn:lnRt}
  \log\frac{R_1}{R_0}=\int_{\alpha_0}^{\alpha_1}\frac{\dd\alpha}{\beta(\alpha)}=\int_{t_1}^{t_0}\dd t\,\hat b(t)=G(t_0)-G(t_1) \;,
 \end{equation}
 where
 \begin{equation}\label{eqn:bhatG}
  \hat b(t)=1+\sum_{k=1}^\infty\frac{\hat b_k}{t^k},\qquad G(t)=t+\hat b_1\log(-\,t)-\sum_{k=2}^\infty\frac{\hat b_k}{(k-1)\,t^{k-1}} \;,
 \end{equation}
 and the first four coefficients relevant for renormalon sum rule applications up to $\Ord(\alpha_s^4)$ are
 \begin{align}\label{eqn:bhatcoeff}
  \hat b_1&=\frac{\beta_1}{2\beta_0^2} \;, & \hat b_3&=\frac{1}{8\beta_0^6}(\beta_1^3-2\,\beta_0\,\beta_1\beta_2+\beta_0^2\,\beta_3),\\
  \hat b_2&=\frac{1}{4\beta_0^4}(\beta_1^2-\beta_0\,\beta_2)\,, & \hat b_4&=\frac{1}{16\beta_0^8}(\beta_1^4-3\,\beta_0\,\beta_1^2\,\beta_2 +\beta_0^2\,\beta_2^2 +2\,\beta_0^2\,\beta_1\,\beta_3-\beta_0^3\,\beta_4) \nonumber\,.
 \end{align}
 One can show the following recursion relation for the $\hat b_k$ coefficients ($\hat b_0 \equiv 1$):
 \begin{equation}\label{eqn:recursiveB}
 \hat b_{n+1} = 2\sum_{i\,=\,0}^n\, \frac{\hat b_{n-i}\,\beta_{i+1}}{(-2\beta_0)^{i+2}}\,,
 \end{equation}
 which can be used for an automated computation. From Eq.~\eqref{eqn:lnRt} one can also derive the known relation
 \begin{equation}\label{eqn:lambda}
  \LQCD=R_i\,\mathrm e^{G(t_i)} \;,
 \end{equation}
that gives 
$\LQCD^{\mathrm{N}^k\mathrm{LL}}$ if the series in $G(t_i)$ is truncated after the $k$-th term.

The matching relations for the strong coupling in the $\nl$ and the $(\nl+1)$-flavor schemes at the scale
$\mbar_Q\equiv\mbar_Q^{(\nl+1)}(\mbar_Q^{(\nl+1)})$ read
\begin{align}
\label{eqn:asmatchingmsbar}
\alpha_s^{(\nl)}(\mbar_Q) =\, &\alpha_s^{(\nl+1)}(\mbar_Q)\bigg[1+0.152778\bigg(\frac{\alpha_s^{(\nl+1)}(\mbar_Q)}{\pi}\bigg)^{\!\!2}\\
&\qquad +\,(0.972057-0.08465\,\nl)\bigg(\frac{\alpha_s^{(\nl+1)}(\mbar_Q)}{\pi}\bigg)^{\!\!3}+\ldots\bigg]\,,\nonumber\\
\alpha_s^{(\nl+1)}(\mbar_Q) = \,& \alpha_s^{(\nl)}(\mbar_Q)\bigg[1-0.152778\bigg(\frac{\alpha_s^{(\nl)}(\mbar_Q)}{\pi}\bigg)^{\!\!2}\\
&\qquad-\,(0.972057-0.08465\,\nl)\bigg(\frac{\alpha_s^{(\nl)}(\mbar_Q)}{\pi}\bigg)^{\!\!3}+\ldots\bigg].\nonumber
\end{align}
The matching relations for the strong coupling in the $\nl$ and the $(\nl+1)$ flavor schemes at the scale
$m_Q^\mathrm{MSRn}\equiv m_Q^{\mathrm{MSRn},(\nl)}(m_Q^{\mathrm{MSRn},(\nl)})$ read 
\begin{align}
\label{eqn:asmatchingmsr}
\alpha_s^{(\nl)}(m_Q^\mathrm{MSRn}) =\,& \alpha_s^{(\nl+1)}(m_Q^\mathrm{MSRn})\bigg[1+0.152778
\bigg(\frac{\alpha_s^{(\nl+1)}(m_Q^\mathrm{MSRn})}{\pi}\bigg)^{\!\!2}\\
&\qquad +\,(0.93753-0.08465\,\nl)\bigg(\frac{\alpha_s^{(\nl+1)}(m_Q^\mathrm{MSRn})}{\pi}\bigg)^{\!\!3}+\ldots\bigg]\,,\nonumber\\
\alpha_s^{(\nl+1)}(m_Q^\mathrm{MSRn}) = \,& \alpha_s^{(\nl)}(m_Q^\mathrm{MSRn})\bigg[1-0.152778
\bigg(\frac{\alpha_s^{(\nl)}(m_Q^\mathrm{MSRn})}{\pi}\bigg)^{\!\!2}\\
&\qquad -\,(0.93753-0.08465\,\nl)\bigg(\frac{\alpha_s^{(\nl)}(m_Q^\mathrm{MSRn})}{\pi}\bigg)^{\!\!3}+\ldots\bigg].\nonumber
\end{align}

The R-anomalous dimension coefficients $\gamma_n^R$ take the following numerical values for the natural MSR mass:
\begin{align}\label{eqn:gammaRn}
\gamma_0^{Rn} & = {\textstyle \frac{16}{3}}\,,\\
\gamma_1^{Rn} & = 96.1039 - 9.55076\, \nl\,,\nonumber\\
\gamma_2^{Rn} & = 1595.75 - 269.953\, \nl - 2.65945\, \nl^2\,,\nonumber\\
\gamma_3^{Rn} & = (12319.\pm417.) - (9103.\pm10.)\, \nl + 610.264\, \nl^2 - 6.515\, \nl^3\,,\nonumber
\end{align}
whereas for the practical MSR mass the coefficients are:
\begin{align}\label{eqn:gammaRp}
\gamma_0^{Rp} & = {\textstyle \frac{16}{3}}\,,\\
\gamma_1^{Rp} & = 97.761 - 9.55076\, \nl\,,\nonumber\\
\gamma_2^{Rp} & = 1632.89 - 264.11\, \nl - 2.65945\, \nl^2\,,\nonumber\\
\gamma_3^{Rp} & = (4724.\pm418.) - (8784.\pm10.)\, \nl + 620.362\, \nl^2 - 6.515\, \nl^3\,.\nonumber
\end{align}
The uncertainties appearing in the coefficients $\gamma_3^{Rn,Rp}$ are from numerical errors in the results of Ref.~\cite{Marquard:2016dcn}. They amount to an uncertainty in the solutions of the R-evolution equation of $1$\,MeV or less for all relevant cases and are smaller than the uncertainty due to missing higher orders. Therefore they can be neglected for all practical purposes.

The coefficients $g_\ell$ defined by the series $\sum_{\ell=0}^\infty g_\ell\,(-t)^{-\ell}\equiv\mathrm e^{G(t)}\,\mathrm e^{-t}\, (-t)^{-\hat b_1}$ relevant for the renormalon sum rule up to $\Ord(\alpha_s^4)$ read
 \begin{equation}\label{eqn:glcoeff}
  g_0\,=\,1 \;, \qquad g_1\,=\,\hat b_2 \;, \qquad g_2\,=\,\frac{1}{2}(\hat b_2^2\,-\,\hat b_3) \;, \qquad g_3\,=\,\frac{1}{6}(\hat b_2^3\,-\,3\,\hat b_2\,\hat b_3\,+\,2\,\hat b_4) \;.
 \end{equation}
One can proof the following recursion relation for $g_\ell$:
\begin{equation}\label{eqn:recursiveG}
	g_{n+1} = \frac{1}{1+n}\sum_{i=0}^n\,(-1)^i\,\hat b_{i+2}\,g_{n-i}\,,
\end{equation}
 suitable for automated computation. The coefficients $g_\ell$ agree with the coefficients $s_\ell$ given in Refs.~\cite{Beneke:1994rs,Beneke:1998ui}.

 The coefficients $S_k$ defined from the series $\sum_{k=0}^\infty S_k\,(-t)^{-k}\equiv -\,t\,\gamma^R(t)\,\hat b(t)\,\mathrm e^{-G(t)}\,\mathrm e^t\,(-t)^{\hat b_1}$ relevant up to $\Ord(\alpha_s^4)$ read [\,$\tilde\gamma_k^R=\gamma_k^R/(2\beta_0)^{k+1}$\,]
 \begin{align}\label{eqn:skcoeff}
  S_0=\,&\tilde\gamma_0^R=\frac{a_1}{2\beta_0} \,,\\
  S_1=\,&\tilde\gamma_1^R-(\hat b_1+\hat b_2)\,\tilde\gamma_0^R=\frac{a_2}{4\beta_0^2}-\frac{a_1}{2\beta_0}(1+\hat b_1+\hat b_2)\, ,\nonumber\\
  S_2=\,&\tilde\gamma_2^R-(\hat b_1+\hat b_2)\,\tilde\gamma_1^R+\bigg[(1+\hat b_1)\,\hat b_2+\frac{1}{2}(\hat b_2^2+\hat b_3)\bigg]\tilde\gamma_0^R\nonumber\\
  =\,&\frac{a_3}{8\beta_0^3}-\frac{a_2}{4\beta_0^2}(2+\hat b_1+\hat b_2)+\frac{a_1}{2\beta_0}\leris{(2+\hat b_1)\,\hat b_2+\frac{1}{2}(\hat b_2^2+\hat b_3)} ,\nonumber\\
  S_3=\,&\tilde\gamma_3^R-(\hat b_1+\hat b_2)\,\tilde\gamma_2^R+\bigg[(1+\hat b_1)\,\hat b_2+\frac{1}{2}(\hat b_2^2+\hat b_3)\bigg]\tilde\gamma_1^R\nonumber\\
  &-\leris{\leri{1+\frac{1}{2}\hat b_1+\frac{1}{6}\hat b_2}\hat b_2^2+\leri{1+\frac{1}{2}\hat b_1+\frac{1}{2}\hat b_2}\hat b_3+\frac{1}{3}\,\hat b_4}\tilde\gamma_0^R\nonumber\\
  =\,&\frac{a_4}{16\beta_0^4}-\frac{a_3}{8\beta_0^3}(3+\hat b_1+\hat b_2)+\frac{a_2}{4\beta_0^2}\leris{(3+\hat b_1)\,\hat b_2+\frac{1}{2}(\hat b_2^2+\hat b_3)}\nonumber\\
  &-\frac{1}{2}\frac{a_1}{2\beta_0}\leris{\leri{3+\hat b_1+\frac{1}{3}\hat b_2}\hat b_2^2+\leri{3+\hat b_1+\hat b_2}\hat b_3+\frac{2}{3}\,\hat b_4} .\nonumber
 \end{align}
The relation between the $S_k$ coefficients and the R-anomalous dimension can be compactly written as follows:
\begin{align}\label{eqn:skcoeff2}
S_k \,=\, & \tilde\gamma_k^R - (1-\delta_{k,0})\,(\hat b_1+\hat b_2)\,\tilde\gamma_{k-1}^R + \sum_{n=0}^{k-2}
\tilde\gamma_{n}^R\,\bigg[\tilde g_{k-n} + (-1)^{k-n} \hat b_{k-n}  \\
& + \sum_{\ell = 1}^{k-n-1}(-1)^{k-n-\ell}\, \tilde g_\ell\,\hat b_{k-n-\ell}\bigg],\nonumber\\
\tilde g_{n+1} \,=\,& -\frac{1}{1+n} \sum_{i=0}^n\,(-1)^i\,\hat b_{i+2} \,\tilde g_{n-i}\,, \qquad \tilde g_0 = 1\,.
\end{align}
In addition one can use Eq.~\eqref{eqn:anintermsofsk} to write a recursion relation for the $S_k$ coefficients, which are then expressed in terms of $a_i$\,:
\begin{equation}\label{eqn:skcoeff3}
  S_k = \frac{a_{k+1}}{(2\beta_0)^{k+1}} - \sum_{n=0}^{k-1}S_n\sum_{\ell=0}^{k-n} g_\ell\,(1+\hat b_1 + n)_{k-\ell-n}\,,
\end{equation}
where $(b)_n=b\,(b+1)\cdots(b+n-1)=\Gamma(b+n)/\Gamma(b)$ is the Pochhammer symbol.
This formula can be used for an automated implementation of $S_k$ once the $g_\ell$ coefficients have been computed.
We note that in order to determine the coefficients $S_k$, one needs all terms up to $k$ loops in the R-evolution equation, and the $(k+1)$-loop QCD $\beta$-function.

\section{Alternative Derivation of the \texorpdfstring{$\Ord(\LQCD)$}{Order Lambda QCD} Renormalon Sum Rule} \label{sec:N12alternative}

In Sec.~\ref{sec:derivation} we have shown how to directly derive the sum rule formula for $N_{1/2}$ displayed in Eq.~\eqref{eqn:P12def} from the computation of the Borel transform of Eq.~\eqref{eqn:borel2} starting from the solution of the R-evolution equation given in Eq.~\eqref{eqn:msrpolefull}. There is an interesting alternative way to determine the sum rule formula which starts from the Borel function $B_{\alpha_s(R)}(u)$ given in Eq.~\eqref{eqn:borel2} without 
knowing the expression for $N_{1/2}$. This expression is equivalent to the  Borel transform of the original series $-R\sum_{i=1}^{\infty} a_i \,[\,\alpha_s(R)/(4\pi)\,]^i$ which has the form:
\begin{equation}\label{eqn:originalB}
 B_{\alpha_s(R)}(u) = -R \sum_{i=1}^{\infty} a_i\, \frac{u^{i-1}}{\Gamma(i)}\, \beta_0^{-i} \;,
\end{equation}
in the fixed-order expansion in powers of the Borel variable $u$.

Consider now the modified Borel function $(\beta_0/4\pi R)(1-2u)^{1+\hat{b}_1} B_{\alpha_s(R)}(u)$. Inserting Eq.~\eqref{eqn:borel2} for $B_{\alpha_s(R)}(u)$ one obtains:
\begin{align}\label{eqn:modBv1}
 \frac{\beta_0}{4\pi R}(1-2u)^{1+\hat{b}_1} B_{\alpha_s(R)}(u) = &-N_{1/2}\sum_{\ell=0}^{\infty}\,g_\ell\;\frac{\Gamma(1+\hat{b}_1 - \ell)}{\Gamma(1+\hat{b}_1)}\,(1-2u)^\ell \\ &+ \frac{\beta_0}{2\pi}(1-2u)^{1+\hat{b}_1}\sum_{\ell=0}^\infty\,g_\ell\;Q_\ell(u)\;,\nonumber
\end{align}
where the role of analytic and non-analytic terms is just reversed compared to Eq.~\eqref{eqn:borel2}. Truncating the series in $\ell$ at order $n$ (which corresponds to including the coefficients $a_i$, $S_k$, $g_\ell$ up to $i=n+1$, $k=n$ and $\ell=n$, respectively), one can see that expanding Eq.~\eqref{eqn:modBv1} in powers of $u$ up to order $n$ and taking the limit $u\rightarrow 1/2$ one singles out $N_{1/2}$ on the RHS:
\begin{align}
 -\,N_{1/2}^{(n)} \;&+\; \frac{\beta_0}{2\pi} \sum_{k=0}^n\; \sum_{m=0}^k \sum_{i=k-m+1}^n \sum_{\ell=0}^i (-1)^m\,g_\ell\, S_{i-\ell}\\
 & \hspace{3cm}\times\frac{\Gamma(2+\hat{b}_1)}{\Gamma(m+1)\,\Gamma(2+\hat{b}_1-m)}\frac{\Gamma(1+\hat{b}_1+k-m-\ell)}
 	{\Gamma(1+\hat{b}_1+i-\ell) \Gamma(k-m+1)} \;,\nonumber
\end{align}
where $N_{1/2}^{(n)}$ refers to the $(n+1)$-loop approximation for $N_{1/2}$.
Applying the same procedure to the Borel transform of Eq.~\eqref{eqn:originalB} and solving for $N_{1/2}^{(n)}$ one obtains:
\begin{align}\label{eqn:N12ordern}
 N_{1/2}^{(n)} ={}& \frac{1}{4\pi}\sum_{k=0}^n\,\sum_{m=0}^k\,\frac{(-1)^m}{(2\beta_0)^{k-m}}\frac{\Gamma(2+\hat{b}_1) a_{k-m+1}}{\Gamma(k-m+1)\Gamma(m+1)\Gamma(2+\hat{b}_1-m)}\\
&+ \frac{\beta_0}{2\pi} \sum_{k=0}^n\; \sum_{m=0}^k \sum_{i=k-m+1}^{n} \sum_{\ell=0}^i (-1)^m\,g_\ell\, S_{i-\ell}\nonumber\\
& \hspace{3cm}\times\frac{\Gamma(2+\hat{b}_1)}{\Gamma(m+1)\Gamma(2+\hat{b}_1-m)}\frac{\Gamma(1+\hat{b}_1+k-m-\ell)}{\Gamma(1+\hat{b}_1+i-\ell)\Gamma(k-m+1)} \;.\nonumber
\end{align}
Although lengthier, it can be checked that this formula agrees exactly with the sum rule of Eq.~\eqref{eqn:P12def} at $(n+1)$-loop order (i.e. when truncated with $k\le n$ as shown). 

In Ref.~\cite{Pineda:2001zq} (see also Ref.~\cite{Lee:1996yk}), a version of the above considerations to determine the normalization of the non-analytic terms in Eq.~\eqref{eqn:borel2}, which we refer to as the Borel method, was proposed. They made the additional assumption that the analytic terms on the RHS of Eq.~\eqref{eqn:borel2} can be neglected because they quickly tend
to zero when multiplied by $(1-2u)^{1+\hat{b}_1}$ in the limit $u\to1/2$. Therefore they did not include the terms related to the polynomials $Q_\ell$. This leads to a formula for the normalization that only contains the first term on the RHS of Eq.~\eqref{eqn:N12ordern}, which they called $N_m$.
After a bit of algebra, the 
double sum of this term can be recast into a single summation, yielding: \footnote{We note that no analytic formula for $N_m^{(n)}$ was provided in Ref.~\cite{Pineda:2001zq}, and that Eq.~\eqref{eqn:Nmv1} correctly encodes the prescription given there. In formula (7) of Ref.~\cite{Pineda:2017uby} the following analytic double series formula was given:
\begin{align}  \label{eq:Nmpineda}
  N_m 
  &= \frac{1}{\nu} \sum_{m,n'=0}^\infty \frac{ \Gamma(2+ b) (-1)^m r_{n'}(\nu)}{\Gamma(m+1)\Gamma(n'+1)\Gamma(2+ b -m)} 
  \bigg(\frac{2\pi}{\beta_0}\bigg)^{n'} \\
  &=  \frac{1}{4\pi}\sum_{m,n'=0}^\infty\frac{ \Gamma(2+\hat b_1) (-1)^m a_{n'+1}}{\Gamma(m+1)\Gamma(n'+1)
\Gamma(2+\hat b_1 -m)} \frac{1}{(2\beta_0)^{n'}} \,, \nonumber
\end{align}
where in the second line we have converted to our conventions for ease of comparison. Eq.~(\ref{eq:Nmpineda}) is not fully specified because it does not provide a prescription how to systematically truncate the two series in order to compute $N_m$ at $(n+1)$-loop order. The sum for $(1-2u)^{1+\hat b_1}=\sum_{m=0}^\infty (2u)^m \Gamma(2+\hat b_1)/[\,\Gamma(m+1) \Gamma(2+\hat b_1-m)\,]$ converges to zero at $u=1/2$, while the other, which is Eq.~\eqref{eqn:originalB}, is divergent for $u=1/2$. To obtain Eq.~(\ref{eqn:Nmv1}) from Eq.~(\ref{eq:Nmpineda}) one switches variable from $(m,n')$ to $(k,m)$ with $k=m+n'$, and then finally truncates with respect to the variable $k$.}
\begin{equation}\label{eqn:Nmv1}
 N_m^{(n)} = \frac{1}{4\pi}\sum_{m=0}^n \frac{(-\hat b_1)_{n-m}\,a_{m+1}}
{(2\beta_0)^m\,m!\,(n-m)!}\;.
\end{equation}
However, the contribution from the second term on the RHS of Eq.~\eqref{eqn:N12ordern} is actually not negligible because it involves the expansion of the $(1-2u)^{1+\hat b_1}$ and setting $u=1/2$ afterwards. In particular, the \mbox{$\beta$-function} coefficients $\beta_{n>1}$ contained in the $g_\ell$ are essential for the cancellation of the \mbox{$\lambda$-dependence} with $n$ beyond 2-loop order, i.e. for $n>1$.

\begin{figure}
\center
 \includegraphics[width=0.53\textwidth]{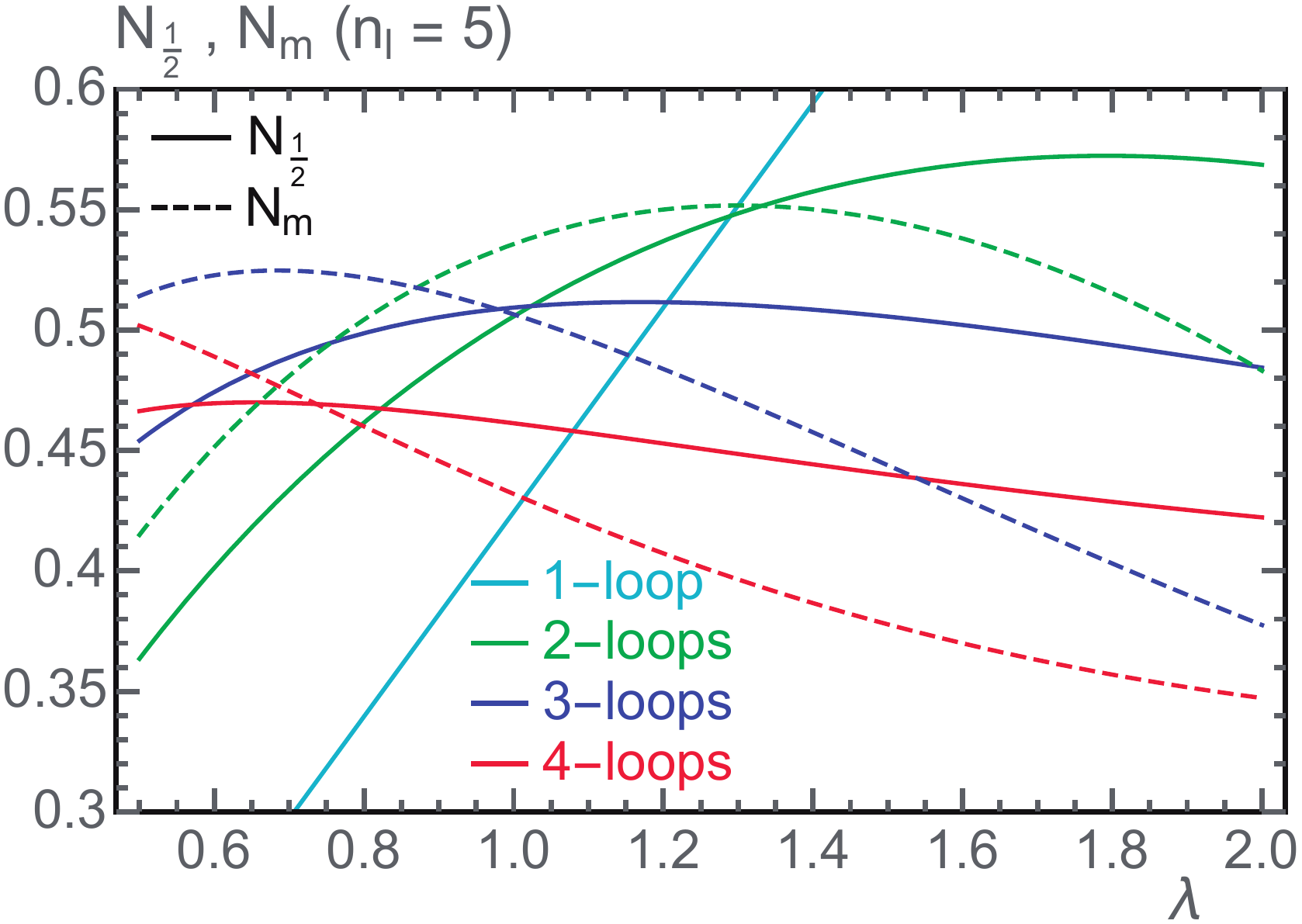}
 \caption{\label{fig:N12vsNm} Comparison of $N_{1/2}^{(n)}$ and $N_m^{(n)}$ for $\nl=5$. Results are shown as a function of $\lambda$ including contributions from one to four loops.}
\end{figure}

This is shown in Fig.~\ref{fig:N12vsNm} where we plot $N_{1/2}^{(n)}$ (solid lines) and $N_{m}^{(n)}$ (dashed lines) obtained from the natural MSR mass for $n=0$ (cyan), $n=1$ (green), $n=2$ (blue) and $n=3$ (red) for $\nl=5$ as a function of $\lambda$ in the interval $[\,0.5,\,2\,]$. We see that the results for $N_m^{(n)}$ differ substantially from $N_{1/2}^{(n)}$ showing that the terms neglected in the approach of Ref.~\cite{Pineda:2001zq} are numerically sizable and, in particular, do not decrease with the order $n$. Moreover, the results for $N_m^{(n)}$
do not appear to show any reduced $\lambda$-dependence beyond 2-loop order, in contrast to the results for $N_{1/2}^{(n)}$.
Interestingly, in Ref.~\cite{Bali:2013pla} it has been shown that when many more terms of the expansion are known [\,they accounted for terms up to $\mathcal{O}(\alpha_s^{20})$ for the quark and gluino QCD static potential\,], Eq.~\eqref{eqn:Nmv1} does eventually converge to the right value and shows reduced scale variation. We have numerically confirmed 
that using series generated from the Borel function of Eq.~(\ref{eqn:borel2}) setting (by hand) explicit expressions for the functions
$Q_\ell(u)$, such as $Q_\ell(u)=\delta_{\ell,0}$.
The eventual convergence at very high orders $n$ can be understood from the fact that the contributions in the asymptotic behavior of the perturbative coefficients $a_n$ that arise from the \mbox{$\beta$-function} coefficients $\beta_{n>1}$ become $1/n$ suppressed and eventually become also numerically small, see Eqs.~(\ref{eqn:anasy}) and (\ref{eqn:anintermsofsk}). But in any case, its very slow convergence renders the Borel method less practical and less precise for most phenomenological applications, for which only a few terms of the perturbative expansion are known.

 \section{Other Short Distance Masses}
 \label{sec:othermasscoeff}
 
 The PS mass \cite{Beneke:1998rk} is defined by the integral of the momentum space color singlet static potential between a quark-antiquark pair, each having infinite mass. The relation of the PS mass to the pole mass has the form
 \begin{equation}\label{eqn:PSmass}
  m_Q^\pole-m_Q^\PS(\mu_f)=\mu_f\sum_{n=1}^\infty a_n^\PS\bigg(\frac{\alpha_s^{(\nl)}(\mu_f)}{4\pi}\bigg)^{\!\!n} ,
 \end{equation}
 where the coefficients are known up to $\Ord(\alpha_s^4)$ based on Refs.~\cite{Fischler:1977yf,Billoire:1979ih,Peter:1996ig,Schroder:1998vy,Anzai:2009tm,Smirnov:2009fh,Lee:2016cgz}, and have the form
 \begin{align}\label{eqn:PSmasscoeff}
  a_1^\PS&={\textstyle \frac{16}{3}}\,,\\
  a_2^\PS&=172.4444 - 13.03704\,\nl \,,\nonumber\\
  a_3^\PS&=11111.55-1522.482\,\nl+41.350\,\nl^2 \,,\nonumber\\
  a_4^\PS&=913336.84-179514.95\,\nl+10535.70\,\nl^2-172.72\,\nl^3 + 22739.57\log\!\bigg(\frac{\mu}{\mu_f}\bigg) \,.\nonumber
 \end{align}
In the standard convention for the PS mass defined in Ref.~\cite{Beneke:2005hg} the term $\log(\mu/\mu_f)$ appearing in $a_4^\PS$ is set to zero.
 
The definition of the 1S mass \cite{Hoang:1998ng,Hoang:1998hm,Hoang:1999ye} in terms of the pole mass is given in Eq.~\eqref{eqn:1Smass} and the coefficients $c_{n,k}$ up to $\Ord(\alpha_s^5)$ read \cite{Hoang:1998ng,Hoang:1998hm,Hoang:1999ye,Penin:2002zv,Kiyo:2014uca,Lee:2016cgz}
 \begin{align}\label{eqn:1Smasscoeff}
  c_{1,0}&=-\,2.09440 \,,\\
  c_{2,0}&=-\,135.438+10.2393\,\nl \,,\nonumber\\
  c_{2,1}&=-\,92.1534+5.5851\,\nl \,,\nonumber\\
  c_{3,0}&=-\,11324.72+1372.745\,\nl-38.9677\,\nl^2 \,,\nonumber\\
  c_{3,1}&=-\,7766.02+1077.92\,\nl-33.5103\,\nl^2 \,,\nonumber\\
  c_{3,2}&=-\,3041.06+368.61\,\nl-11.1701\,\nl^2 \,,\nonumber\\
  c_{4,0}&=-\,1005116.33+176714.27\,\nl-10088.35\,\nl^2+168.57\,\nl^3-63574.35\,\log(\alpha_s^{(\nl)}(\mu)) \,,\nonumber\\
  c_{4,1}&=-\,901778.56+162559.51\,\nl -9263.14\,\nl^2+163.15\,\nl^3 \,,\nonumber\\
  c_{4,2}&=-\,303000.33+61184.26\,\nl-3823.90\,\nl^2+74.47\,\nl^3 \,,\nonumber\\
  c_{4,3}&=-\,89204.48+16219.00\,\nl-982.97\,\nl^2+19.86\,\nl^3 \,.\nonumber
 \end{align}

Employing the $\Upsilon$-expansion (relativistic order counting) the formula for the MSR masses as a function of the 1S mass up to $\Ord(\alpha_s^4)$ reads [\,$M_{B}^{\rm 1S}=C_F\,\alpha_s^{(\nl)}(R)\,m_Q^\OS$, $A_R=C_F\,\alpha_s^{(\nl)}(R)$, $a_s=\alpha_s^{(\nl)}(R)/(4\pi)$, $L=\log(R/M_{B}^{\rm 1S})$\,]
\begin{align}\label{eqn:MSRin1Supsilon}
  &m_Q^\MSR(R)-m_Q^\OS=\,-\left[\,R \,a_1+M_{B}^{\rm 1S}\, c_{1,0}\,\right]a_s\\
  &-\Big[R\, a_2-M_{B}^{\rm 1S}\Big(A_R\, c_{1,0}^2- c_{2,0} - c_{2,1}L\Big)\Big]\,a_s^2\nonumber\\
  &-\Big[R \,a_3+M_{B}^{\rm 1S}\Big(A_R^2 \,c_{1,0}^3-A_R\, c_{1,0}\Big( 2\,c_{2,0}-c_{2,1}+2\,c_{2,1} L\Big) +
  c_{3,0}+ c_{3,1}L + c_{3,2}L^2\Big) \Big]a_s^3\nonumber\\
  &-\Big[R \,a_4-M_{B}^{\rm 1S}\Big(A_R^3 \,c_{1,0}^4-A_R^2\, c_{1,0}^2\Big(3\,c_{2,0}-\big({\textstyle\frac{5}{2}}-3\,L\big)\,c_{2,1}\Big) \nonumber\\
  &\qquad +A_R \Big( c_{2,0}\, \big(c_{2,0}-(1-2\,L)\,c_{2,1} \big)-(1-L)\, c_{2,1}^2 L + c_{1,0} \big(\,2\,c_{3,0}\nonumber\\
  &\qquad\; -(1-2\,L)\,c_{3,1}-2\,(1-L)\, c_{3,2}\,L \big) \Big)
  - c_{4,0} - c_{4,1} L - c_{4,2} L^2- c_{4,3} L^3\Big) \Big]a_s^4\,.\nonumber
 \end{align}
Employing the non-relativistic order counting the formula for the MSR masses as a function of the 1S mass up to $\Ord(\alpha_s^5)$ reads [\,$M_B^{1S}=C_F\,\alpha_s^{(\nl)}(R)\,m_Q^\OS$,  
 $a_s=\alpha_s^{(\nl)}(R)/(4\pi)$, \mbox{$L=\log(R/M_B^{1S})$}\,]
 \begin{align}\label{eqn:MSRin1Snonrel}
 m_Q^\MSR(R)-m_Q^\OS= \,& -\left[\,R \,a_1+M_{B}^{\rm 1S}\, c_{1,0}\,\right]a_s\\
 &-\Big[R\, a_2+M_{B}^{\rm 1S}\Big(c_{2,0} + c_{2,1}L\Big)\Big]\,a_s^2\nonumber\\
 &-\Big[R \,a_3-M_{B}^{\rm 1S}\Big(4\pi\,C_F\,c_{1,0}^2 - c_{3,0} - c_{3,1} L - c_{3,2} L^2\Big) \Big]a_s^3\nonumber\\
 &-\Big[R \,a_4-M_{B}^{\rm 1S}\Big(4\pi\,C_F\,c_{1,0}\big(2\,c_{2,0}-\,(1-2\,L)\,c_{2,1}\big)\nonumber\\
 &\qquad - c_{4,0} - c_{4,1} L - c_{4,2} L^2 - c_{4,3}L^3\Big) \Big]a_s^4\,.\nonumber
 \end{align}

\end{appendix}

\bibliography{./sources}
\bibliographystyle{JHEP}

\end{document}